\pdfoutput=1

\documentclass[12pt,chapterheads]{ucsd}
\usepackage{amsmath,amssymb,amsthm,mathrsfs,amsfonts,dsfont}
\newtheorem{definition}{Definition}

\newtheorem{theorem}{Theorem}
\newtheorem{lemma}{Lemma}
\usepackage{enumerate}
\usepackage{fancyvrb}
\usepackage{comment}
\usepackage{graphicx}
\usepackage{hyperref,url}
\usepackage{algorithm} 
\usepackage{algpseudocode} 
\usepackage{framed}
\usepackage{caption}
\usepackage{subcaption}

\DeclareMathOperator*{\argmin}{arg\,min}
\DeclareMathOperator*{\argmax}{arg\,max}

\begin{document}

%
%
%
%


\title{Aren't we all nearest neighbors?}

\author{Mark Michel Saroufim}
\degreeyear{2014}

\degreetitle{Master of Science} 

\field{Computer Science}
\chair{Professor Sanjoy Dasgupta}
%

\othermembers{
Professor Charles Elkan\\ 
Professor Yoav Freund\\
Professor Shachar Lovett\\
}
\numberofmembers{4} 

%
\begin{frontmatter}

%
%
\makefrontmatter 

%
%
%
%
%
%
\begin{dedication} 
  To my parents, for the genes and the other stuff too!
\end{dedication}

%
%

%
%
\begin{epigraph} 
  \emph{If people never did silly things \\ nothing intelligent would ever get done.}\\
  ---Ludwig Wittgenstein
\end{epigraph}

%

%
\tableofcontents
\listoffigures  
\listoftables   

%
%
\begin{acknowledgements} 
My experience in graduate school has been one of the most rewarding experiences in my life, who knew that there would be so much to learn across the Atlantic Ocean! I was incredibly fortunate to have interacted with so many great researchers that taught me what it meant to do good research. I'd like to first thank my advisor Sanjoy Dasgupta for mentoring and guiding me with patience through the numerous revisions that this thesis has undertaken, this document would not exist if it wasn't for him (Hopefully, I've inherited an $\epsilon$ of his clarity). I'd also like to thank Charles Elkan for helping me navigate through my first year of graduate school and for being the first person to teach me what machine learning is all about. Finally, I'd like to thank Yoav Freund and Shachar Lovett for readily sharing problems and insight ranging from Quantum Interactive Proofs to revisions of the axioms of Probability Theory using Game Theory. Low Hanging Fruit is good but definitely not as good as their higher counterparts.

\end{acknowledgements}

%
%


%
%
\begin{abstract}
We start with a review of the pervasiveness of the nearest neighbor search problem and techniques used to solve it along with some experimental results. In the second chapter, we show reductions between two different classes of geometric proximity problems: the nearest neighbor problems to solve the Euclidean minimum spanning tree problem and the farthest neighbor problems to solve the $k$-centers problem. In the third chapter, we unify spatial partitioning trees under one framework the meta-tree. Finally, we propose a dual tree algorithm for Bichromatic Closest Pair and measure the complexity of batch nearest neighbor search.
\end{abstract}

\end{frontmatter}

\chapter{Nearest Neighbor}

\section{Introduction}

This chapter will focus almost exclusively on the nearest neighbor problem. The problem is defined by two objects: the first is an ordered pair $(X,d)$ where $X$ is a set of points $x_1, \dots , x_n = X \subset \mathds{R}^D$ and $d$ is a distance function $d : X \times X \rightarrow \mathds{R}$. The second is a query point $q \in \mathds{R}^D$. The nearest neighbor search problem is then formulated as
 \[ \argmin_{x \in X} d(q,x) \]
  Notice for most problems we restrict $d$ to have some sort of structure we call $d$ a metric if for any $p, q, u \in X$, $d$ satisfies the following conditions. 

\begin{itemize}
\item $d(p,q) \geq 0$ with equality if and only if $p = q$
\item $d(p,q) = d(q,p)$
\item $d(p,q) + d(q,u) \geq d(p,u)$
\end{itemize}
If $d$ is a metric we then call the ordered pair $(X,d)$ a metric space

Examples of metrics are the Minkowski norm where 
\[d(q,u) = (\sum_{i = 1}^d |q_i - u_i|^p )^{\frac{1}{p}} \]

By setting $p = 2$ we recover the standard $l_2$ norm otherwise known as the Euclidean norm 
\[d(q,u) = \sqrt{\sum_{i = 1}^D |q_i - u_i|^2} \]
Often we will also be also be interested in using a wider class of distance functions called Bregman Divergences (see appendix). Of particular interest is what is called the KL (Kullback-Leibler) divergence which is a natural distance measure between distributions.

\[d_{KL}(p,q) = \sum_{i = 1}^{D} p_i \ln \frac{p_i}{q_i} \]

It is also worth noting that since good algorithms with guarantees have evaded researchers for nearest neighbor search, a problem that has become increasingly interesting is that of approximate nearest neighbor search or $c$-nearest neighbor. We now ask for a point $x'$ that is not too far ($c$-approximate) from the optimal nearest neighbor of $q$. 

\[ d(q,x') < c  \argmin_{x \in X} d(q,x) \]

In what follows we will make it clear from context which of the two problems we are referring to.

\section{Applications}

\subsection{Traveling Salesman}
One of the first successes of applying nearest neighbor search was in finding an efficient algorithm for the Euclidean traveling salesman problem. As a brief reminder, TSP is one of the quintissential $NP$-hard problems where given an undirected weighted graph $G = (V,E)$ we'd like to find the shortest path that traverses all the vertices $v_i \in V$ without ever revisiting the same vertex twice except for the first one. The complexity of this problem is trivially $\mathcal{O}(n! n)$, $n!$ to enumerate all possible tours and $n$ to check whether they are indeed tours. The greedy nearest neighbor \ref{alg:greedyTSP} for TSP guarantees quickly finding a solution to TSP but has no guarantees on how bad it is from the optimal one.

\begin{algorithm}
\begin{algorithmic}[1]
    \State $Tour = \{ \}$
    \State Pick Arbitrary vertex $v \in V$
    \While{$V \neq \Phi$ }
        \State $v' = \argmin_{v_i \in V - v}d(v',v_i)$
        \State $V = V - v$
        \State $v = v'$ 
        \State $Tour.append(v)$

    \EndWhile
    \State \Return{ $Tour$}

\end{algorithmic}
\caption{Greedy Nearest Neighbor TSP}
\label{alg:greedyTSP}
\end{algorithm}

\subsection{k-nn Classification}
Suppose  you've trained a model on a training set consisting of a dataset $x_i, \dots , x_n = X \in \mathds{R}^d$ where every point $x_i$ is associated with some label $f(x_i) = y_i \in \mathds{N}$. Now you are given a new point $q$ that is not yet associated with a label $f(q)$. A natural way to classify $q$ is then to find its nearest neighbor:
 \[ \argmin_{x \in X} d(q,x) \]
and set $f(q) = f(x)$. The process we've just described is nearest neighbor classification \cite{Alt92}, it's simple and takes $\mathcal{O}(n)$ if we choose to trivially search for a nearest neighbor. Nearest neighbor classification could be very sensitive to outliers but it is easy to make it more robust if we repeat the process $k$ times and use a simple voting scheme (majority) to decide the label of $q$. More generally we can associate a prior weight $w_i$ on the $i$'th label and multiply that by the number of nearest neighbors $n_i$ of $q$ that were labeled $i$.
\[f(q) = \argmax_i \{w_i n_i \} \]

There also exists schemes where the voting power of a point is inversely proportional to its distance from the query point $q$. As an example below in figure \ref{fig:iris} we project the iris dataset onto a two dimensional plane, set the number of nearest neighbors $k = 18$ and then use a kd tree to find them, the voting power of every point is inversely proportional to the distance from the query point $q$.

\begin{figure}[h!]
  \centering
   \includegraphics[scale=0.5]{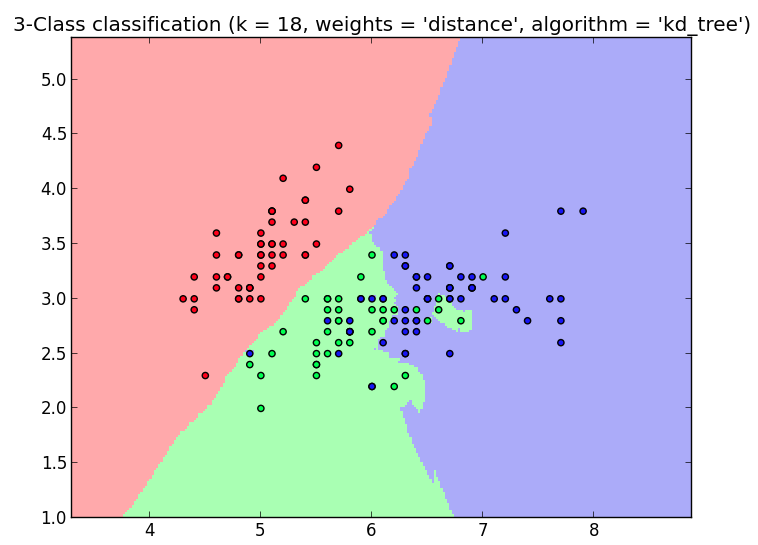}
      \caption{$k$-nn classification on iris dataset}
      \label{fig:iris}
\end{figure}

\subsection{N-body problems}
N-body problems have fascinating origins in Newtonian mechanics, specifically suppose we are trying to understand the interaction between $N$ spatial bodies. Newton's law of universal gravitation tells us that the bodies with masses $m_1$ and $m_2$ at a distance $r$ from each other, attract each other with a force \[F =G \frac{m_1 m_2}{r^2} \] where $G$ is a gravitational constant. This seems easy enough but now suppose we have more than two bodies and now have to deal with $N$ such bodies. Since the force $F$ between two bodies is stronger the closer they are we can choose to look at the $k$ closest bodies and compute $F$ between a body $q$ and its $k$ nearest neighbors. Why not just compute $F$ to all bodies? Astrophysicists estimate the number of stars in the milky way alone to be a 100 billion, this puts us in a range where $\mathcal{O}$ starts to matter.

\subsection{Single Linkage Clustering}
Clustering is a fairly ubiquitous problem in machine learning it can be thought of as a dimensionality reduction problem where we try to reduce the size of a dataset from $n$ to $k$ where $k << n$. The remaining $k$ points are called cluster centers. An approach to clustering uses nearest neighbor search as its main subroutine, UPGMAM \cite{LL98} clustering is an agglomerative hierarchical clustering technique where we start with $n$ cluster centers (one for each data point) we then find $\arg_y min_{x \in X} d(x,y)$ and merge $y$ to the cluster center that $x$ is assigned to and repeat this process until all points belong to the same cluster. We can also choose to terminate before all the points belong to the same cluster to get $k$ cluster centers instead of 1.

\section{Algorithms}
Let's restate the nearest neighbor problem \[ \argmin_{x \in X} d(q,x) \] The trivial solution will compare $d(q,x)$ for all $x \in X$ and pick the smallest one. This approach takes $\mathcal{O}(n)$ time and is fine given that we are only doing a nearest neighbor query once. However, assuming that we have not just one but $q_1 , \dots , q_m = Q$, then the naive approach will now take $\mathcal{O}(n^2)$ which is extremely large for our purposes. Instead we'd like to take a similar approach to the one seen in sorting algorithms where we incur some sort of cost $P(n)$ to build a nearest neighbor data structure and then answer nearest neighbor queries in $T(n) = o(n)$. This results in a total running time of $\mathcal{O}(P(n) + |Q|T(n))$ where $|Q|$ is the size of the query set.

\subsection{Tree Based Techniques}

In what follows we will introduce several spatial partitioning schemes for the purpose of fast nearest neighbor retrieval. All the below examples are binary spatial partitioning trees where the two subtrees of any given node are determined using some sort splitting rule to hierarchichally divide up the dataset $X$ into a binary spatial tree $T_X$ using a $BuildTree(X)$ routine after which we answer nearest neighbor queries of points $q$ using $NNS(q,T_X)$

\begin{enumerate}
\item Comprehensive Search (shown in  \ref{alg:cmpnns}) where we are conservative about pruning out subtrees and potentially could visit all $\mathcal{O}(n)$ points.
\item Defeatist Search where at every iteration we prune out an entire subnode of the binary partition tree so that we visit at most $\mathcal{O}(\log n)$ points
\end{enumerate}

Even though Comprehensive Search always finds the true nearest neighbor (Defeatist Search has no such guarantees), its time complexity is the same as a trivial linear scan. For that reason Defeatist search is instead used in practice and unless explicitly mentioned all nearest neighbor search schemes in this text will be of the Defeatist nature. The difference as far as implementation is concerned is minimal, below is the Comprehensive Nearest Neighbor shown in algorithm \ref{alg:cmpnns} .To recover Defeatist Nearest Neighbor Search we simply omit the the last return statement.

\begin{algorithm}

\begin{algorithmic}[1]
\Procedure{NNS}{$q,T_X$}

\If{$T_X$ is a leaf}
    \State \Return{ $\argmin_{x' \in X}d(q,x')$ }
\EndIf

\If{$x \in Left(T_X)$ and $x \not \in Right(T_X)$}
    
    \State \Call{NNS}{$q,Left(T_X)$}

\ElsIf{$x \not \in Left(T_X) == NULL$ and $x \in$ $Right(T_X)$}
    
    \State \Call{NNS}{$q, Right(T_X)$}

\Else
     \State \Return{ $\argmin_{x \in X}$ (\Call{NNS}{$q,Left(T_X)$} , \Call{NNS}{$q,Right(T_X)$}}

\EndIf

\EndProcedure

\end{algorithmic}
\caption{Comprehensive Nearest Neighbor Search}
\label{alg:cmpnns}
\end{algorithm}

\subsubsection{kd-trees}

 The canonical example of a spatial partitioning tree is the $kd$-tree \cite{Bent75} ($k$ dimensional tree) which divides the dataset by the median of one of the coordinates essentially recursively splitting the size of the data-structure by 2 at every level of the tree \ref{alg:kd}. With the data so cleanly separated, it's easy to navigate the tree $T_X$ looking for the nearest neighbor of a query point $q$, pseudocode here \ref{alg:kdnns}. Because $q$ can only be on one or the other side of the median of one of the dimensions, the size of the search space is halved at every iteration. Once we reach a leaf node we can naively compare the distance between $q$ and all points $p \in leaf$ and return the smallest such distance.  Constructing the $k$-d tree takes $T(n) = \mathcal{O}(n) +  2T(n/2) = \mathcal{O}(n \log n)$. In the pseudocode below we will use $x^i$ to denote the i'th coordinate value of $x$.

\begin{algorithm}

\begin{algorithmic}[1]
        \State Find Median dimension of some dimension $i$ $med^i(X)$ (typically max-variance) 
\Repeat
        \State For all $x \in X$

        \If{$x^i > med^i(X)$}
            \State Add $x$ to $Right(T_X)$
        \Else 
            \State Add $x$ to $Left(T_X)$
        \EndIf
\Until{All $x \in X$ have been considered}
        \State $kd(Left(X))$
        \State $kd(Right(X))$
\end{algorithmic}
\caption{Constructing a $k$-d tree $kd(X)$}
\label{alg:kd}
\end{algorithm}

Queries for nearest neighbor of a point $q \in Q$ from $R$ can then be answered in $\mathcal{O}(\log n)$ using  algorithm \ref{alg:kdnns}.

\begin{algorithm}
\caption{Nearest Neighbor Search using $k$-d tree $NNS(q,T_X)$}
\label{alg:kdnns}
\begin{algorithmic}[1]
\State Sort $med^i(X)$ for all $X$
\State Set $i = 0$
\Repeat
    \If{$T_X$ is a leaf}
    \State return $\argmin_{x' \in T_X}d(q,x')$
\EndIf
\If{$q^i > med^i(X)$}
    \State $i = i + 1$
    \State $X = Right(T_X)$
    \State $NNS(q,Right(T_X))$
 \Else
     \State $i = i + 1$
     \State X = $Left(T_X)$
    \State $NNS(q,Left(T_X))$

\EndIf
\Until{$i = D$ or $T_X$ is a leaf}

\end{algorithmic}
\end{algorithm}

Unfortunately the analysis above is flawed in fact there is no prior guarantee that our data can be so cleanly separated into halves! In the worst case we can expect to recurse on all nodes in the tree, bumping up the query time for NNS from $\mathcal{O}(\log n)$ to $\mathcal{O}(n)$. Might as well just naively search for the nearest neighbor. The problems stem for the inadequacy of $k$-d trees to give any structure to high dimensional spaces. More generally, any spatial tree that will use coordinate directions will be inadequate for nearest neighbor search. We will consider an example proposed in \cite{DS14}, $q$ is our query point and the dataset is $x_1, \dots , x_n = X$. Take $x_1 = (1 , \dots , 1)$ and for the other points $x_i \in X - \{x_1 \}$ pick a random coordinate uniformly at random and set its value to $M$ where $M$ is some large constant and set the other points to random values picked uniformly from $[0,1]$, we can see here that any coordinate split will create a large separation between $q$ and $x_1$

\subsubsection{PCA Trees}  
The idea behind PCA trees \cite{Ich12} is similar to that of $k$-d trees but instead of splitting according to the medians of the dimensions we split according to the principal eigenvectors of the data's covariance matrix. In fact we can simply use the same construction scheme as $k$-d trees but change the split rule to the location of our points relative to the principal eigenvector and of course apply it recursively to the left and right children as shown in algorithm \ref{alg:pcatree}.

\begin{algorithm}
\caption{Nearest Neighbor Search using a PCA tree $NNS(q,T_X)$}
\label{alg:pcatree}
\begin{algorithmic}[1]
        \State Sort in descending order the principal components $\lambda_1, \dots , \lambda_k$ with corresponding eigenvectors $u_1, \dots, u_k$ of covariance matrix $\Sigma$
 
\State Sort $med^i(X)$ for all $i[$
\State Set $i = 0$
\Repeat
\If{$q . u_i > 0$}
    \State $i = i + 1$
    \State $NNS(q,Right(X))$
 \Else
     \State $i = i + 1$
    \State $NNS(q,Left(T_X))$

\EndIf
\Until{$i = k$}

\end{algorithmic}
\end{algorithm}

PCA trees unfortunately can be fooled even by relatively simple datasets, consider for instance an arrangement of points $x_1 , \dots , x_n \in \mathds{R}^2$ organized in two parallel lines with the first coordinate axis. The distance between two successive points on the first line is 2 and the distance between two successive points on the second line is 1. The distance between the two lines is $4$ Since there is a large amount of data parallel to the first axis, the first axis will indicate the direction of the first principal component. Once we project our data onto this principal component we will interleave points from the two parallel lines onto the same line and if we set $q$ to be any point in the dataset we will always get an incorrect nearest neighbor.

\subsubsection{Random Projections Trees}
Random Projection \cite{DF08} Trees are essentially $k$-d trees where the splitting rule is done according to a random direction in the dataset (as shown in algorithm \ref{alg:rptree}) instead of the max variance coordinate like in $k$-d trees. The construction of the datastructure is outlined below and nearest neighbor search is identical to search in a $k$-d tree.

\begin{algorithm}

\begin{algorithmic}[1]
        \State i = 1
        \State Draw uniformly at random a direction $w$ from $D - i$ dimensional sphere
        \Repeat
\If{$q . w > 0$}
    \State $i = i + 1$
    \State $NNS(q,Right(T_X))$
 \Else
     \State $i = i + 1 $
    \State $NNS(q,Left(T_X))$

\EndIf
\Until{$i = k$}
  
\end{algorithmic}
\caption{Nearest Neighbor Search using rp tree $NNS(q,T_X)$ }
\label{alg:rptree}
\end{algorithm}

In the paper, the authors show how rp-trees can adapt to the intrinsic dimension of a dataset where the intrinsic dimension is defined using the local covariance dimension which is a measure of how well the covariance of the data is captured using $d$ eigenvalues of the covariance matrix where $d << D$ and $D$ is the actual dimension of the dataset

\subsubsection{2-means Trees}
This tree has a fairly simple splitting rule \cite{BB95}, instead of splitting along a random diretion or the max variance direction, we divide up the dataset into two clusters and set the split rule as the midpoint between the two cluster centers. To obtain the two clusters $C_1$ and $C_2$, their respective centers $\mu_1$ and $\mu_2$ and the points points assigned to then $x \in C_j$ we run the $k$-means algorithm on our dataset and set $k = 2$
\[ \argmin_{C_1,C_2} \sum_{j = 1}^{2} \sum_{x \in C_j} \| x - \mu_j \| \]

\subsubsection{Spill Trees}
A spill tree is not a spatial partitioning tree in of itself because any of the trees we've discussed so far can also be spill trees. A common problem among spatial partitioning trees is that points near the decision boundary of splits can be seperated from their neighbors \cite{DS14} \cite{ML11}. However, by allowing spill i.e overlap between the right and left subtree of a given node we can limit the problems associated with a hard partitioning (See algorithm \ref{alg:spill} . We can create a soft partitioning by maintaining two decision boundaries $split + \tau$ and $split - \tau$. If a given datapoint is to the left of $split - \tau$ then we assign it to the left subtree. The interesting case is when a datapoint lies between the two decision boundaries, when that happens we simply assign the point to both subtrees. It is worth noting that allowing spill means we will have duplicate points across different leaves meaning we will slow down nearest neighbor queries. Spill trees serve no real purpose if we're doing a full search but can dramatically improve the results of defeatist search at an extra time cost. So adjusting the size of spill essentially gives us an easy way to set a tradeoff between the running time of nearest neighbor queries and the quality of the found nearest neighbors.

\begin{algorithm}

\begin{algorithmic}[1]
        \State i = 1
        \State Draw uniformly at random a direction $w$ from $D - i$ dimensional sphere
        \Repeat
\If{$q . w > \tau$}
    \State $i = i + 1$
    \State $NNS(q,Right(T_X))$
 \ElsIf{$q . w < - \tau$}
     \State $i = i + 1 $
    \State $NNS(q,Left(T_X))$
  \Else
    \State $i = i + 1$
    \State \Return{$\argmin_x NNS(q,Left(T_X)),NNS(q,Right(T_X))$}
\EndIf
\Until{$i = k$}
  
\end{algorithmic}
\caption{Nearest Neighbor Search using spill rp tree $NNS(q,T_X,\tau)$ }
\label{alg:spill}
\end{algorithm}

We summarize the above techniques by first stating the general spatial partitioning tree algorithm regardless of the splitting rule used. As a reminder, a splitting rule is a function
 $f: X \rightarrow \{0 , 1 \}$ that assigns a $D$ dimensional point to one of two subsets of nodes Left or Right. We include table \ref{summary} that shows the different partitioning rules as they were presented in \cite{ML11}

\begin{table}
    \caption {Summary of Split rules for different spatial trees}
    \label{summary}
    \centering
    \begin{tabular}{l l l l}
    $k$d-tree & pca-tree & rp-tree & 2means-tree \\ \hline
    $\argmax_{e_i} \sum_{x} (e_i^T(x - \mu))^2$           & $\argmax_v v^T \Sigma v $ s.t $\|v_2 \| = 1$       &  $S^{D - 1}$      & $ \mu_1 - \mu_2 $            \\
    \end{tabular}
    
\end{table}

\subsection{Hashing Based Techniques}
Another completely different approach to solving nearest neighbor problems is Locality Sensitive Hashing (LSH) \cite{IM98}. Before we introduce the framework we will introduce some basic terminology.
Locality Sensitive Hashing is defined over a family of hash functions.

\begin{definition}
We call a family $\mathcal{H}$, $(R,cR,P_1,P_2)$-sensitive if given two points $p,q \in \mathds{R}^D$
\begin{enumerate}
\item if $\|p-q \| \leq R$ then $\Pr_{\mathcal{H}}[h(q)=h(p)] \geq P_1$
\item if $\|p-q \| \geq cR$ then $\Pr_{\mathcal{H}}[h(q) = h(p)] \leq P_2$
\end{enumerate}

\end{definition}

The actual \ref{alg:lsh} then chooses $L$ hash functions composed of a concatenation of $k$ hash functions from the family $\mathcal{H}$ to achieve performance close to a constant factor away from the theoretical optimal performance of hash based schemes. Now analogously to the tree based approaches we first construct a datastructure in this case the hash tables

\begin{algorithm}[!h]
\caption{Locality Sensitive Hashing $LSH(\mathcal{H} , L , R , c )$}
\label{alg:lsh}
\begin{algorithmic}[1]
\State Draw $h_{i,j}$ from a family of hash function $\mathcal{H}$
\State Choose $L$ hash functions $g_1, \dots , g_L$ where $g_j = (h_{1,j}, \dots h_{k,j})$
\State For every point $x_i \in X \subset \mathcal{R}^D$, hash it into $L$ different hash tables by evaluating $g_j(x_i)$
\State $j = 0$
\While{$j \leq L$}
    \State Retrieve points hashed into $g_j(q)$
    \State Compute the distance to all retrieved points to $q$, if any of the points is a $cR$ nearest neighbor then return it and terminate
    \State $j = j + 1$

\EndWhile

\end{algorithmic}

\end{algorithm}

\section{Experiments}
In this section we validate experimentally which spatial tree data structures perform well on real data. We will use $n$ to denote the number of samples, $D$ to denote the dimensionality of the dataset and $c$ to denote the number of possible labelings.

\begin{enumerate}
\item Pima Indians diabetes dataset $n = 768$, $d = 10$, $c = 2$
\item OptDigits dataset $n = 5620$,$d = 64$, $c = 10$ 
\end{enumerate}

\begin{figure}
        \centering
        \begin{subfigure}[b]{0.3\textwidth}
                \includegraphics[width=\textwidth]{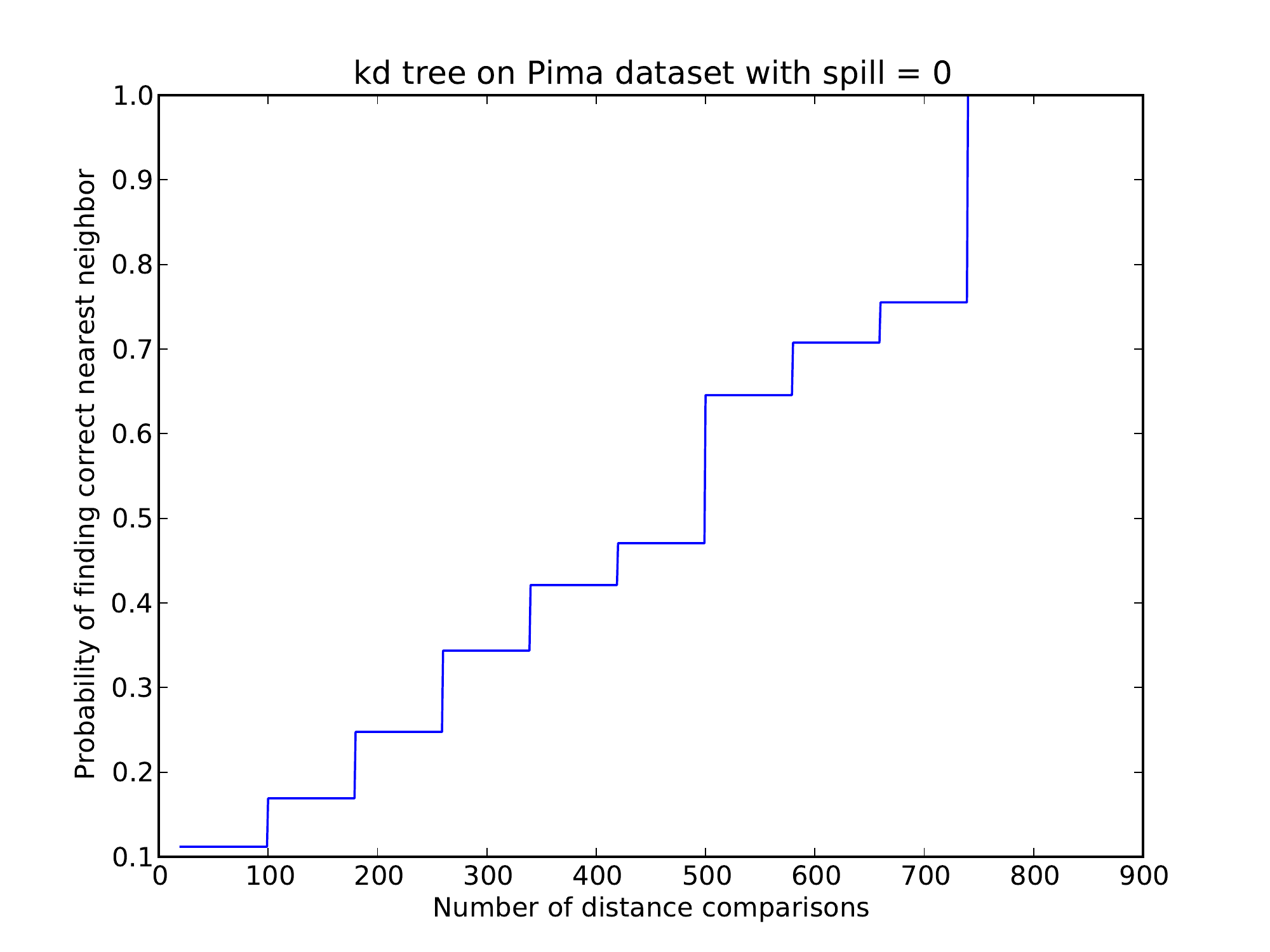}
        \end{subfigure}%
        ~ 
        \begin{subfigure}[b]{0.3\textwidth}
                \includegraphics[width=\textwidth]{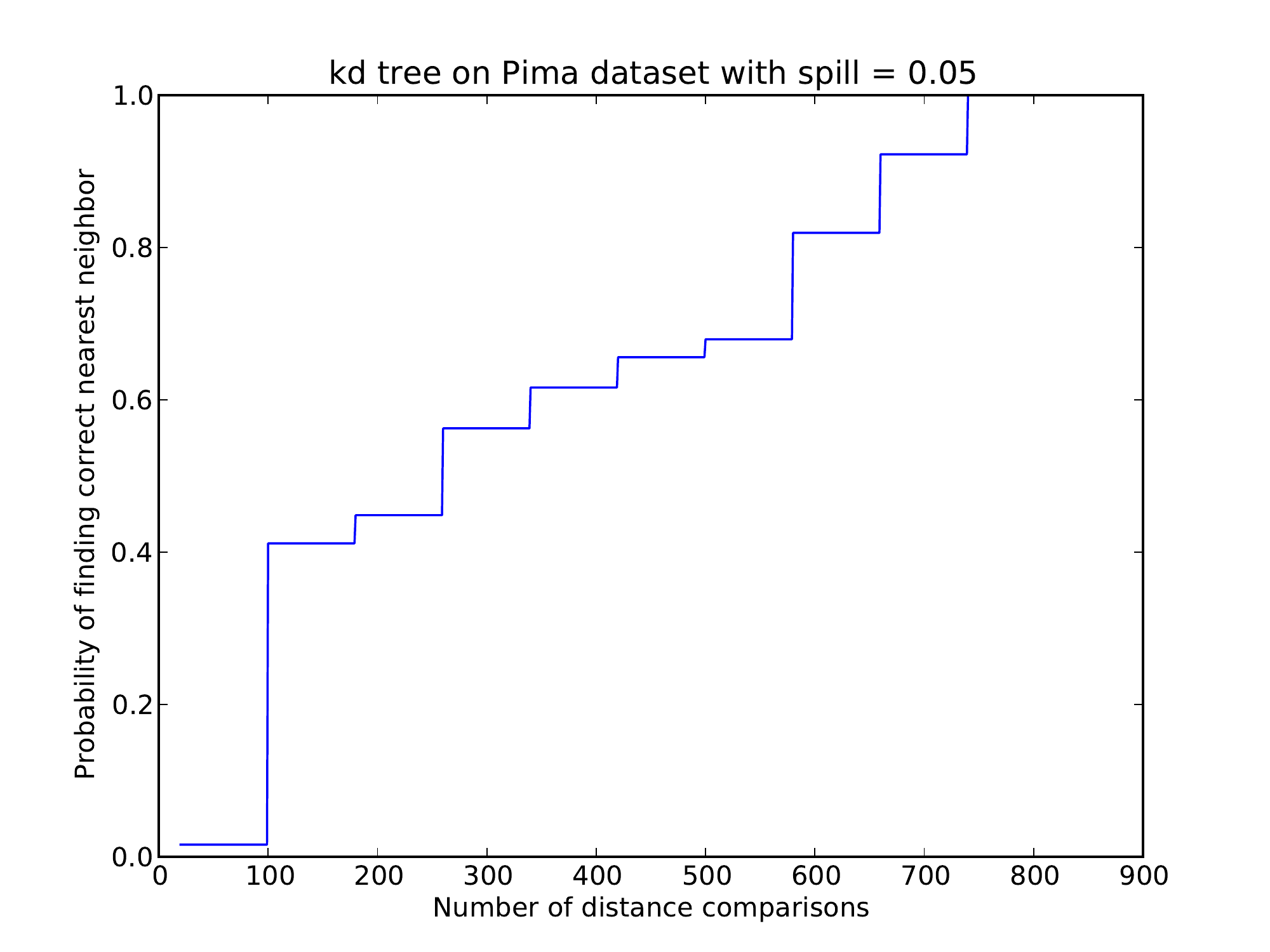}
              
        \end{subfigure}
        ~ 
        \begin{subfigure}[b]{0.3\textwidth}
                \includegraphics[width=\textwidth]{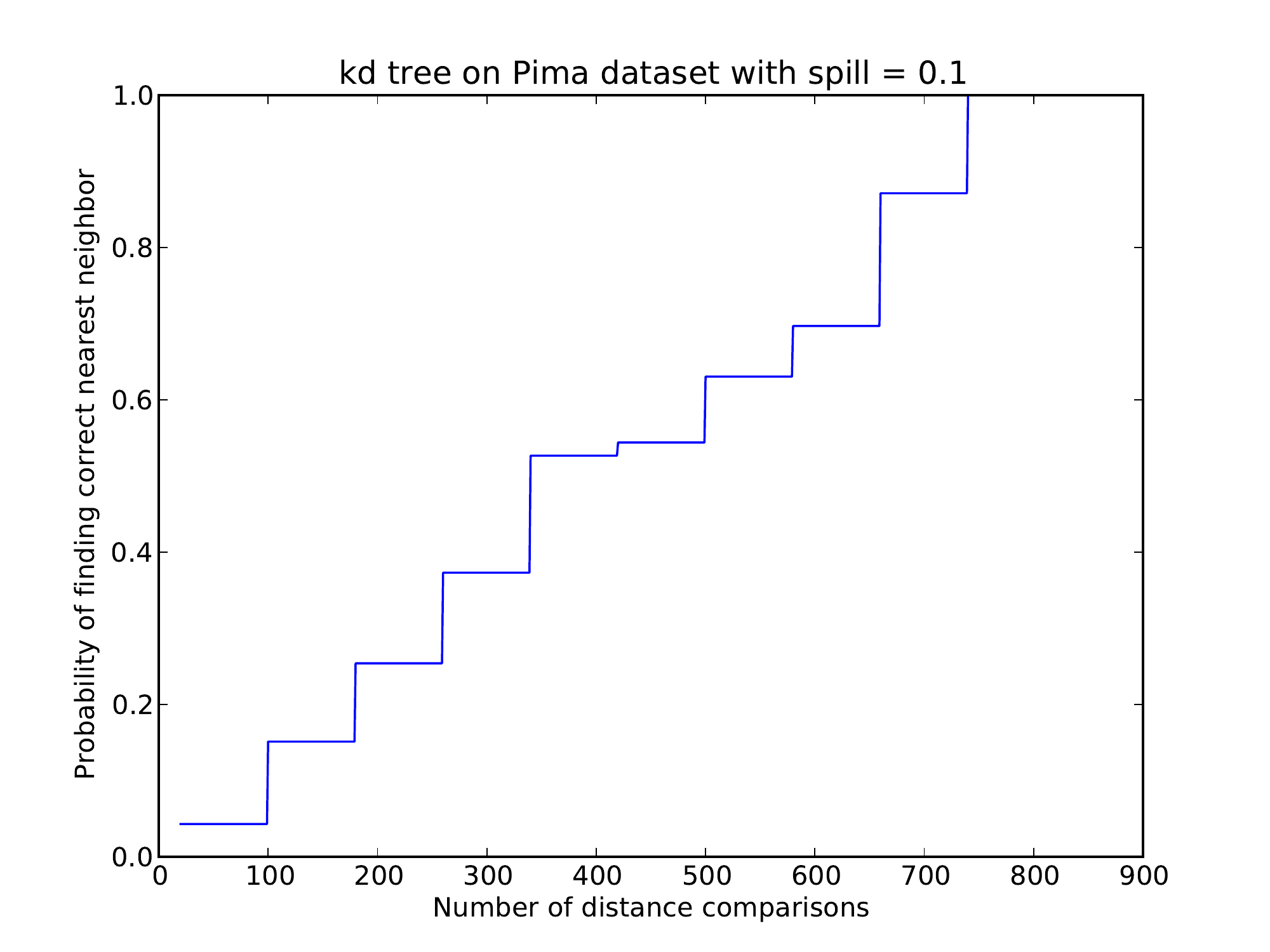}
             
        \end{subfigure}

        \begin{subfigure}[b]{0.3\textwidth}
                \includegraphics[width=\textwidth]{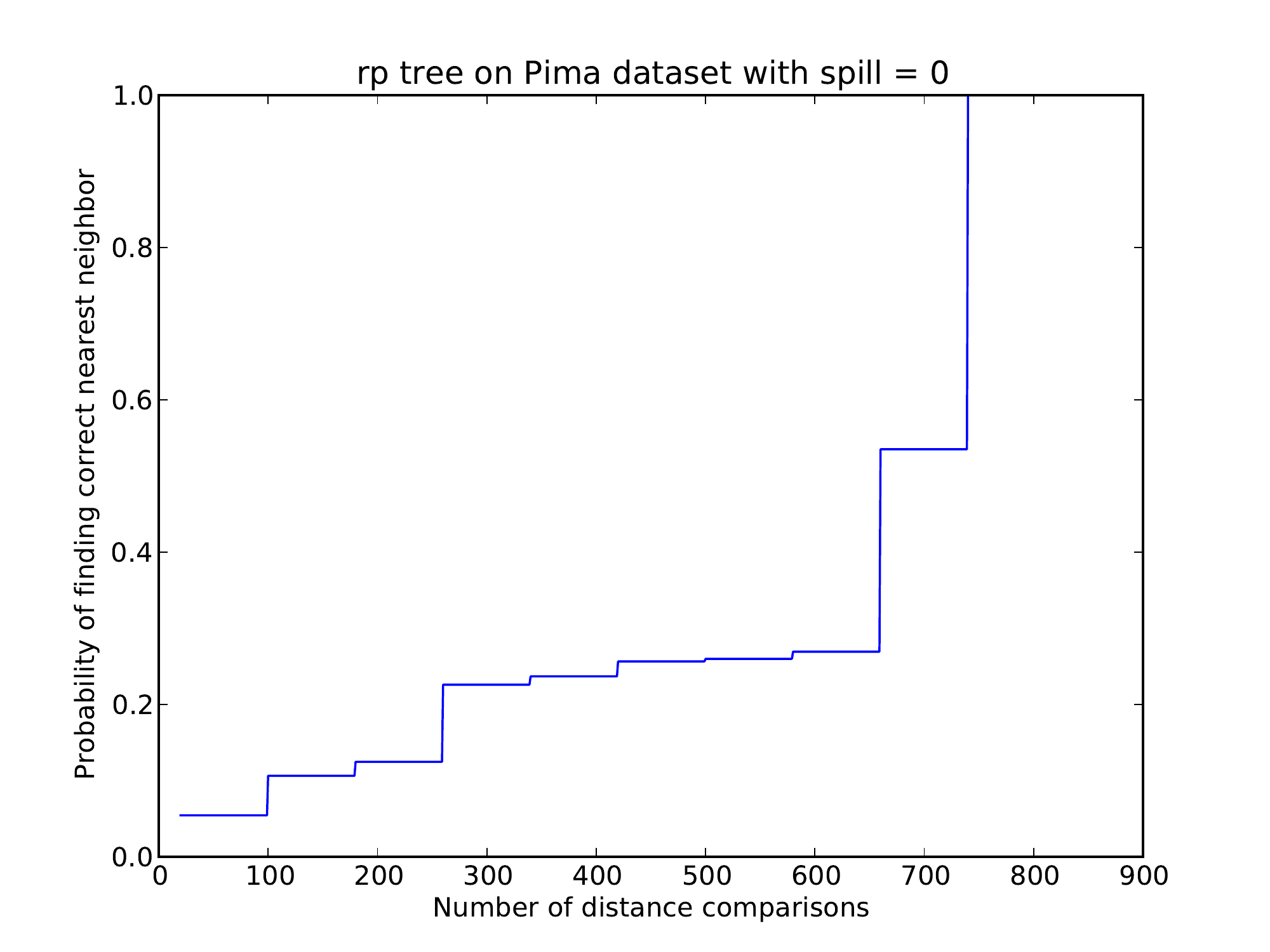}
             
        \end{subfigure}
        ~
        \begin{subfigure}[b]{0.3\textwidth}
                \includegraphics[width=\textwidth]{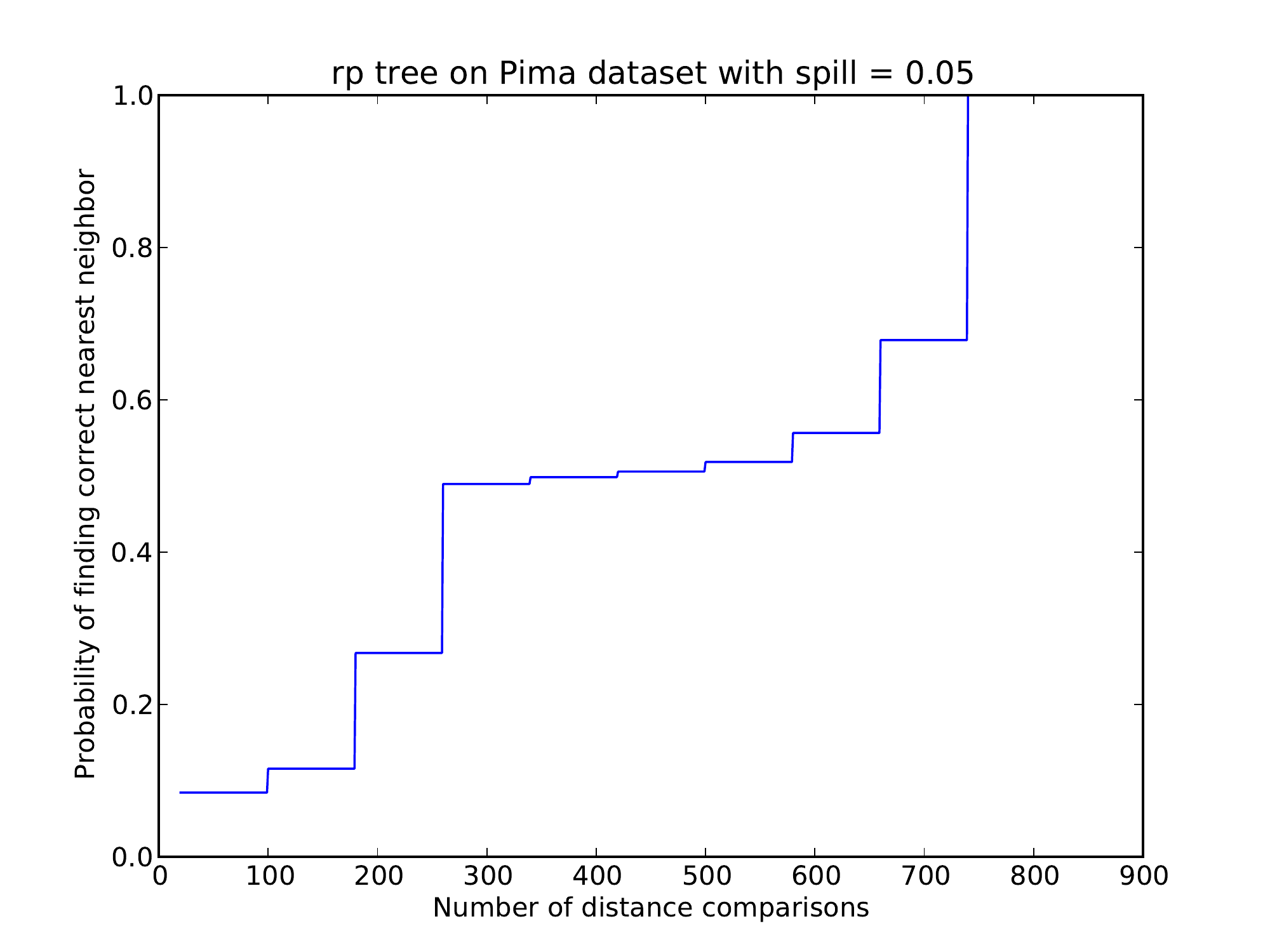}
             
        \end{subfigure}
        ~
        \begin{subfigure}[b]{0.3\textwidth}
                \includegraphics[width=\textwidth]{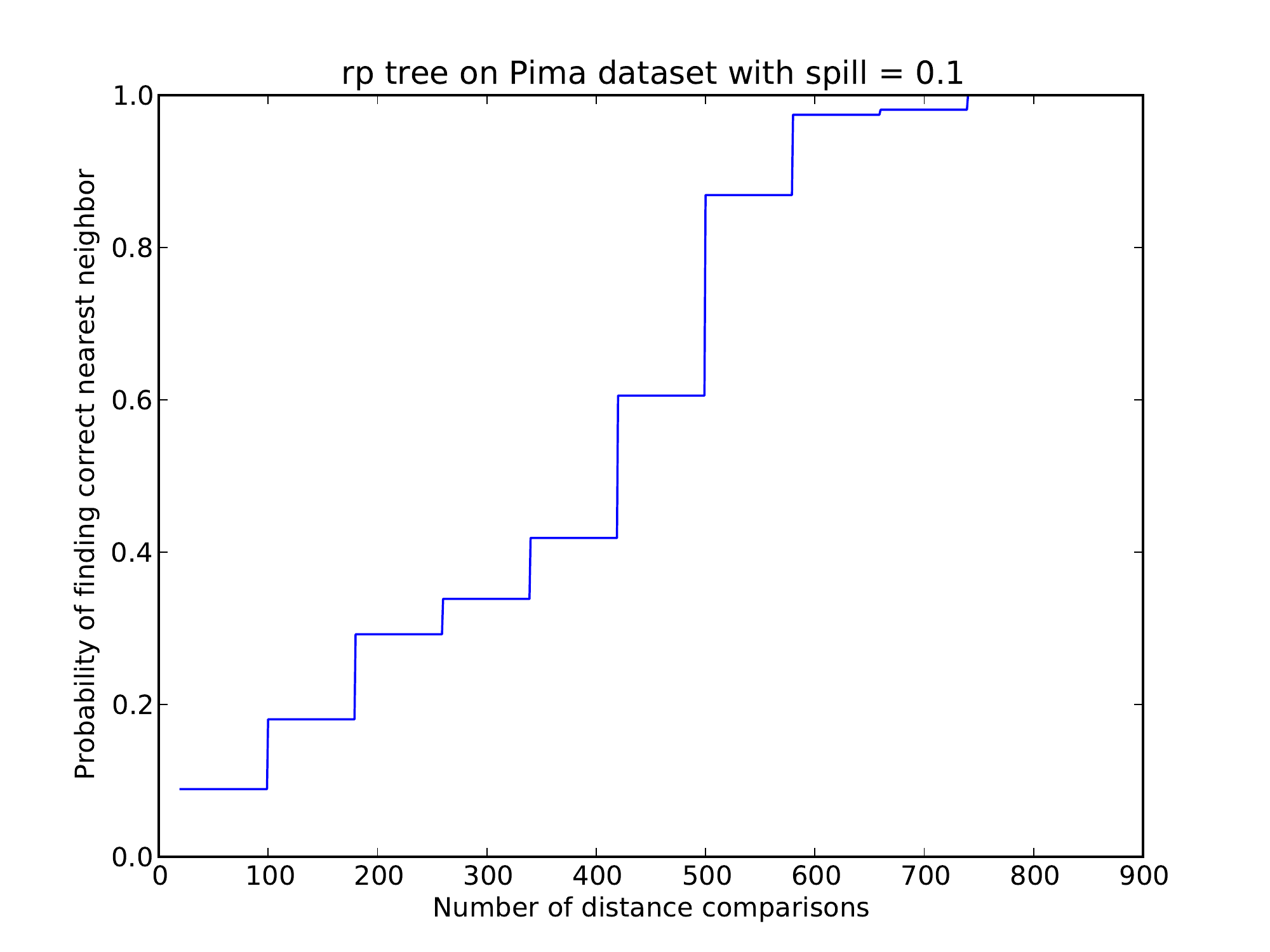}
             
        \end{subfigure}

        \begin{subfigure}[b]{0.3\textwidth}
                \includegraphics[width=\textwidth]{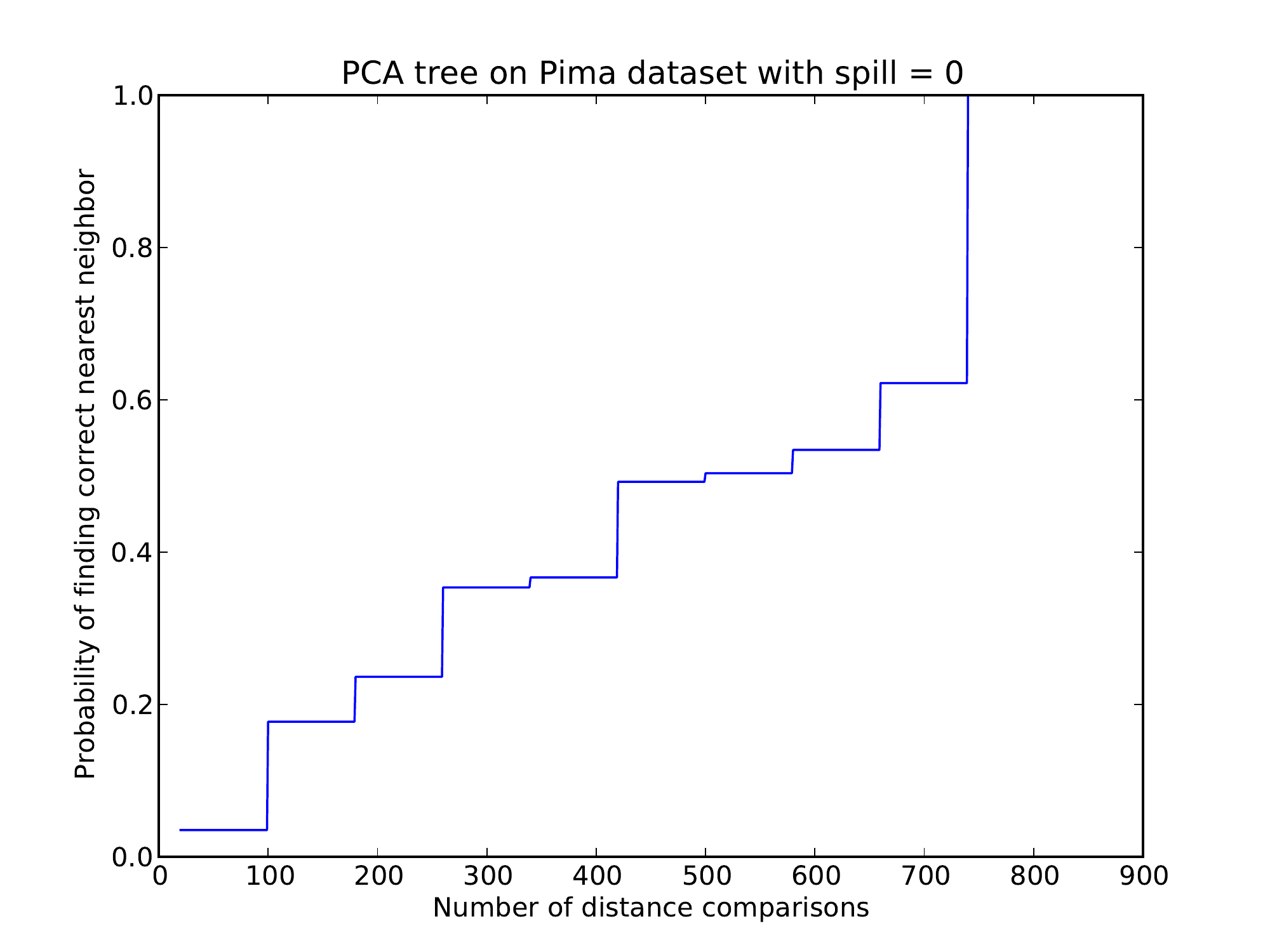}
        \end{subfigure}%
        ~ 
        \begin{subfigure}[b]{0.3\textwidth}
                \includegraphics[width=\textwidth]{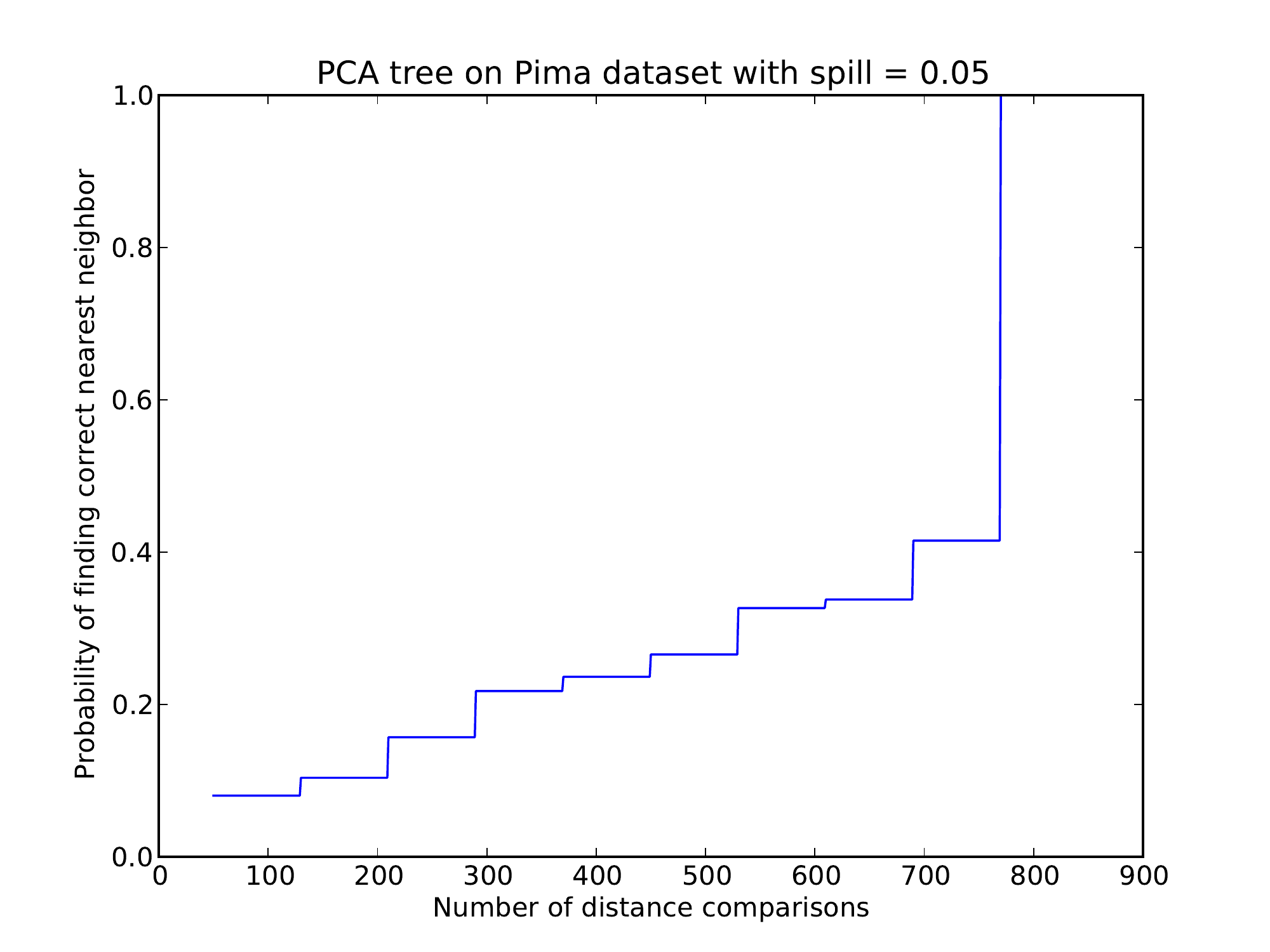}
              
        \end{subfigure}
        ~ 
        \begin{subfigure}[b]{0.3\textwidth}
                \includegraphics[width=\textwidth]{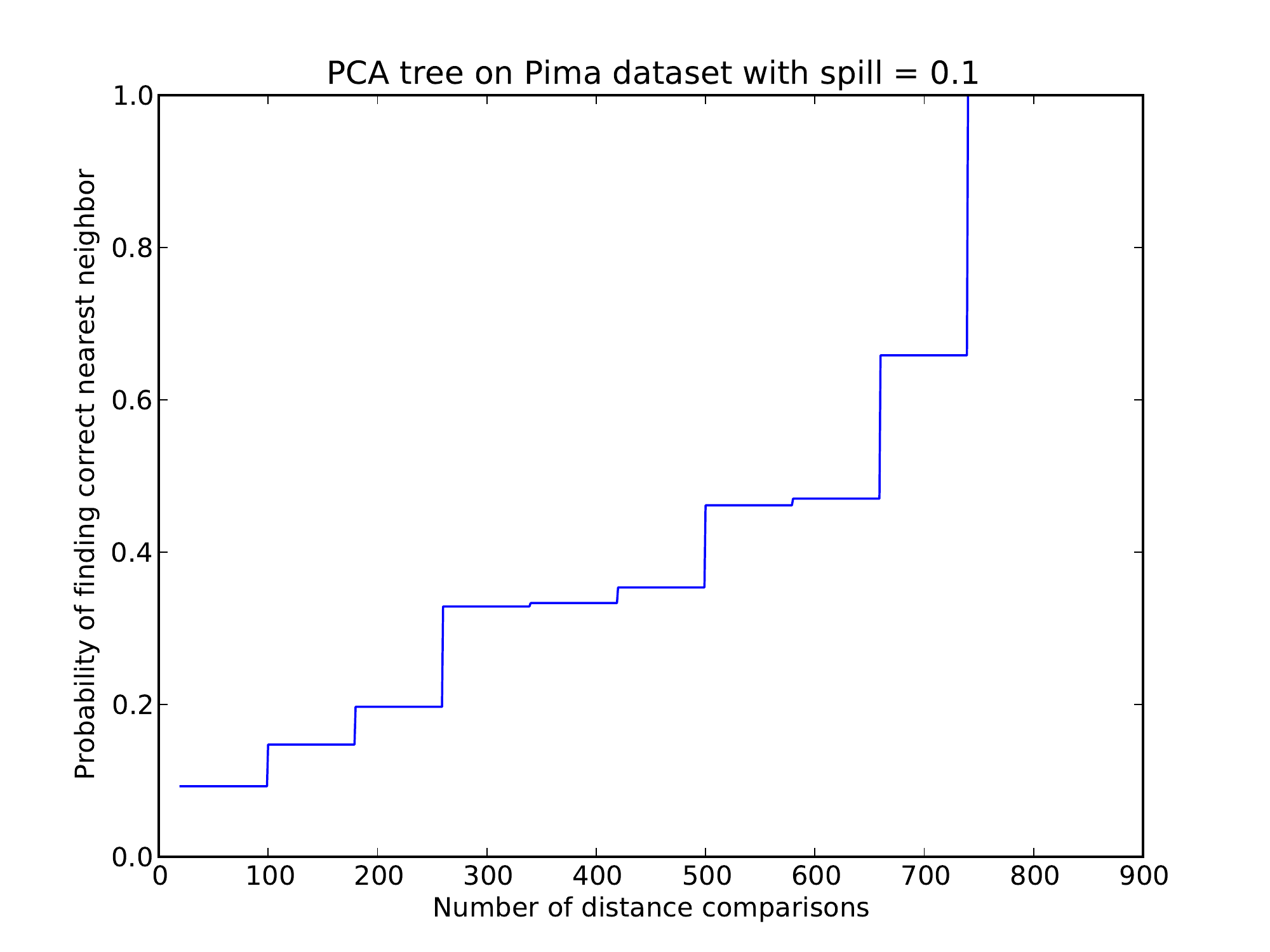}
             
        \end{subfigure}

        \begin{subfigure}[b]{0.3\textwidth}
                \includegraphics[width=\textwidth]{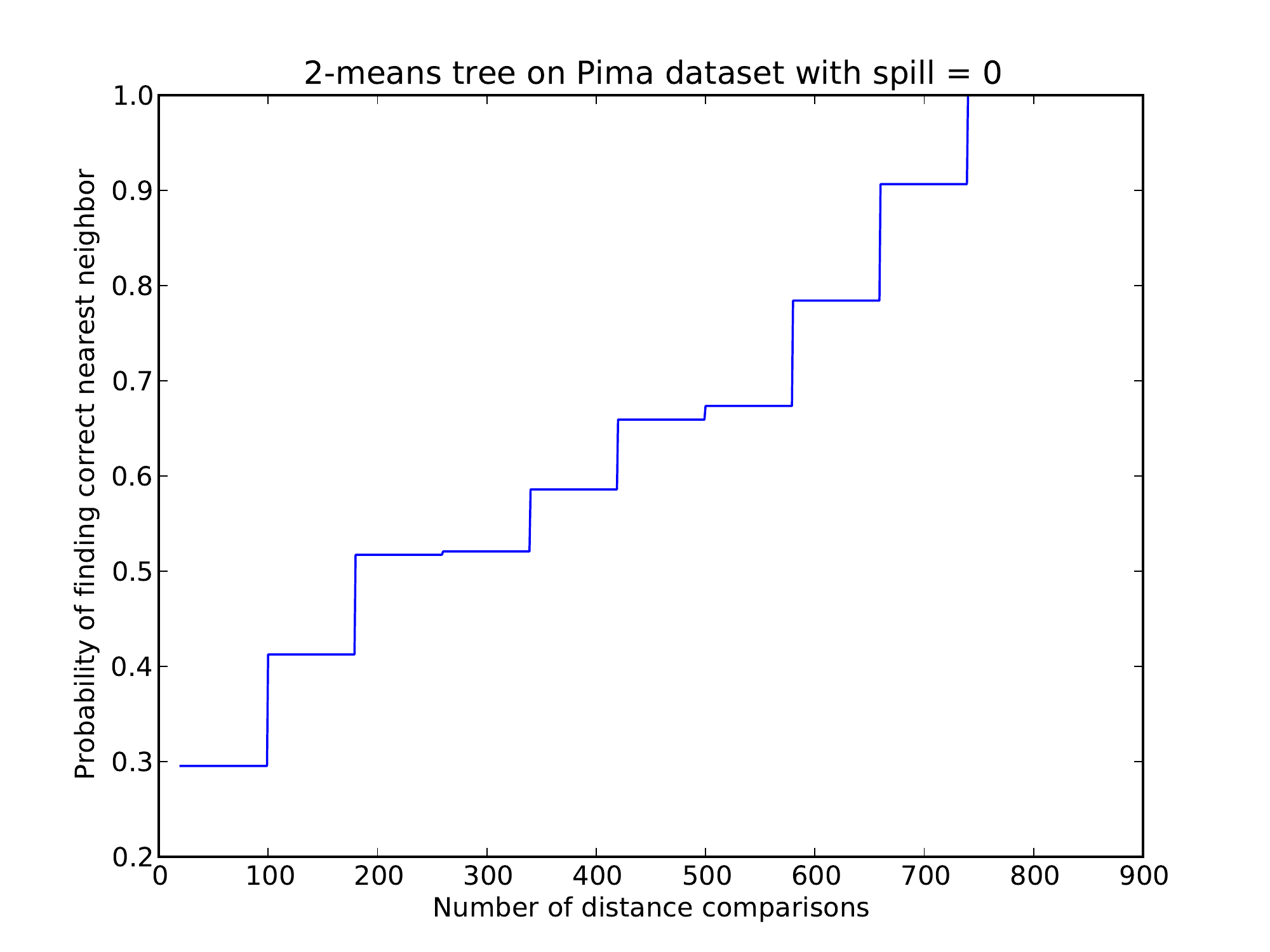}
             
        \end{subfigure}
        ~
        \begin{subfigure}[b]{0.3\textwidth}
                \includegraphics[width=\textwidth]{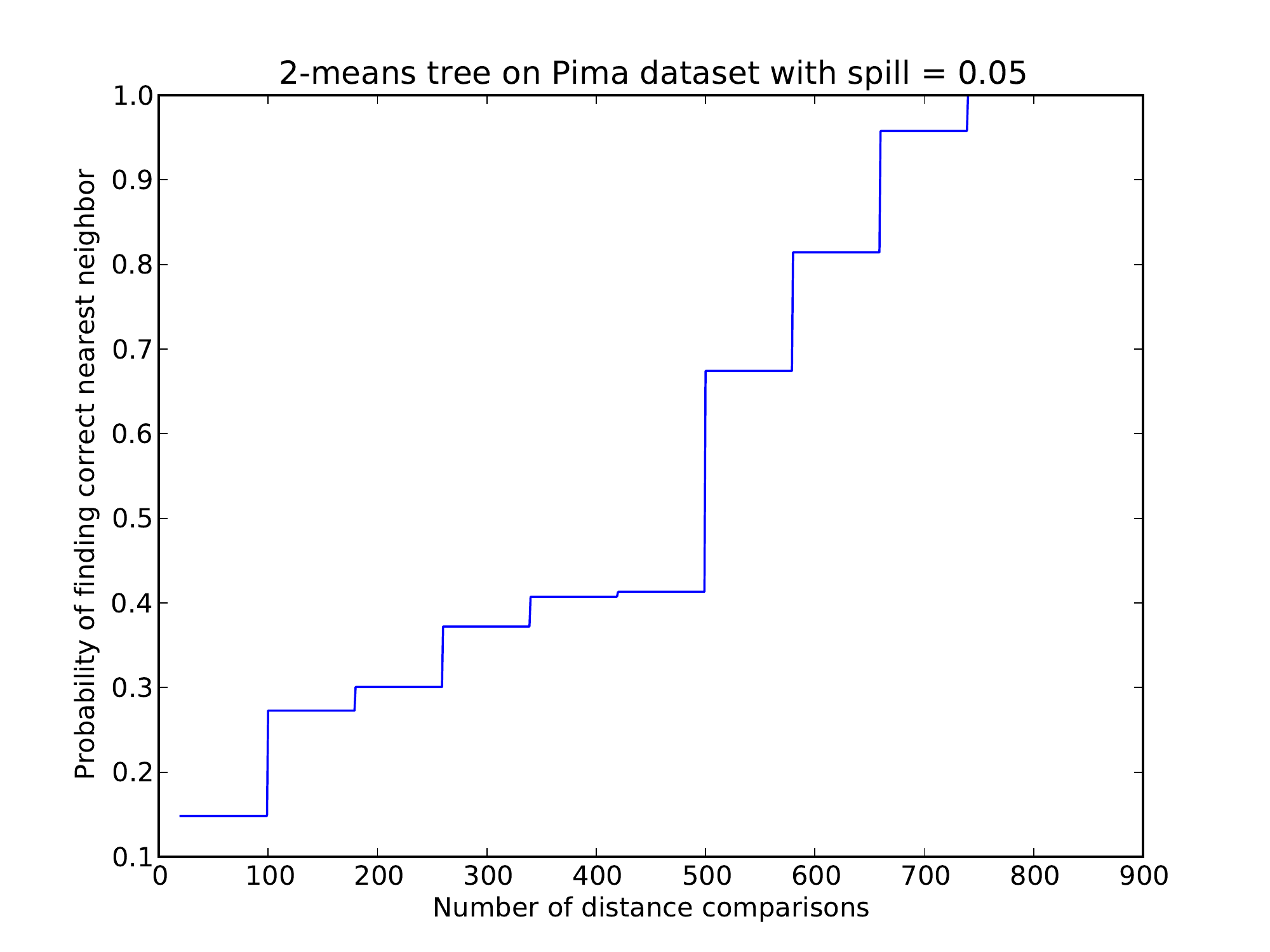}
             
        \end{subfigure}
        ~
        \begin{subfigure}[b]{0.3\textwidth}
                \includegraphics[width=\textwidth]{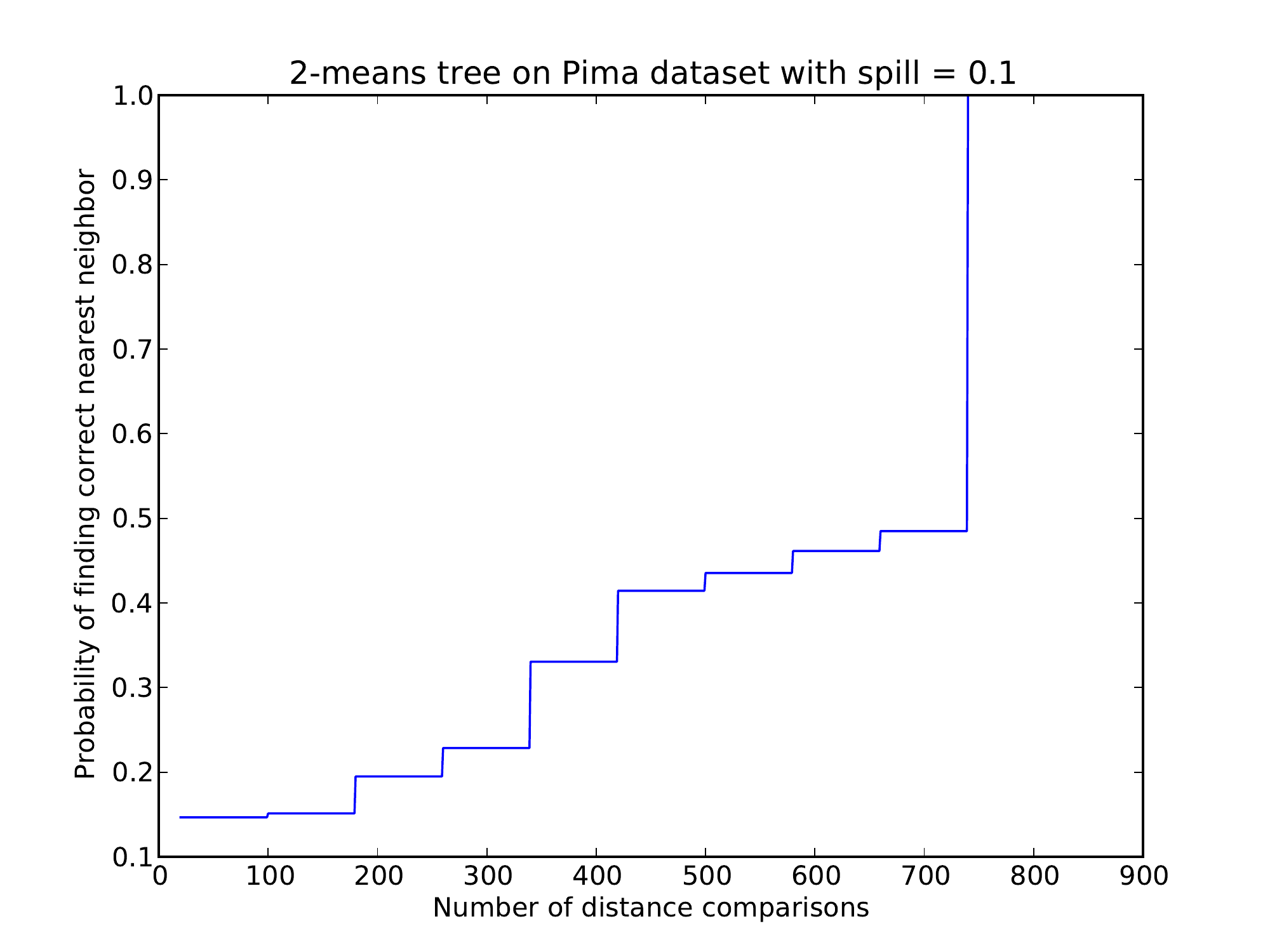}
             
        \end{subfigure}

        \caption{Probability of finding nearest neighbor vs number of comparisons for different spatial trees and tree configurations for the Pima dataset}
        \label{fig:pima}
\end{figure}

\begin{figure}
        \centering
        \begin{subfigure}[b]{0.3\textwidth}
                \includegraphics[width=\textwidth]{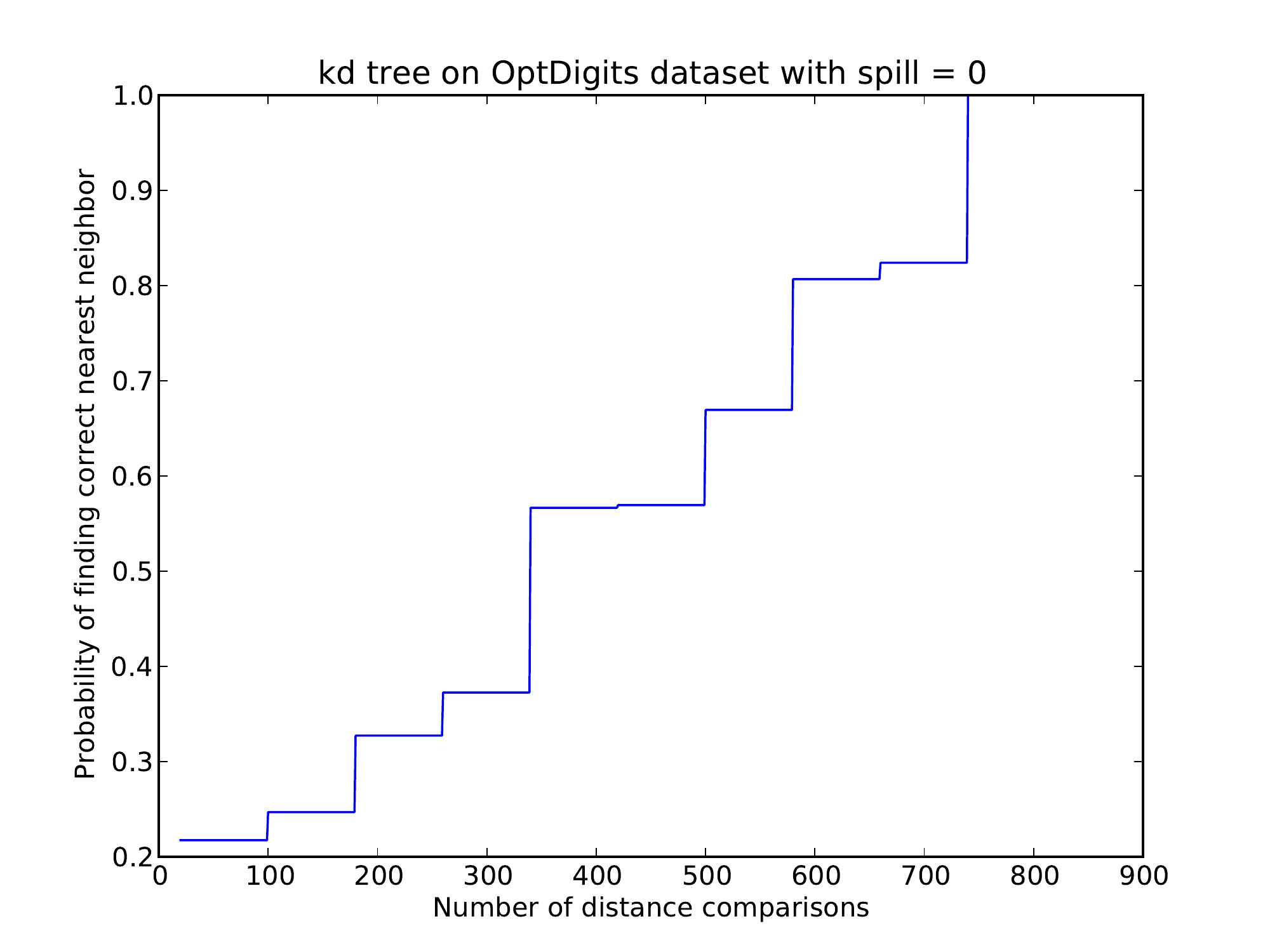}
        \end{subfigure}%
        ~ 
        \begin{subfigure}[b]{0.3\textwidth}
                \includegraphics[width=\textwidth]{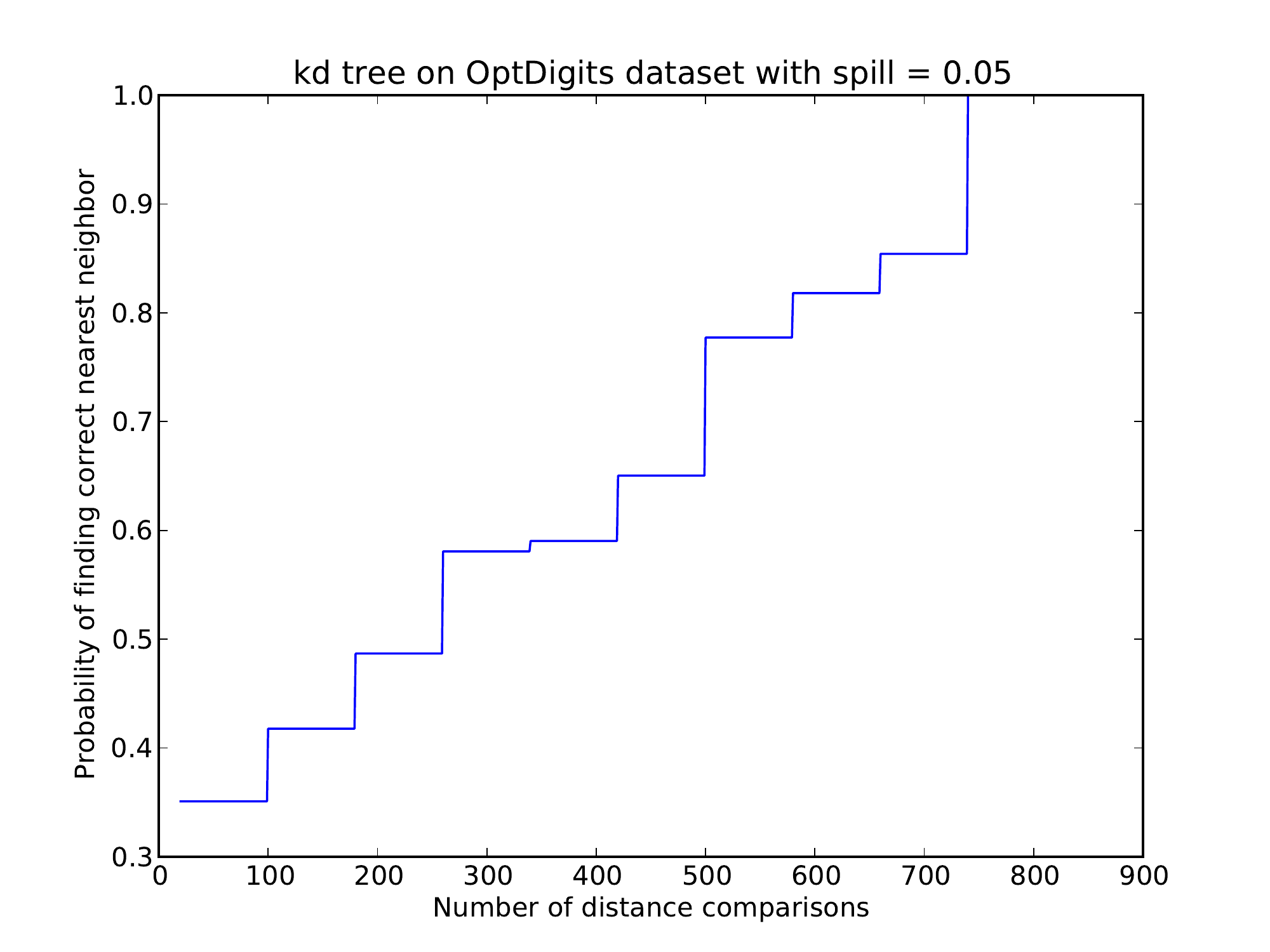}
              
        \end{subfigure}
        ~ 
        \begin{subfigure}[b]{0.3\textwidth}
                \includegraphics[width=\textwidth]{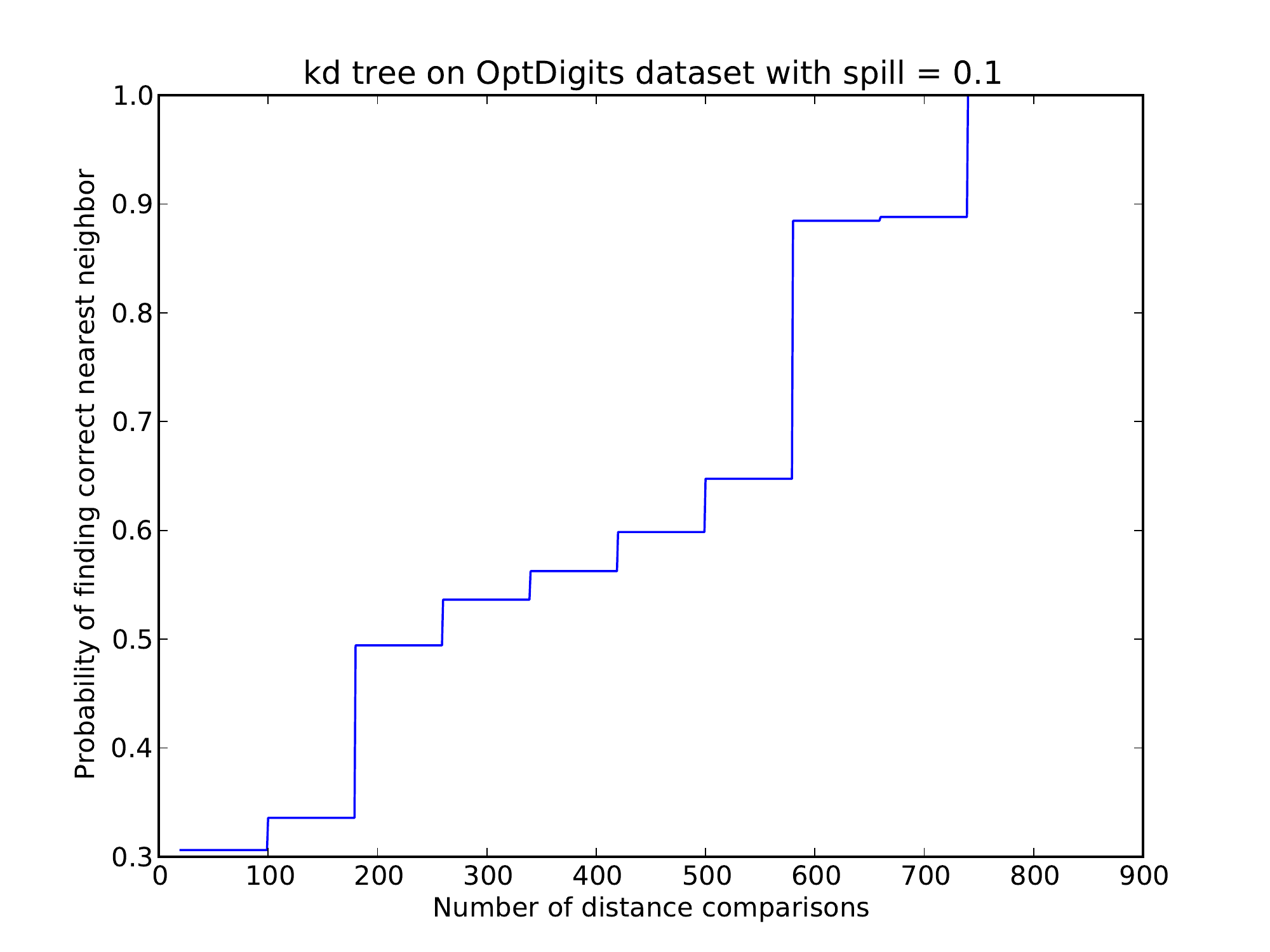}
             
        \end{subfigure}

        \begin{subfigure}[b]{0.3\textwidth}
                \includegraphics[width=\textwidth]{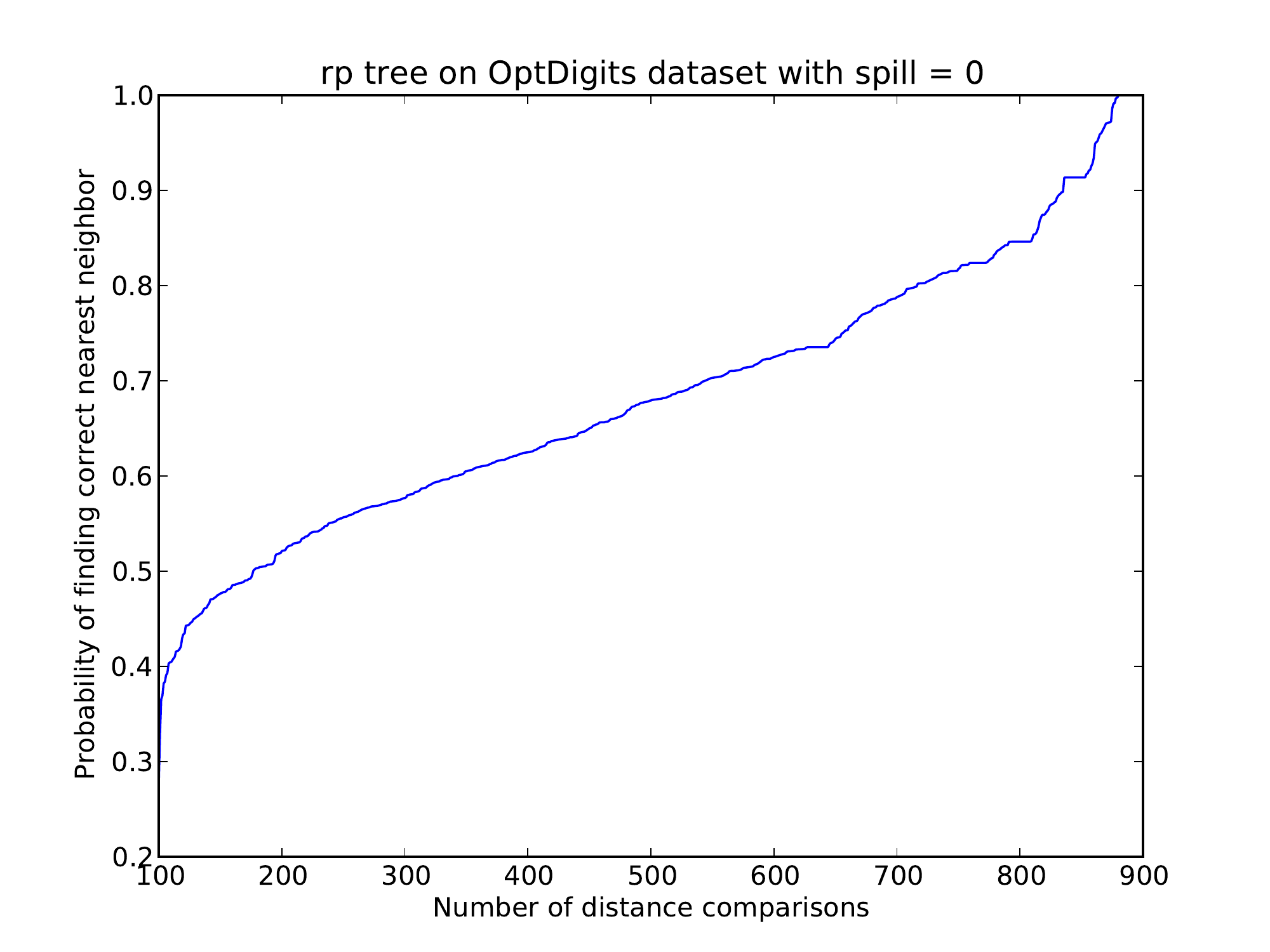}
             
        \end{subfigure}
        ~
        \begin{subfigure}[b]{0.3\textwidth}
                \includegraphics[width=\textwidth]{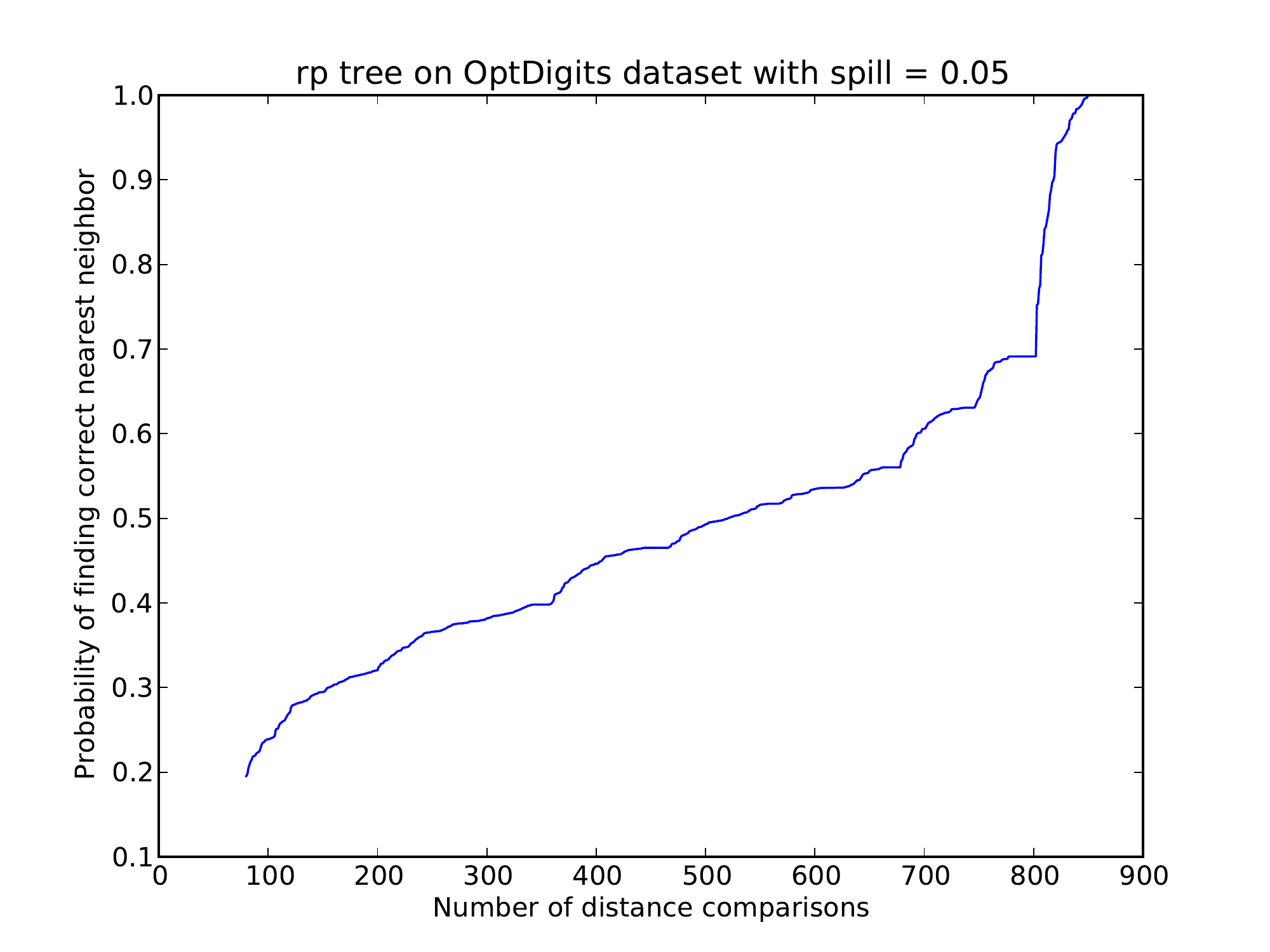}
             
        \end{subfigure}
        ~
        \begin{subfigure}[b]{0.3\textwidth}
                \includegraphics[width=\textwidth]{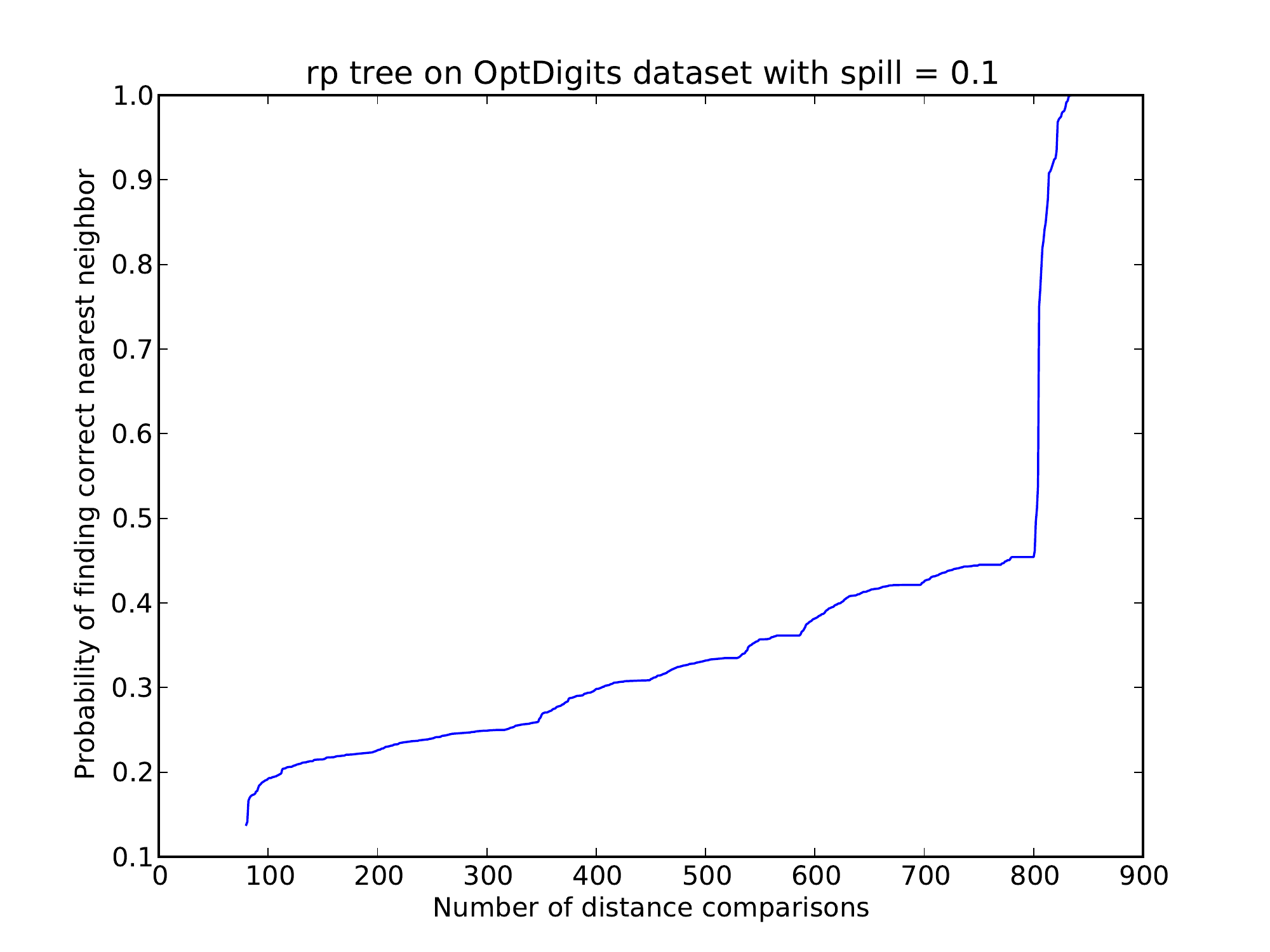}
             
        \end{subfigure}

        \begin{subfigure}[b]{0.3\textwidth}
                \includegraphics[width=\textwidth]{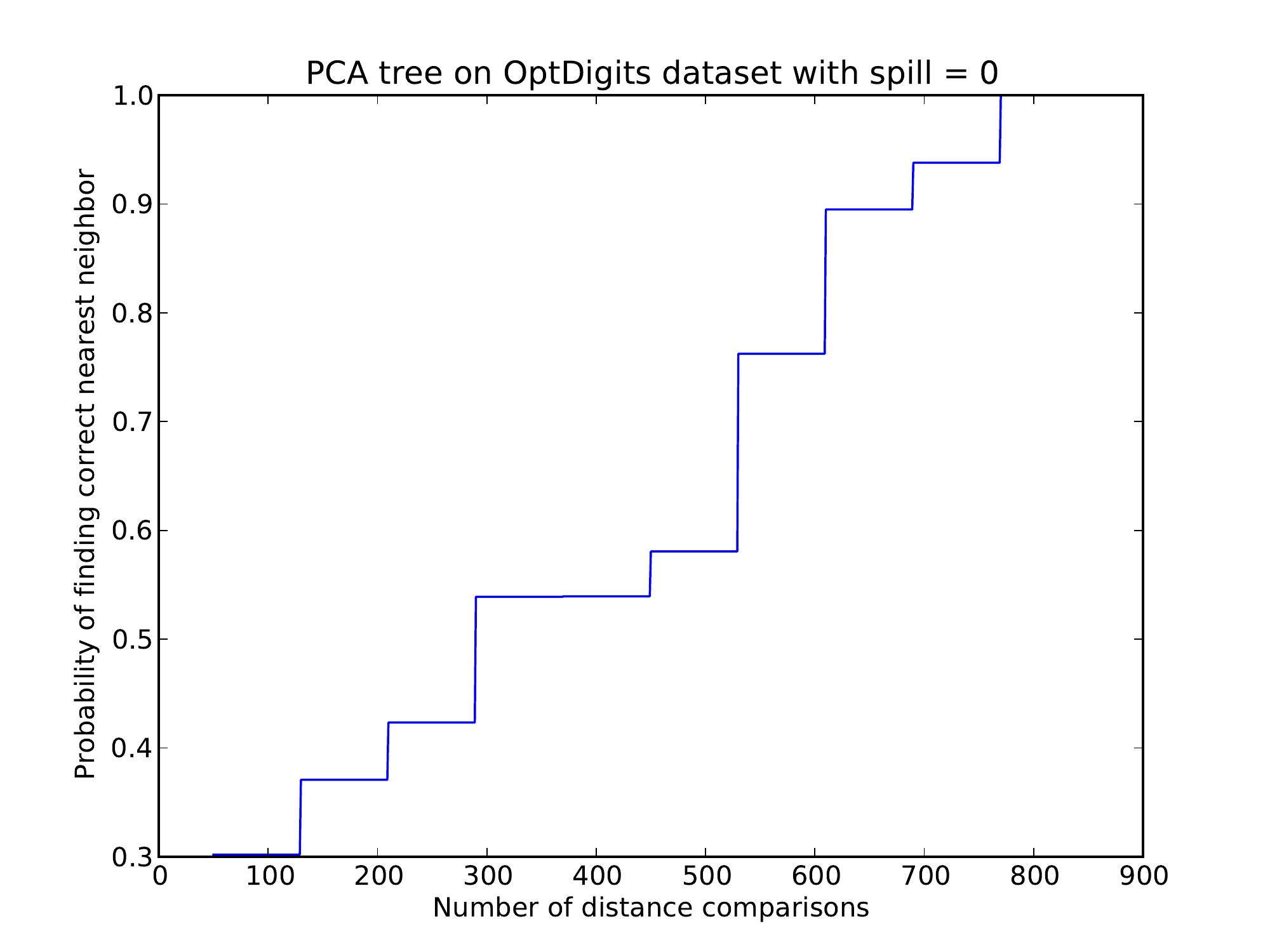}
        \end{subfigure}%
        ~ 
        \begin{subfigure}[b]{0.3\textwidth}
                \includegraphics[width=\textwidth]{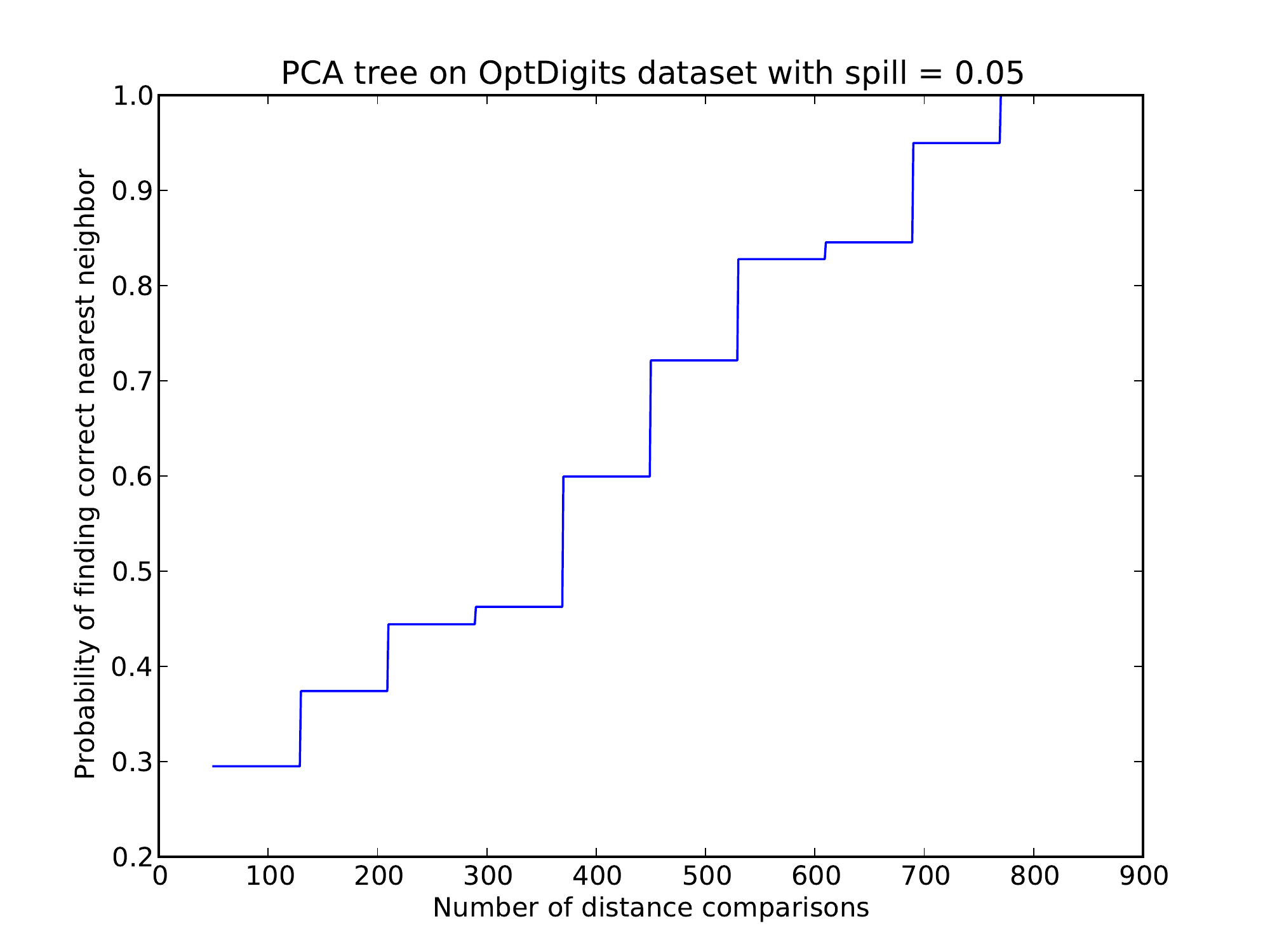}
              
        \end{subfigure}
        ~ 
        \begin{subfigure}[b]{0.3\textwidth}
                \includegraphics[width=\textwidth]{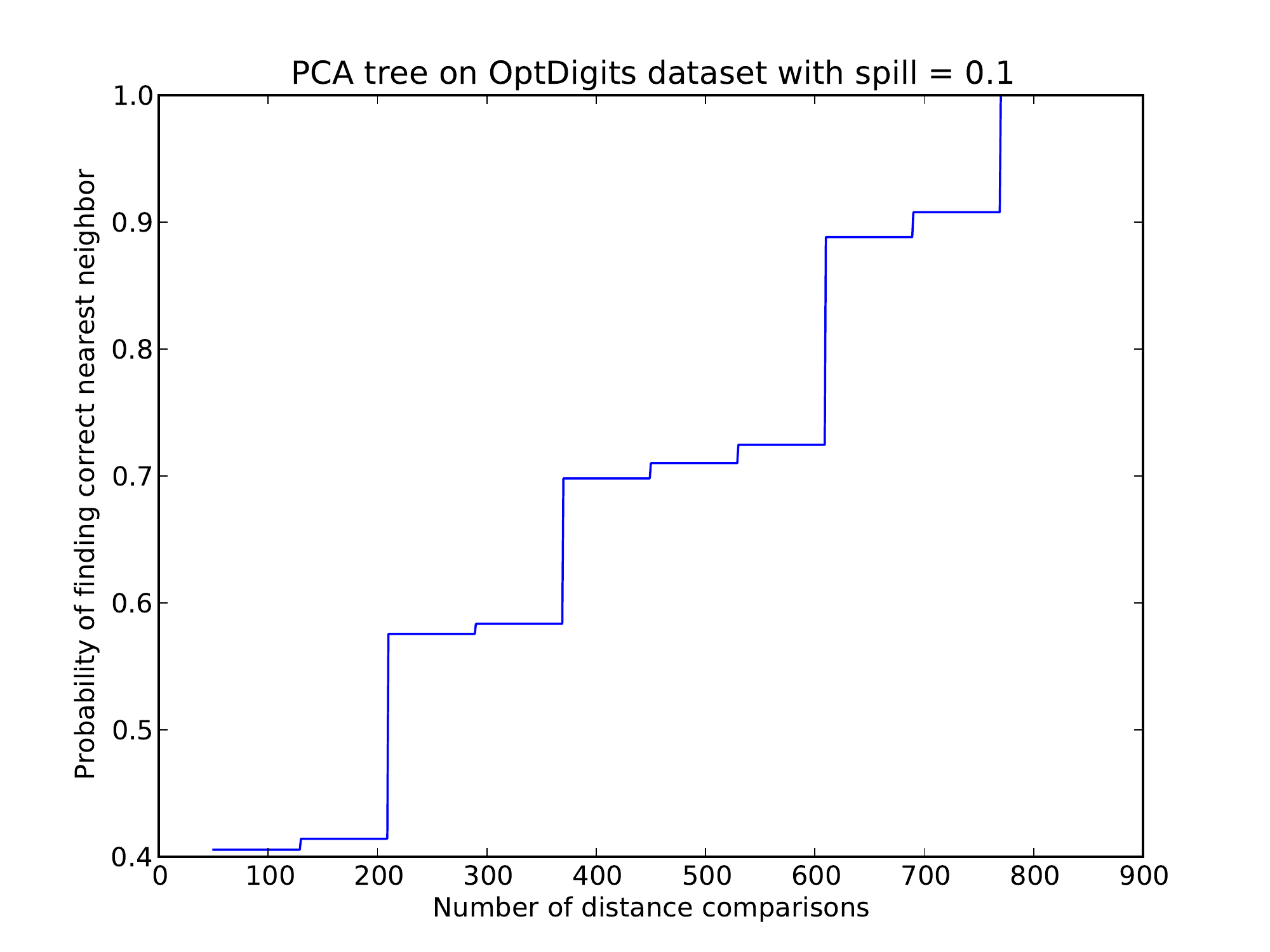}
             
        \end{subfigure}

        \begin{subfigure}[b]{0.3\textwidth}
                \includegraphics[width=\textwidth]{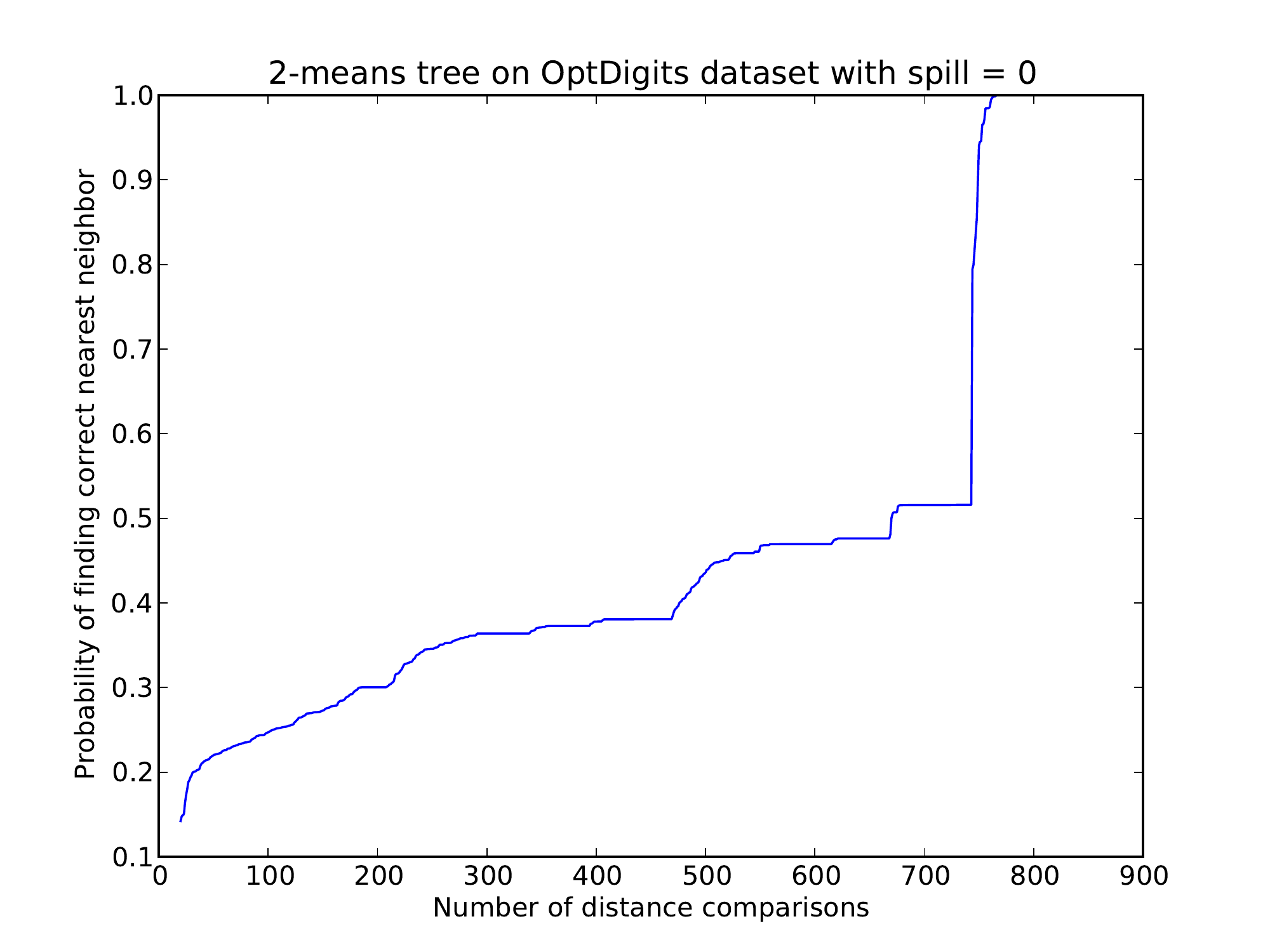}
             
        \end{subfigure}
        ~
        \begin{subfigure}[b]{0.3\textwidth}
                \includegraphics[width=\textwidth]{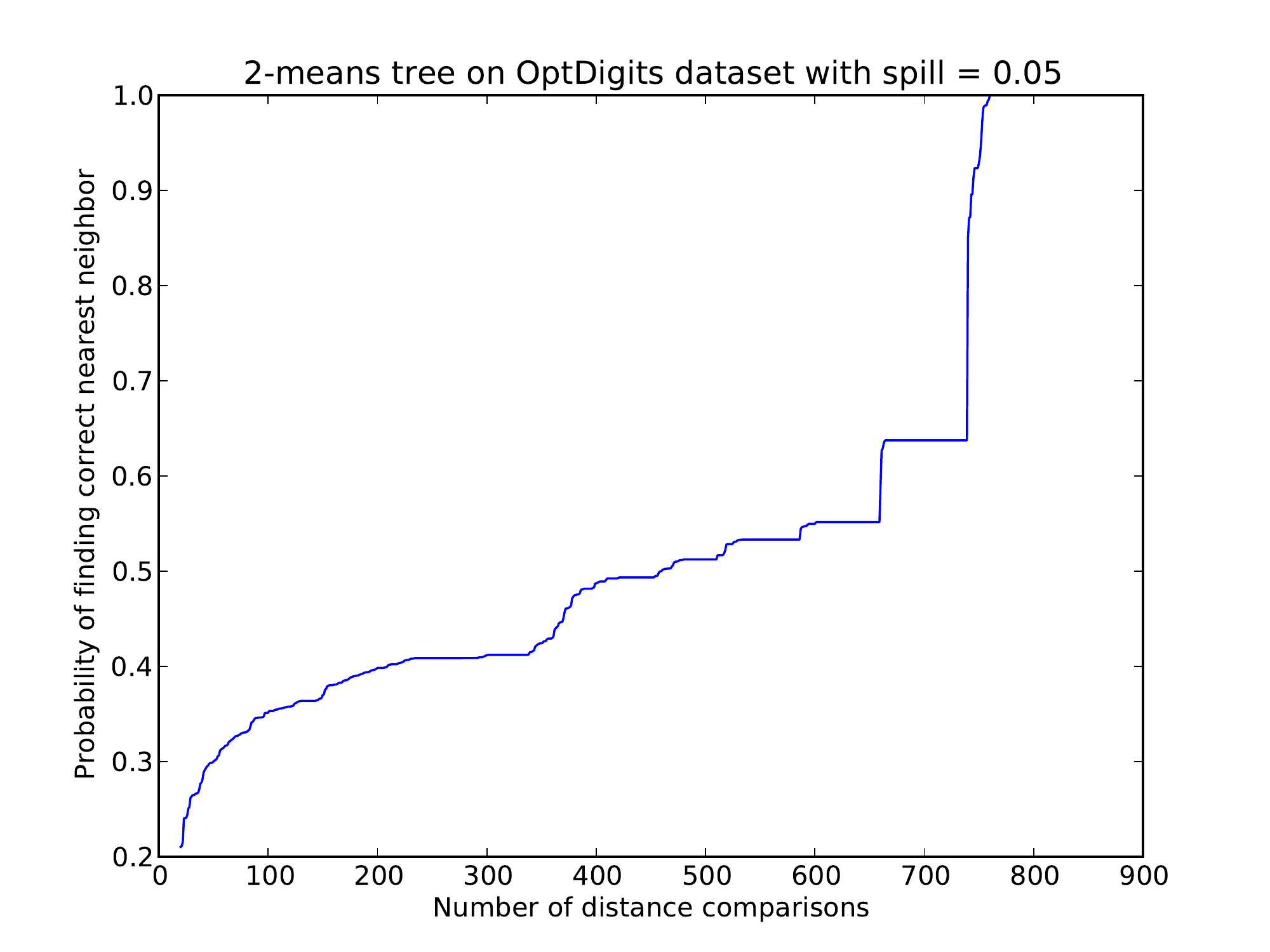}
             
        \end{subfigure}
        ~
        \begin{subfigure}[b]{0.3\textwidth}
                \includegraphics[width=\textwidth]{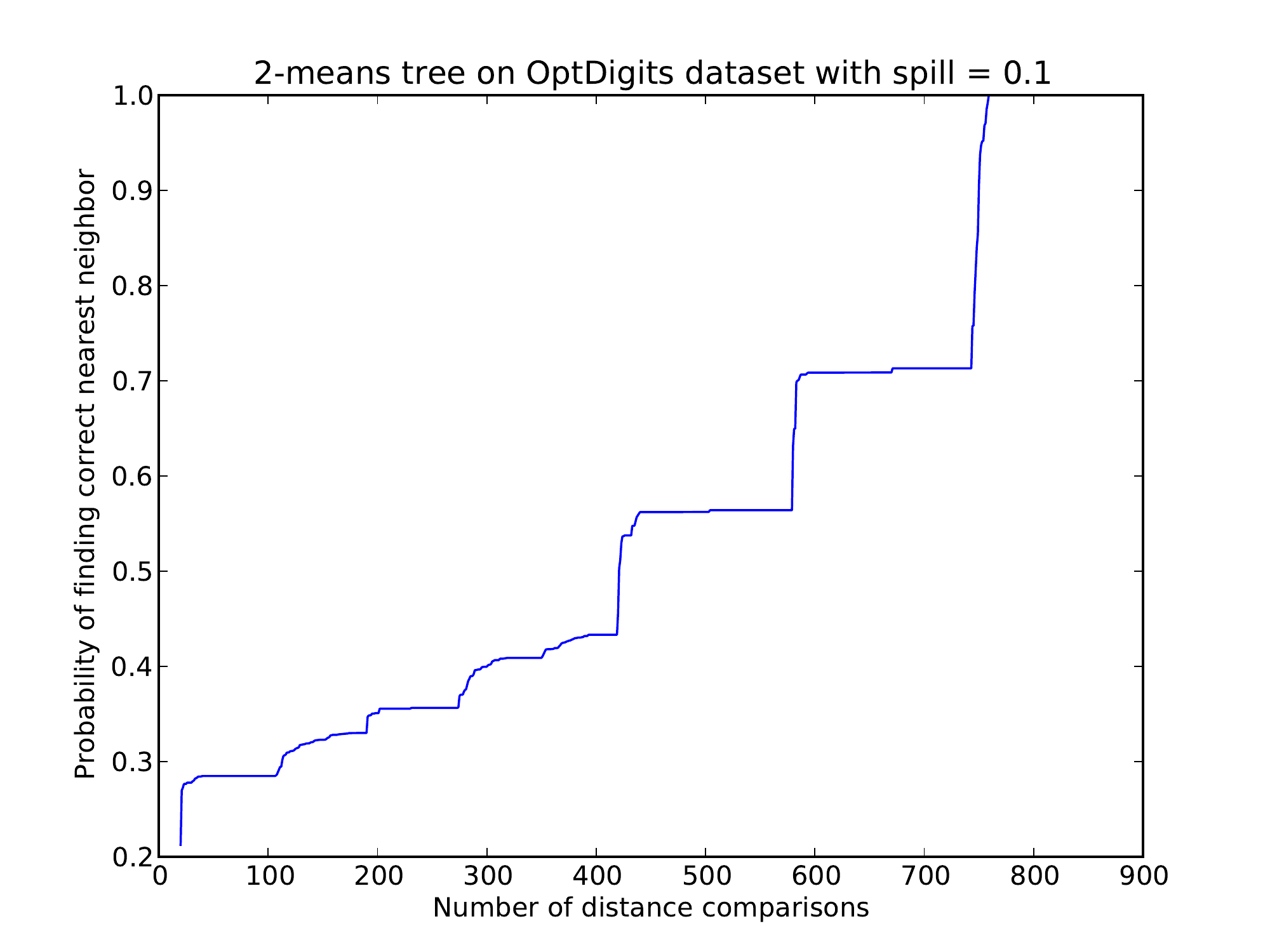}
             
        \end{subfigure}

        \caption{Probability of finding nearest neighbor vs number of comparisons for different spatial trees and tree configurations for the OptDigits dataset}
        \label{fig:opt}
\end{figure}

\begin{figure}
        \centering
        \begin{subfigure}[b]{0.3\textwidth}
                \includegraphics[width=\textwidth]{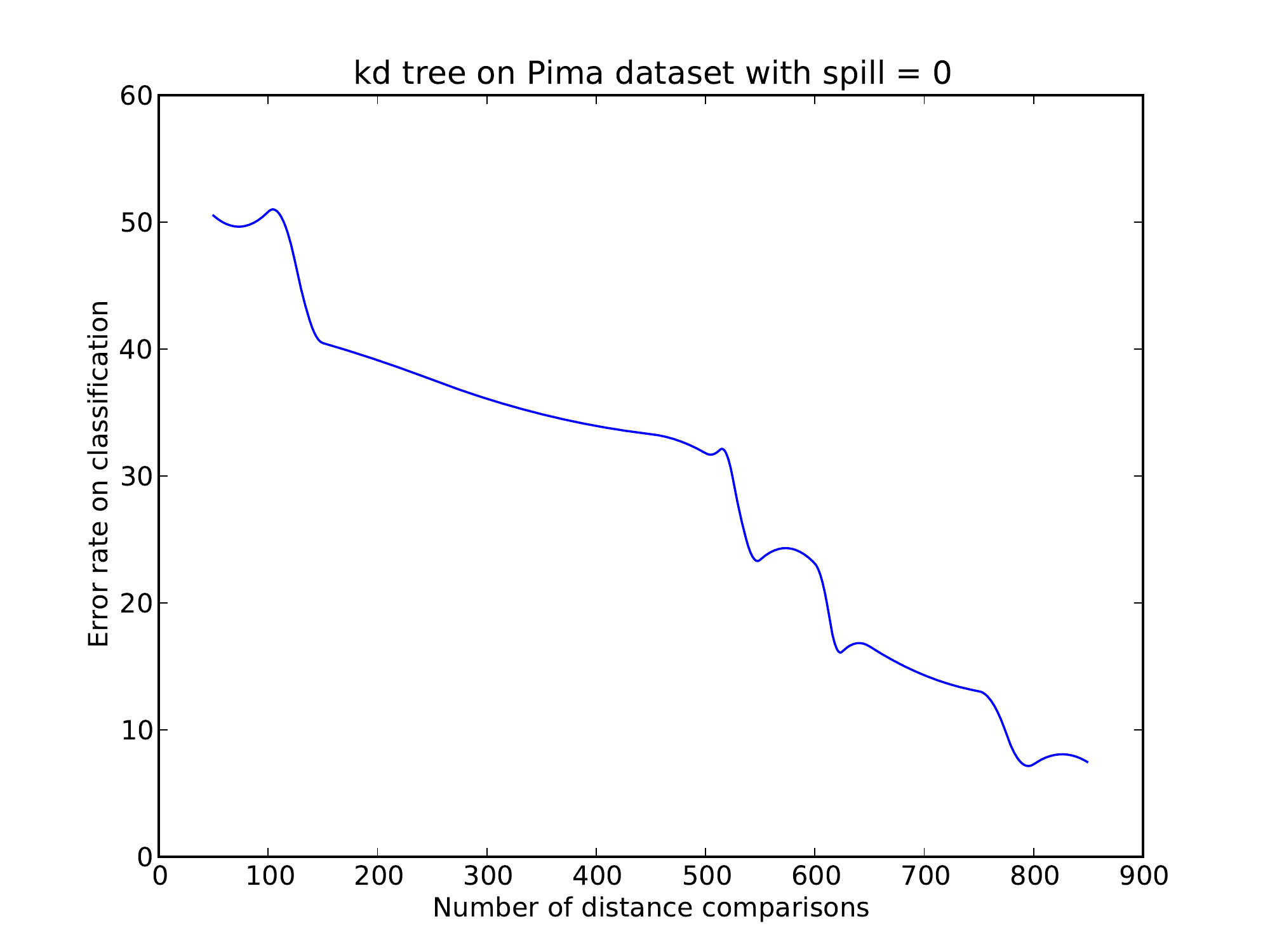}
        \end{subfigure}%
        ~ 
        \begin{subfigure}[b]{0.3\textwidth}
                \includegraphics[width=\textwidth]{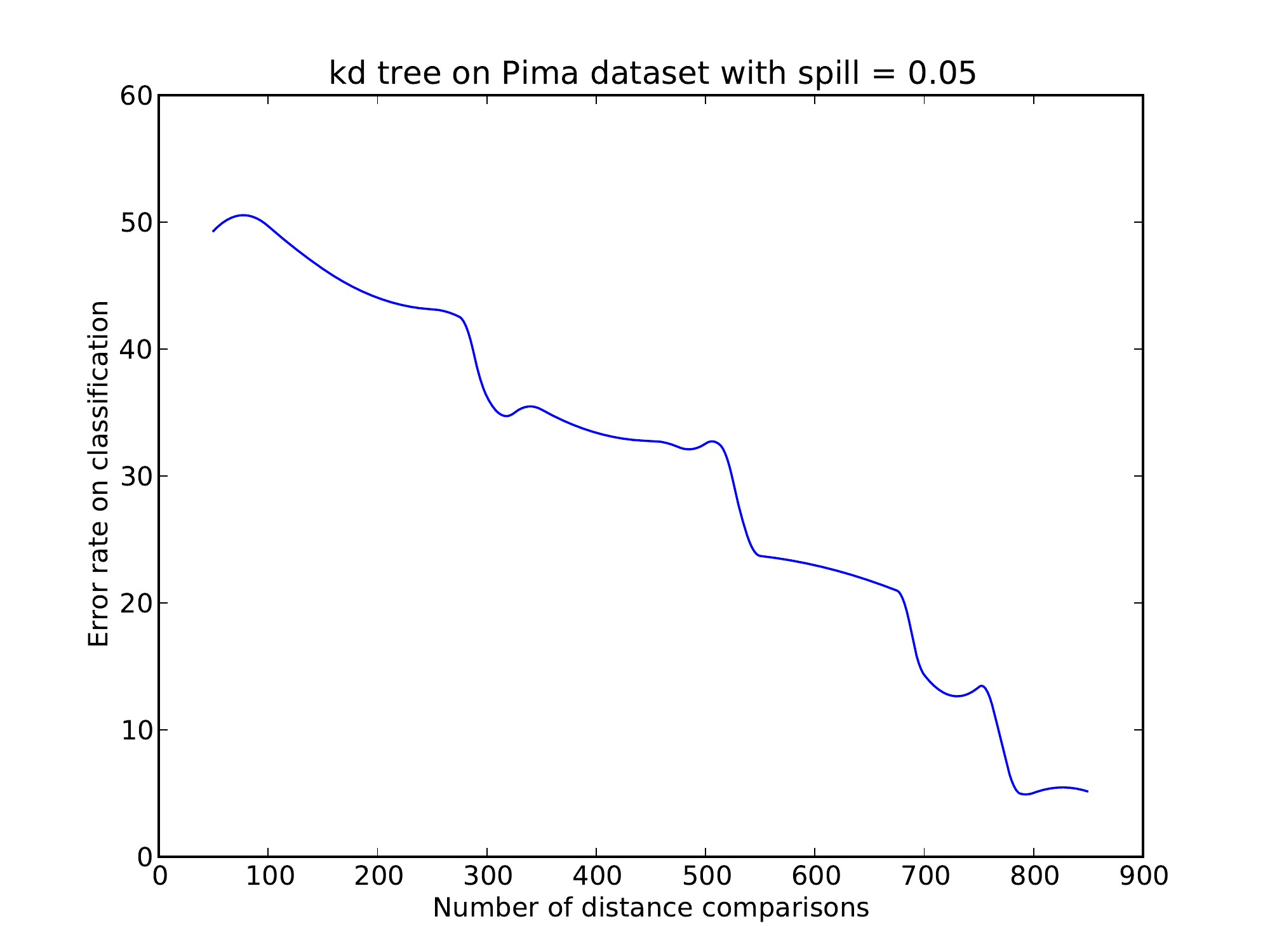}
              
        \end{subfigure}
        ~ 
        \begin{subfigure}[b]{0.3\textwidth}
                \includegraphics[width=\textwidth]{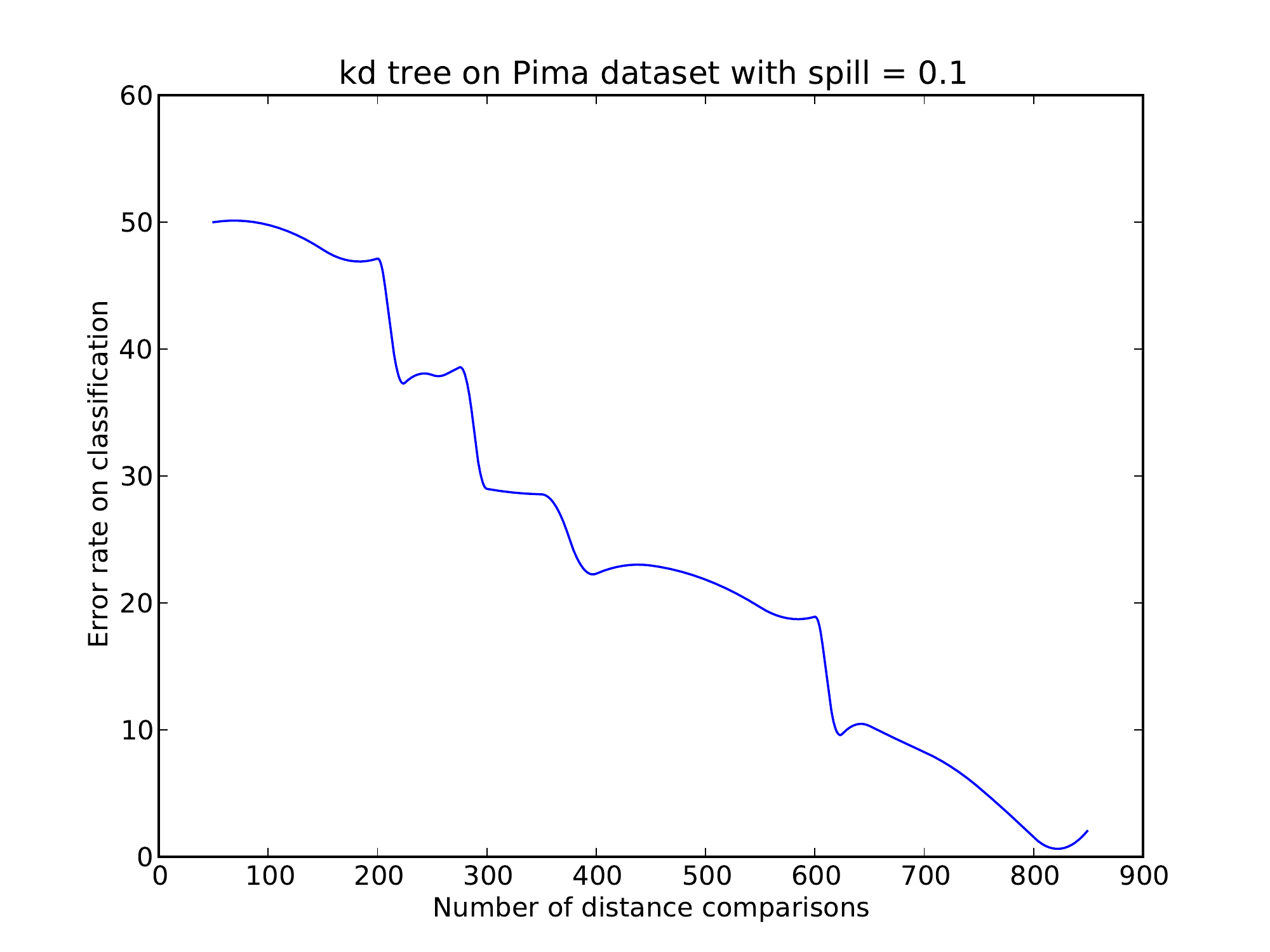}
             
        \end{subfigure}

        \begin{subfigure}[b]{0.3\textwidth}
                \includegraphics[width=\textwidth]{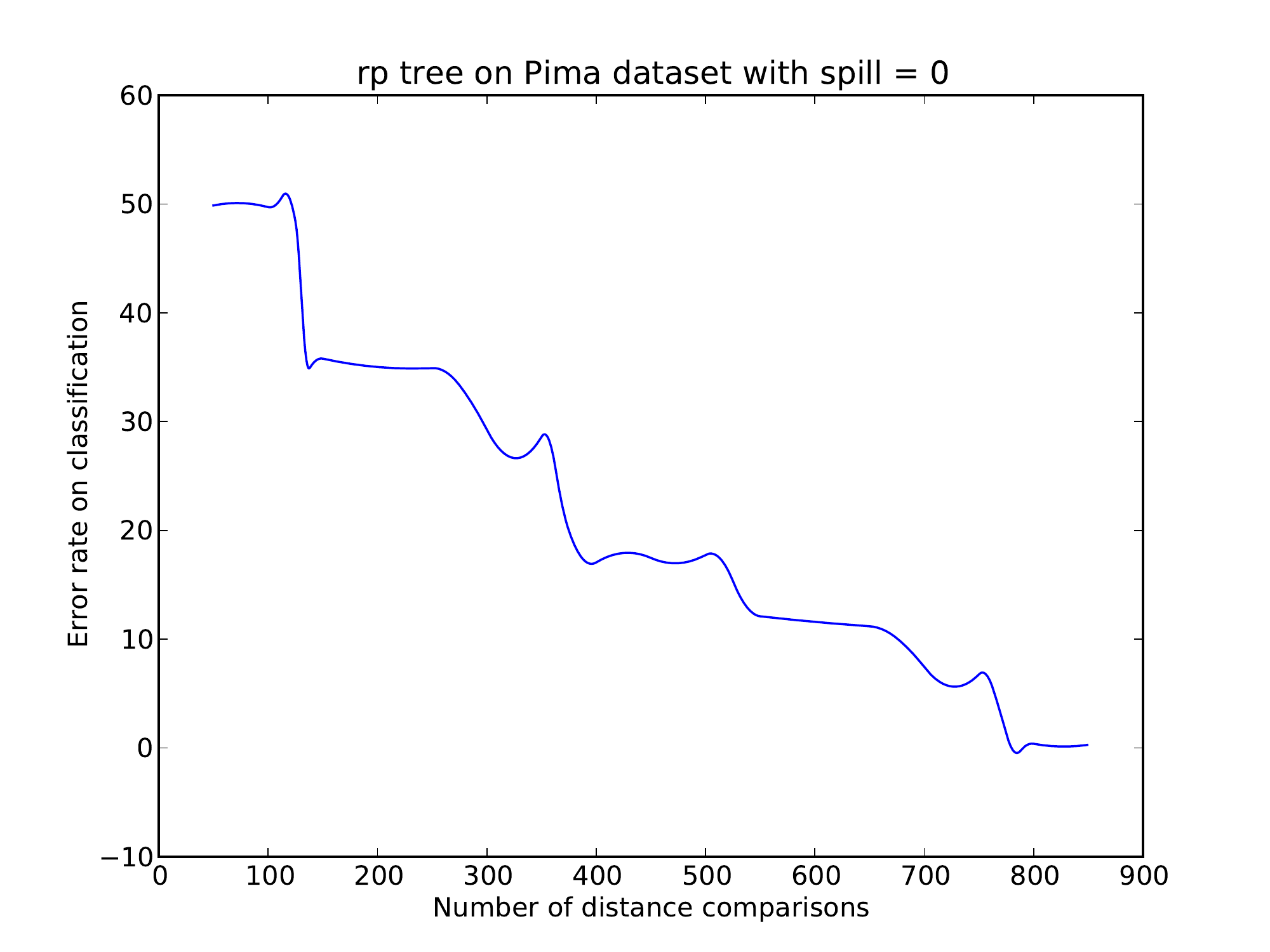}
             
        \end{subfigure}
        ~
        \begin{subfigure}[b]{0.3\textwidth}
                \includegraphics[width=\textwidth]{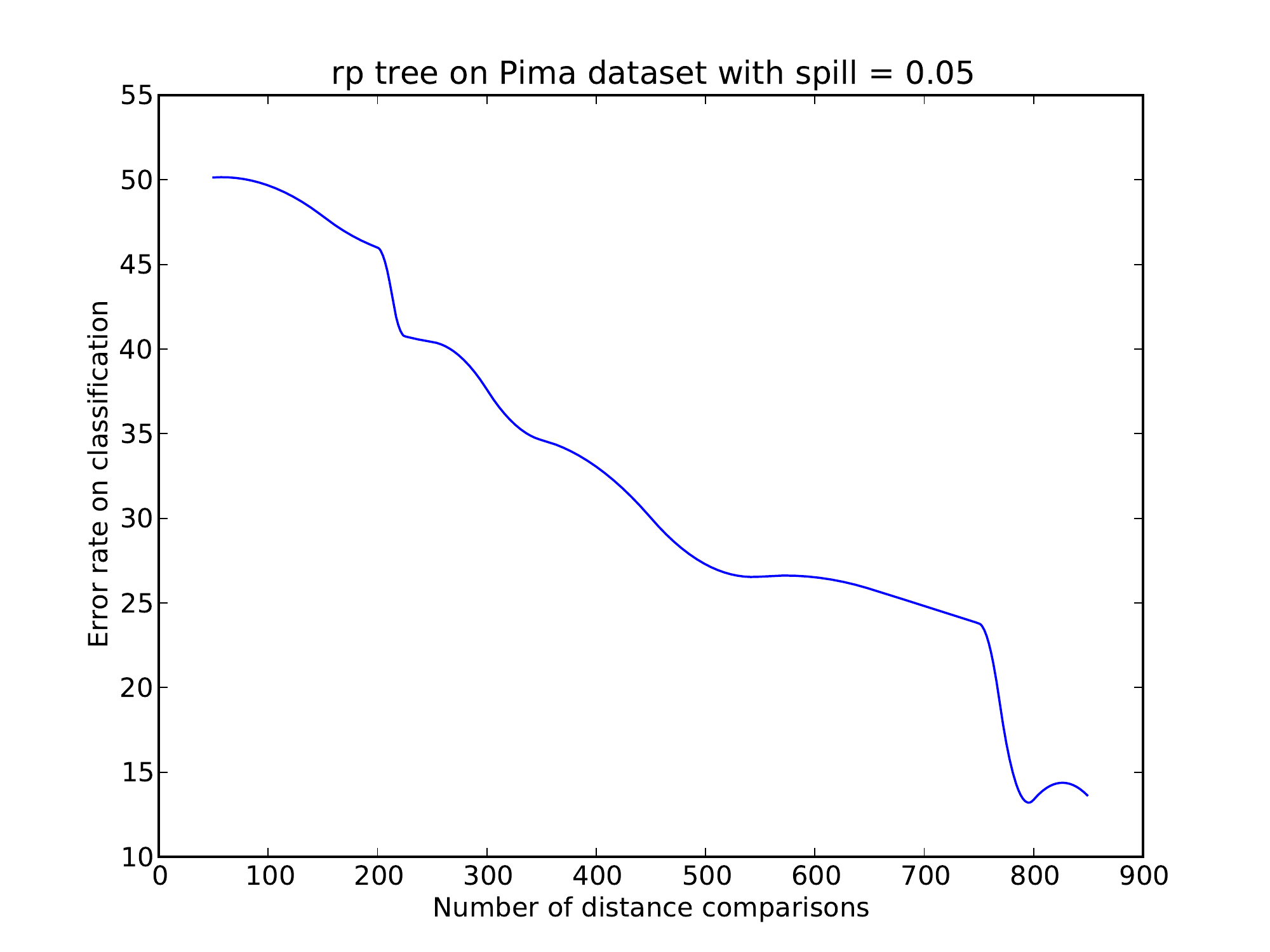}
             
        \end{subfigure}
        ~
        \begin{subfigure}[b]{0.3\textwidth}
                \includegraphics[width=\textwidth]{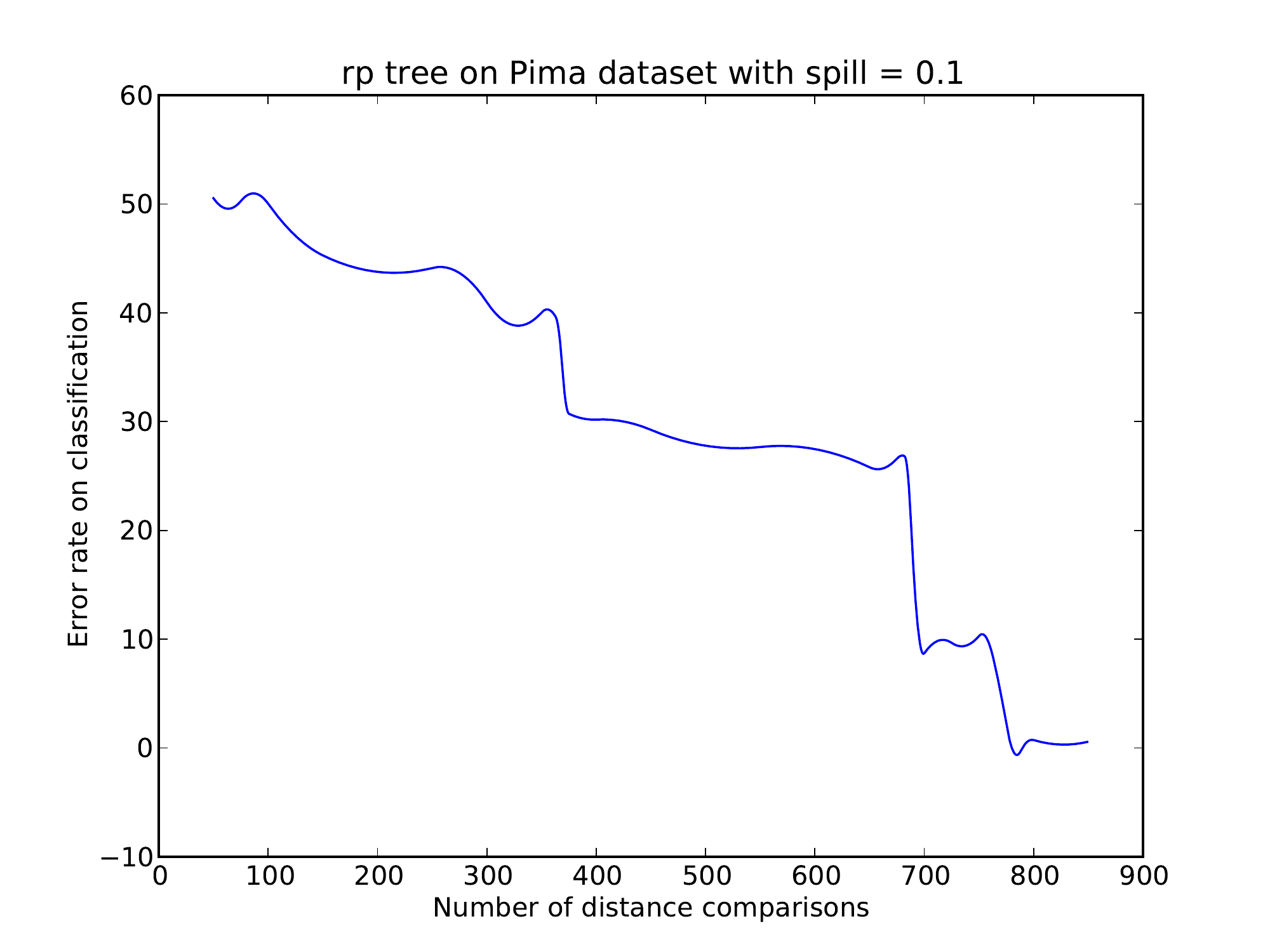}
             
        \end{subfigure}

        \begin{subfigure}[b]{0.3\textwidth}
                \includegraphics[width=\textwidth]{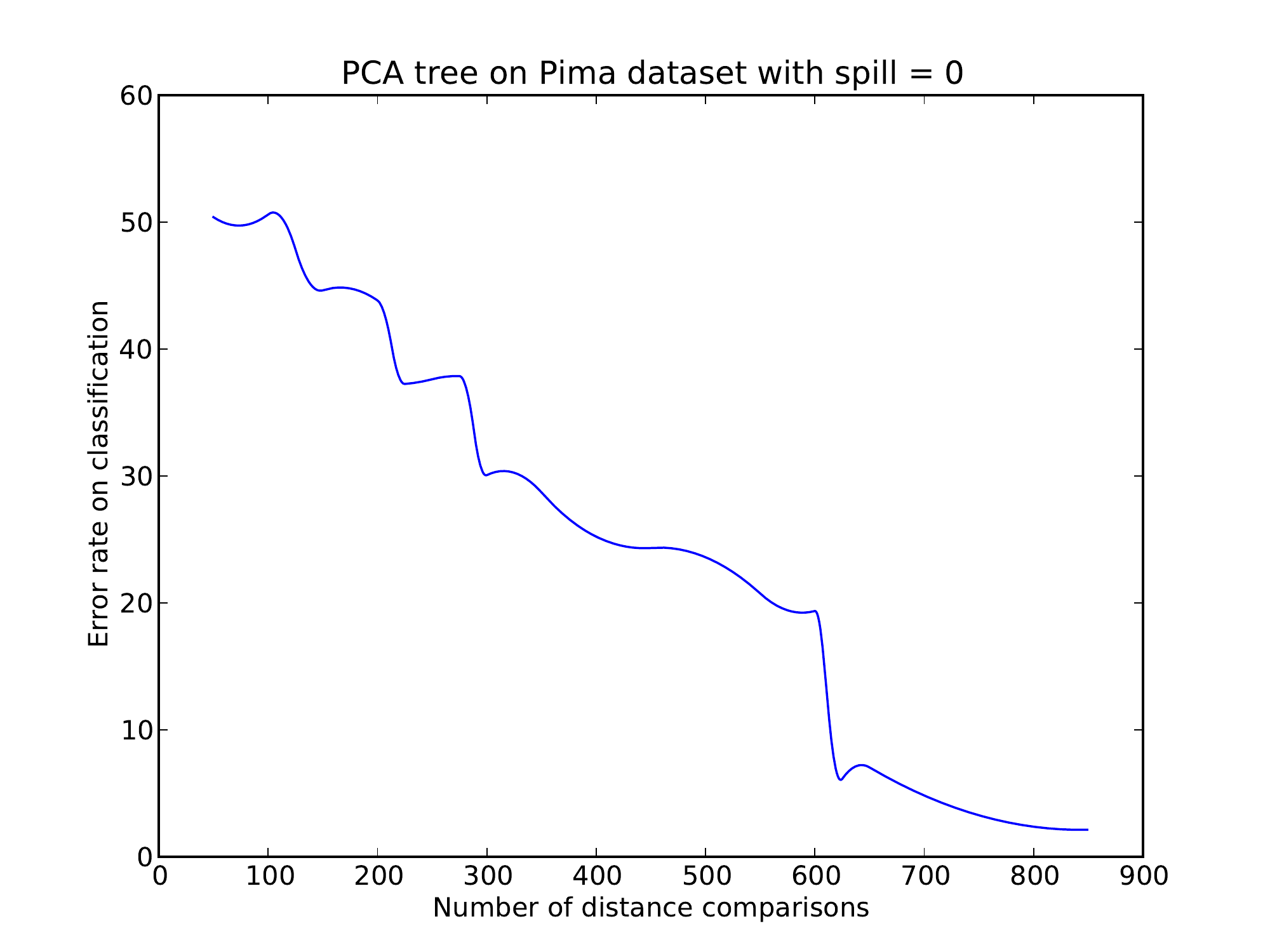}
        \end{subfigure}%
        ~ 
        \begin{subfigure}[b]{0.3\textwidth}
                \includegraphics[width=\textwidth]{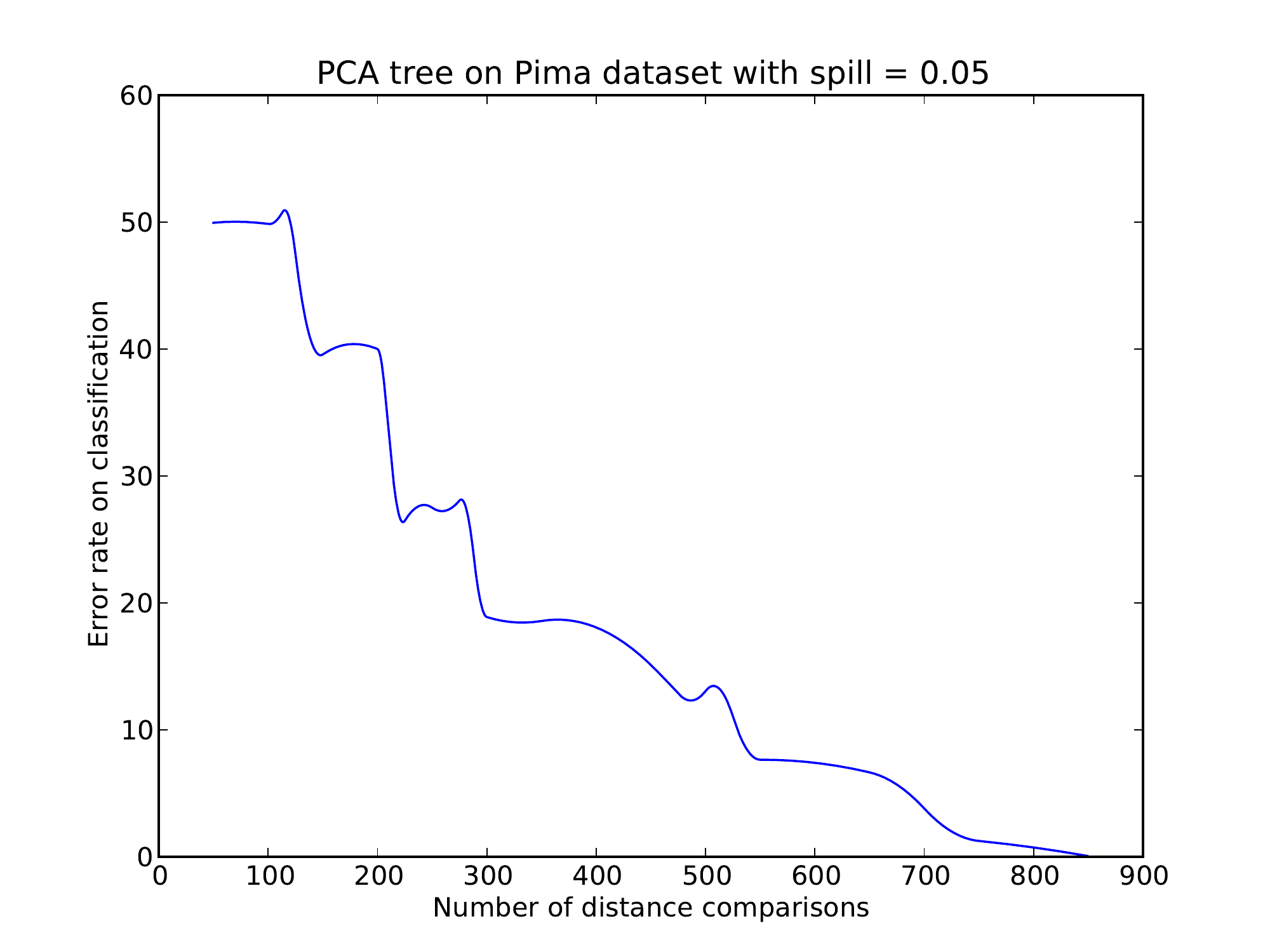}
              
        \end{subfigure}
        ~ 
        \begin{subfigure}[b]{0.3\textwidth}
                \includegraphics[width=\textwidth]{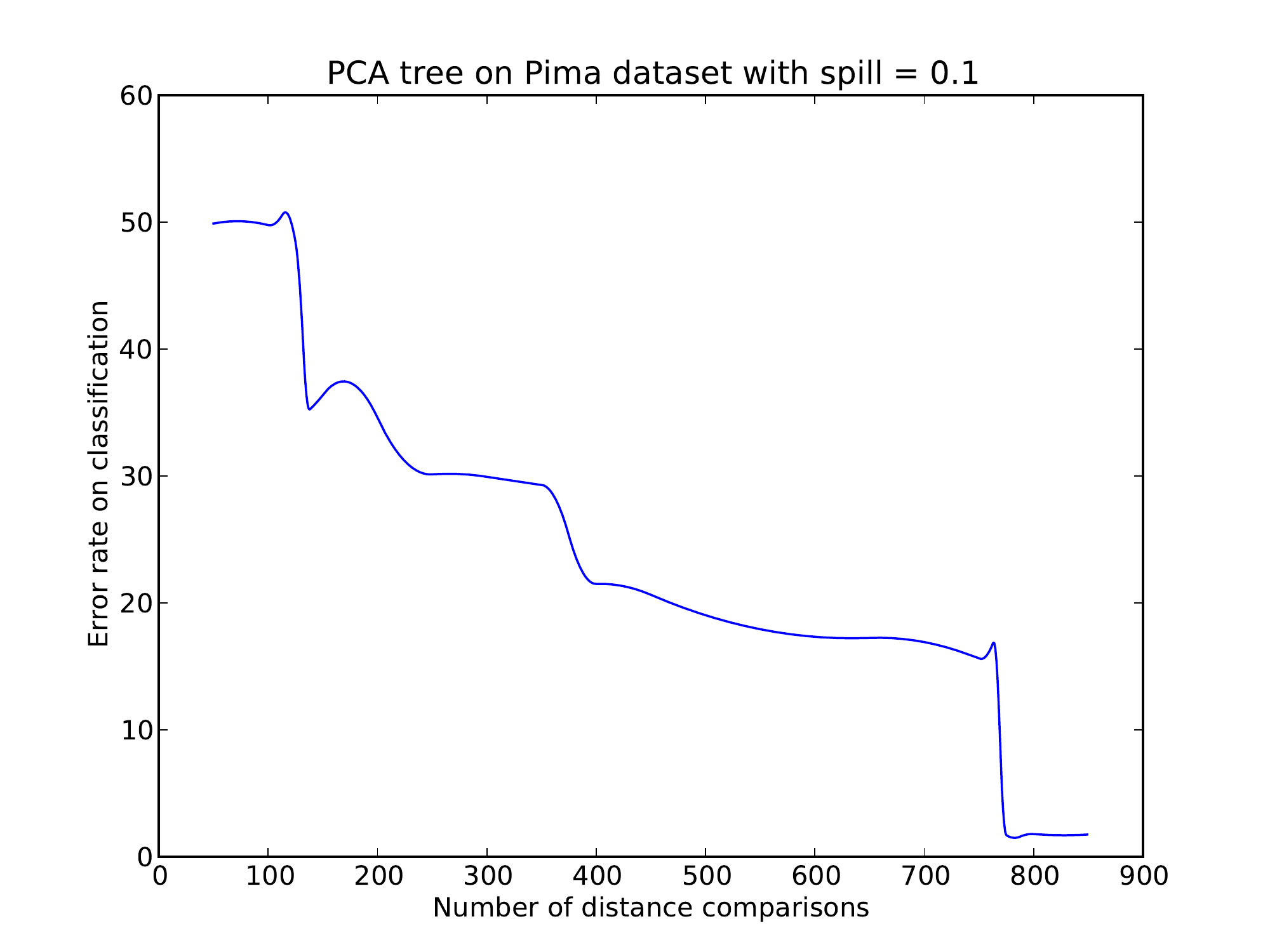}
             
        \end{subfigure}

        \begin{subfigure}[b]{0.3\textwidth}
                \includegraphics[width=\textwidth]{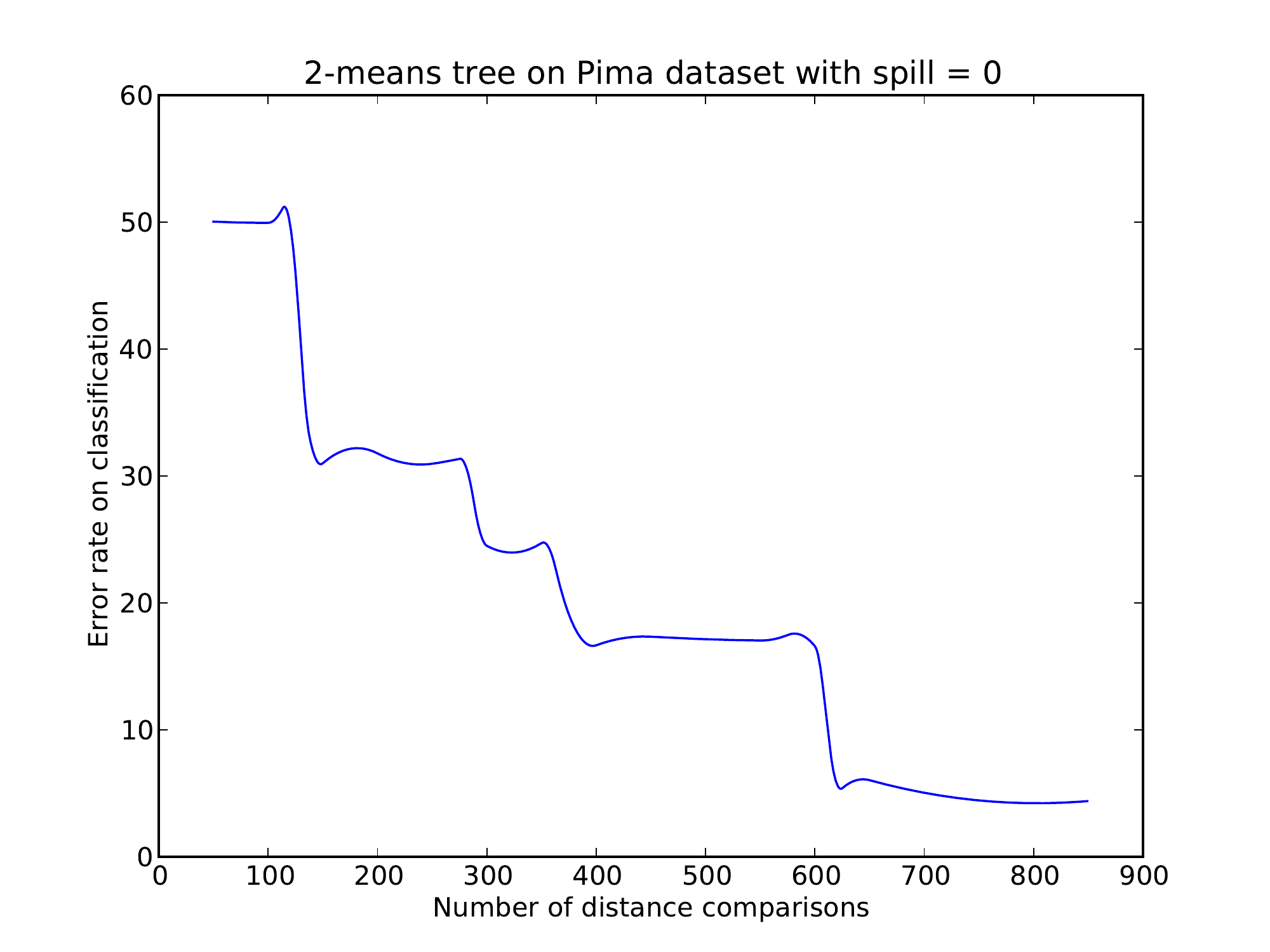}
             
        \end{subfigure}
        ~
        \begin{subfigure}[b]{0.3\textwidth}
                \includegraphics[width=\textwidth]{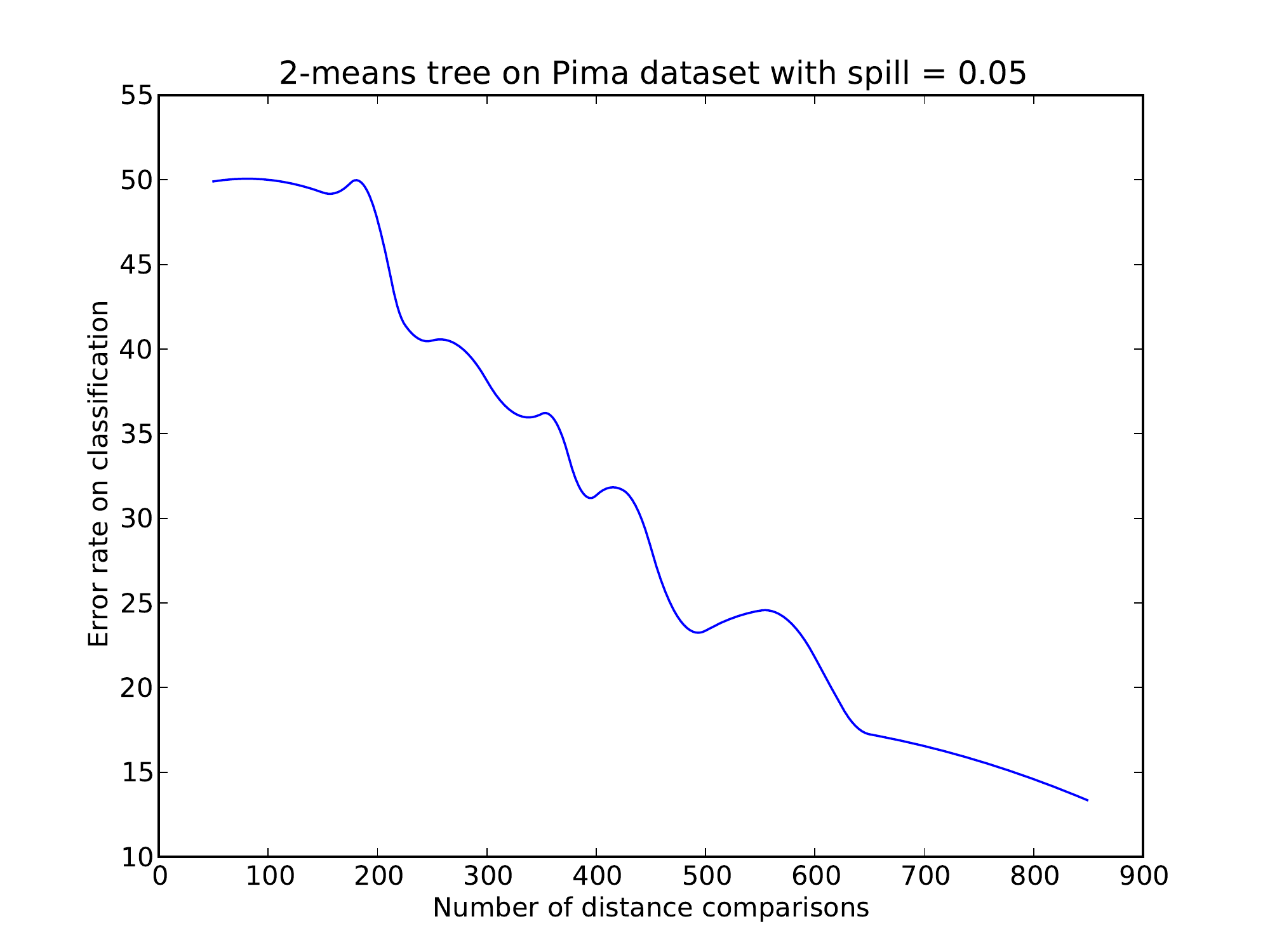}
             
        \end{subfigure}
        ~
        \begin{subfigure}[b]{0.3\textwidth}
                \includegraphics[width=\textwidth]{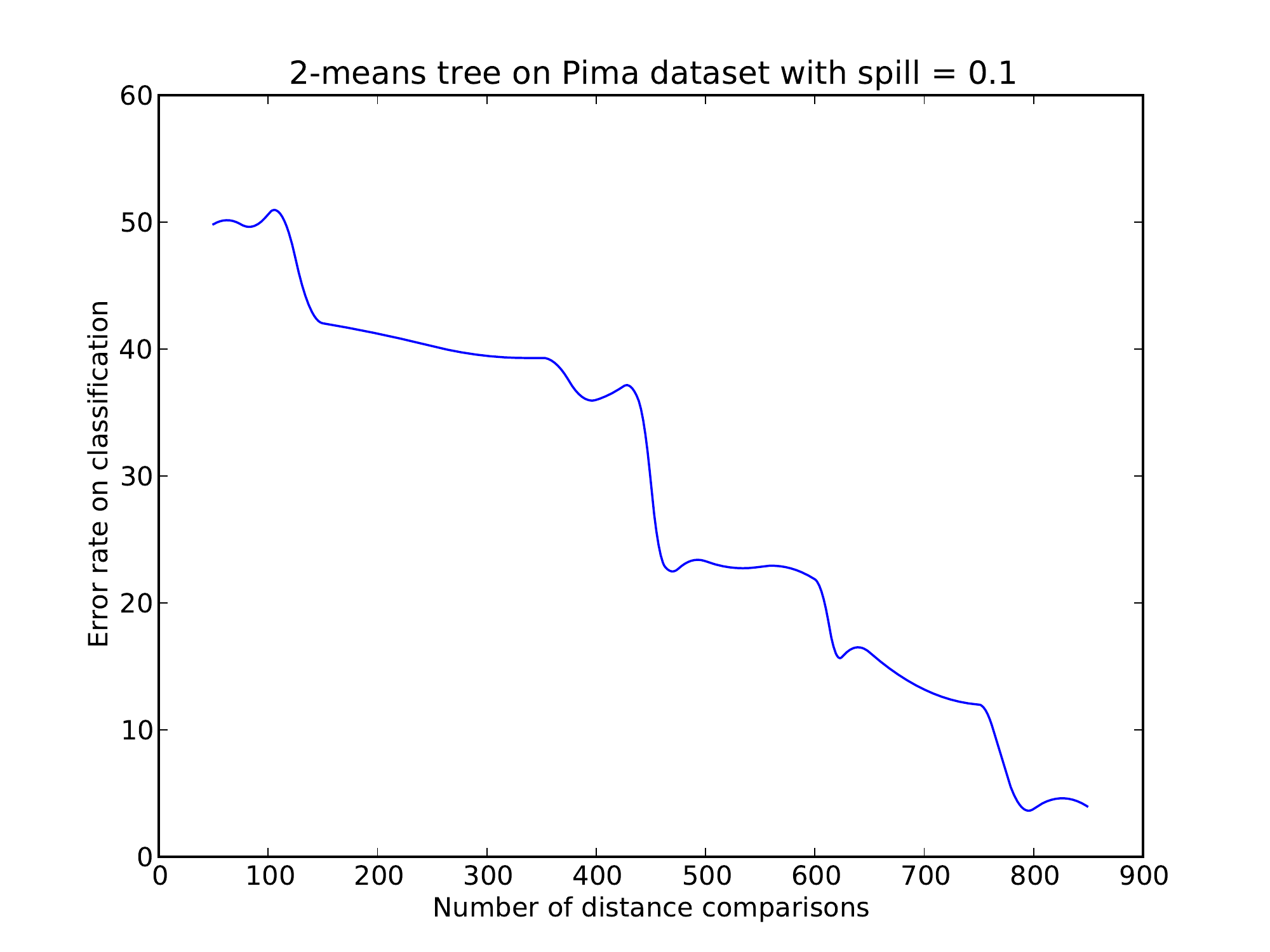}
             
        \end{subfigure}

        \caption{Error Rate on Classification vs number of comparisons for different spatial trees and tree configurations for the Pima dataset}
        \label{fig:pimac}
\end{figure}

\begin{figure}
        \centering
        \begin{subfigure}[b]{0.3\textwidth}
                \includegraphics[width=\textwidth]{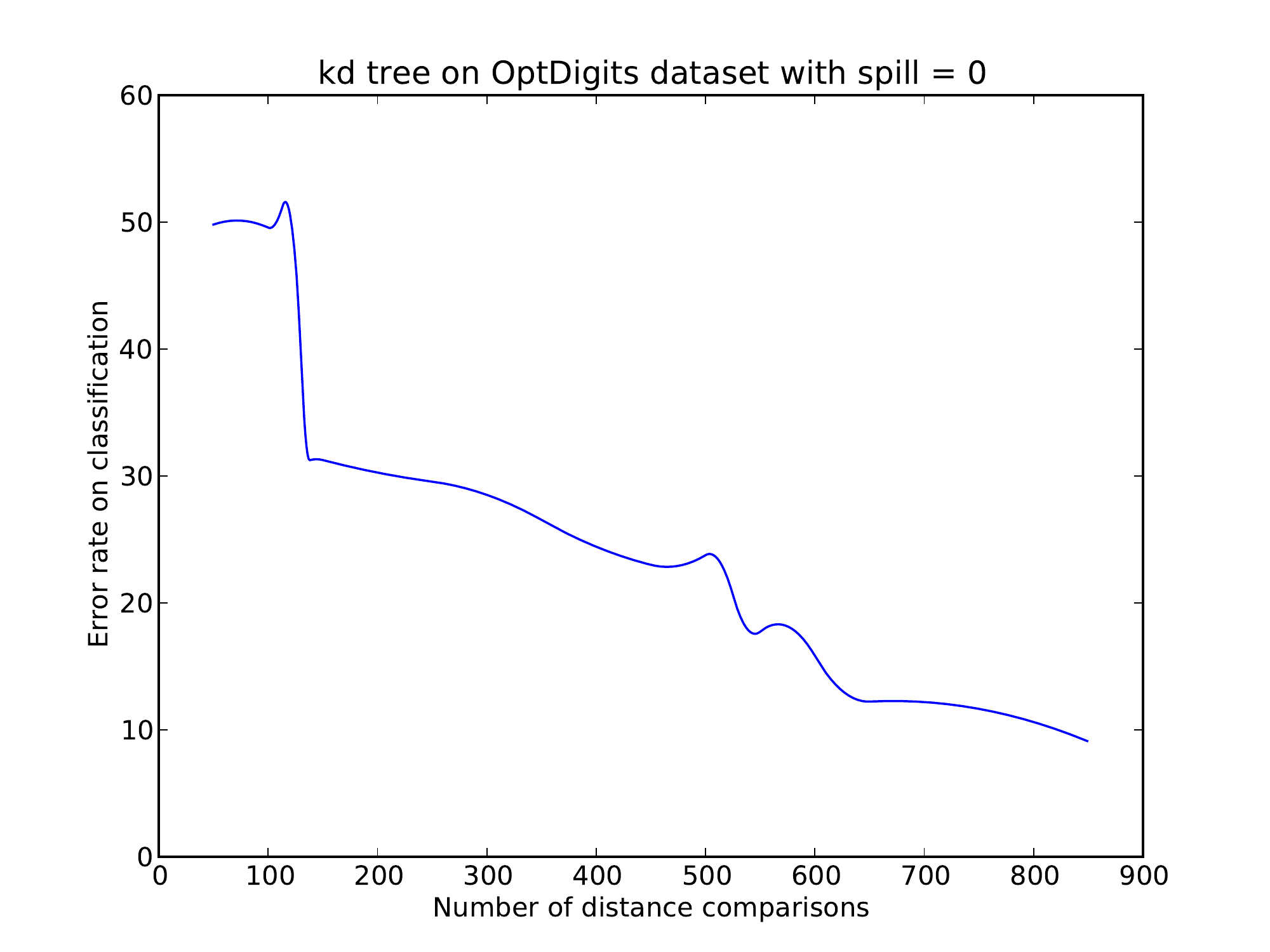}
        \end{subfigure}%
        ~ 
        \begin{subfigure}[b]{0.3\textwidth}
                \includegraphics[width=\textwidth]{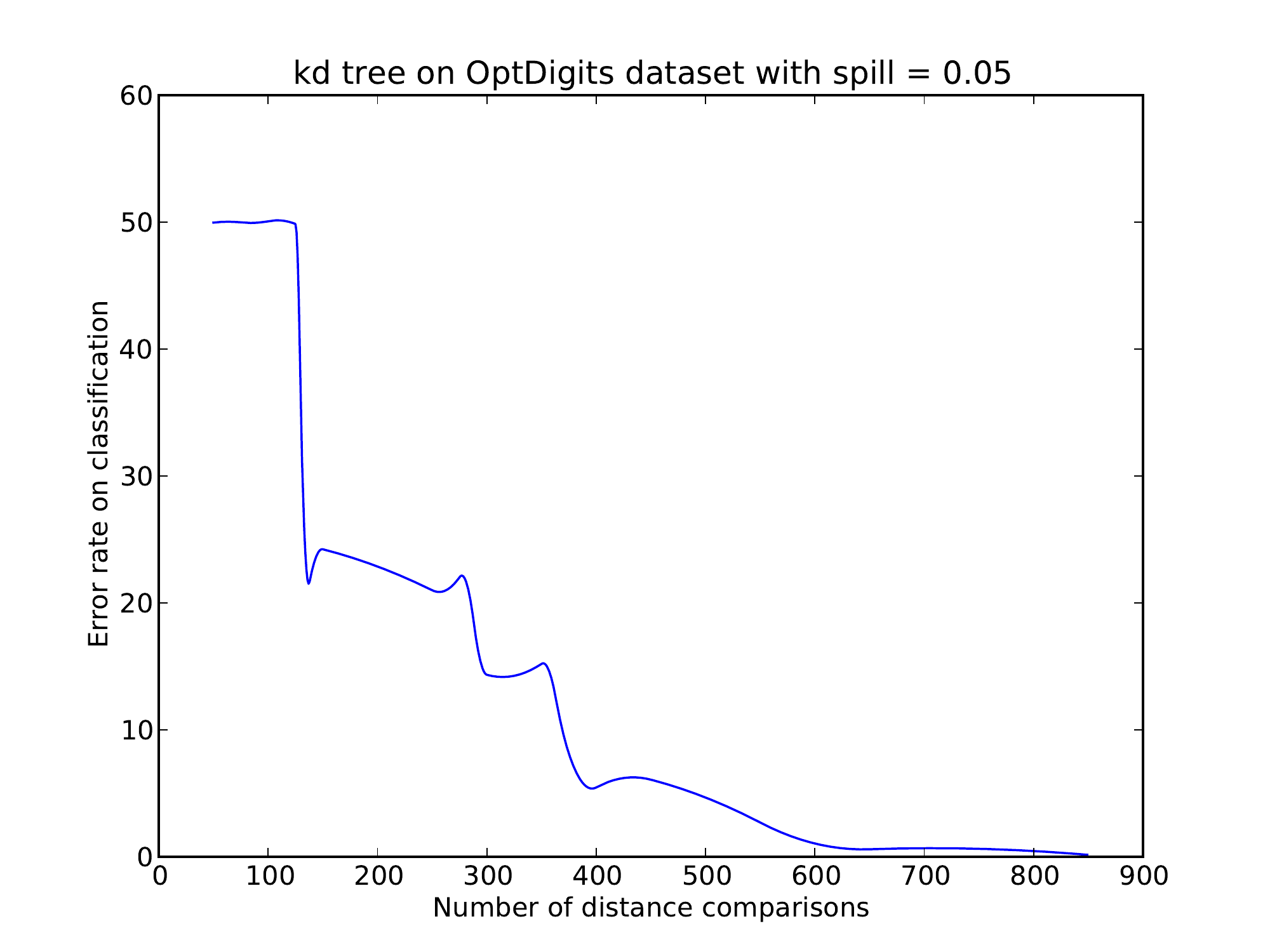}
              
        \end{subfigure}
        ~ 
        \begin{subfigure}[b]{0.3\textwidth}
                \includegraphics[width=\textwidth]{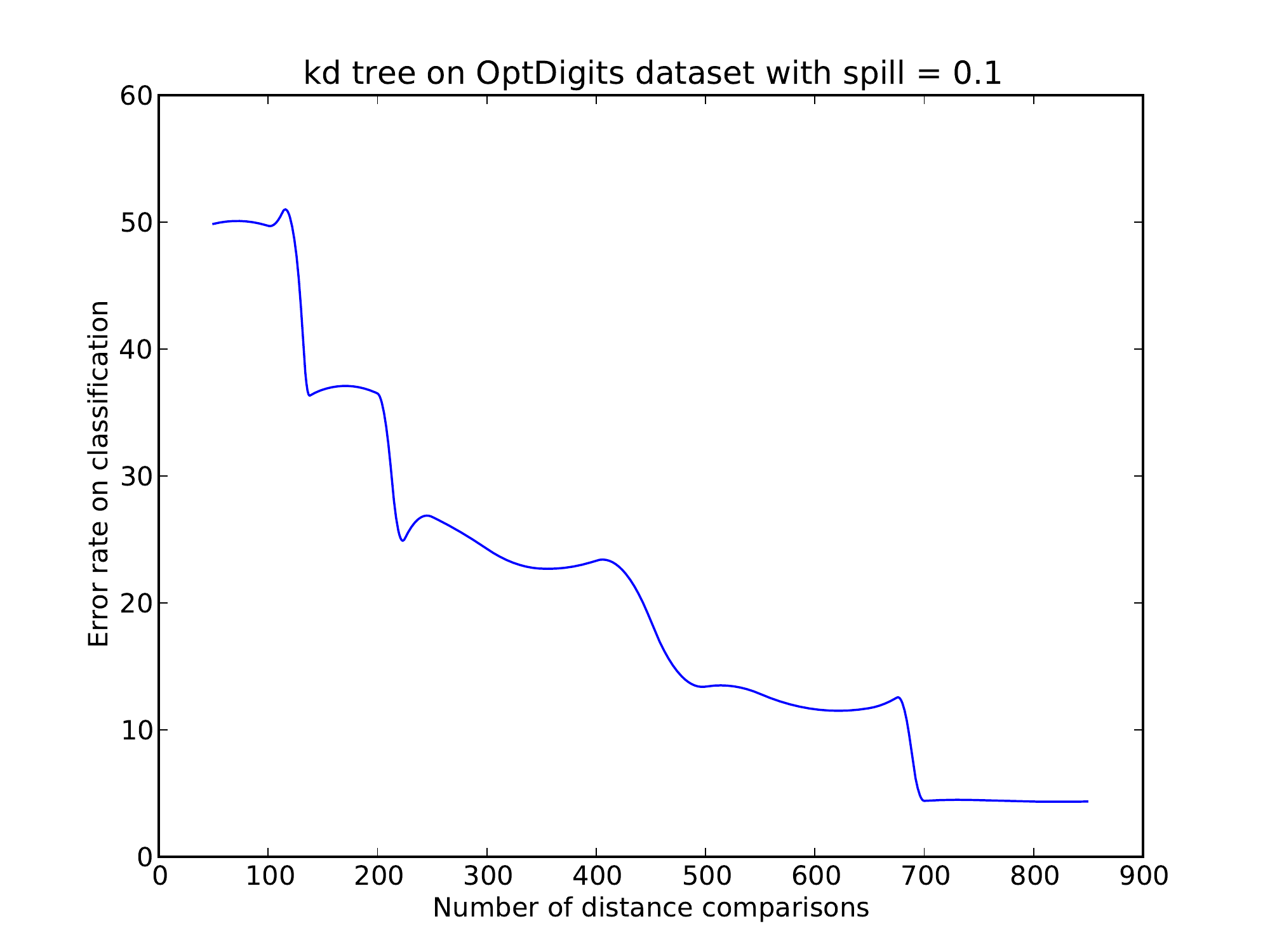}
             
        \end{subfigure}

        \begin{subfigure}[b]{0.3\textwidth}
                \includegraphics[width=\textwidth]{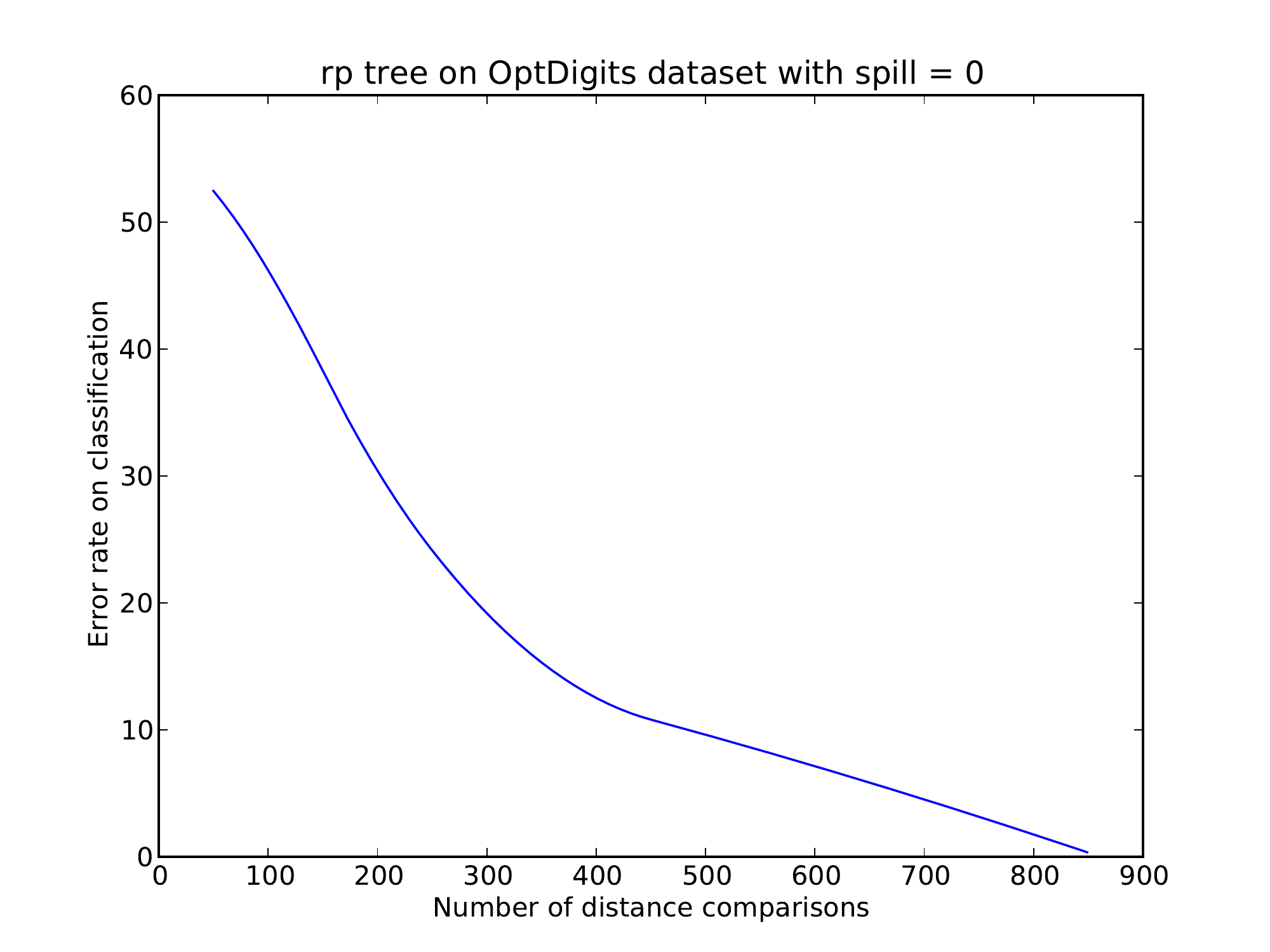}
             
        \end{subfigure}
        ~
        \begin{subfigure}[b]{0.3\textwidth}
                \includegraphics[width=\textwidth]{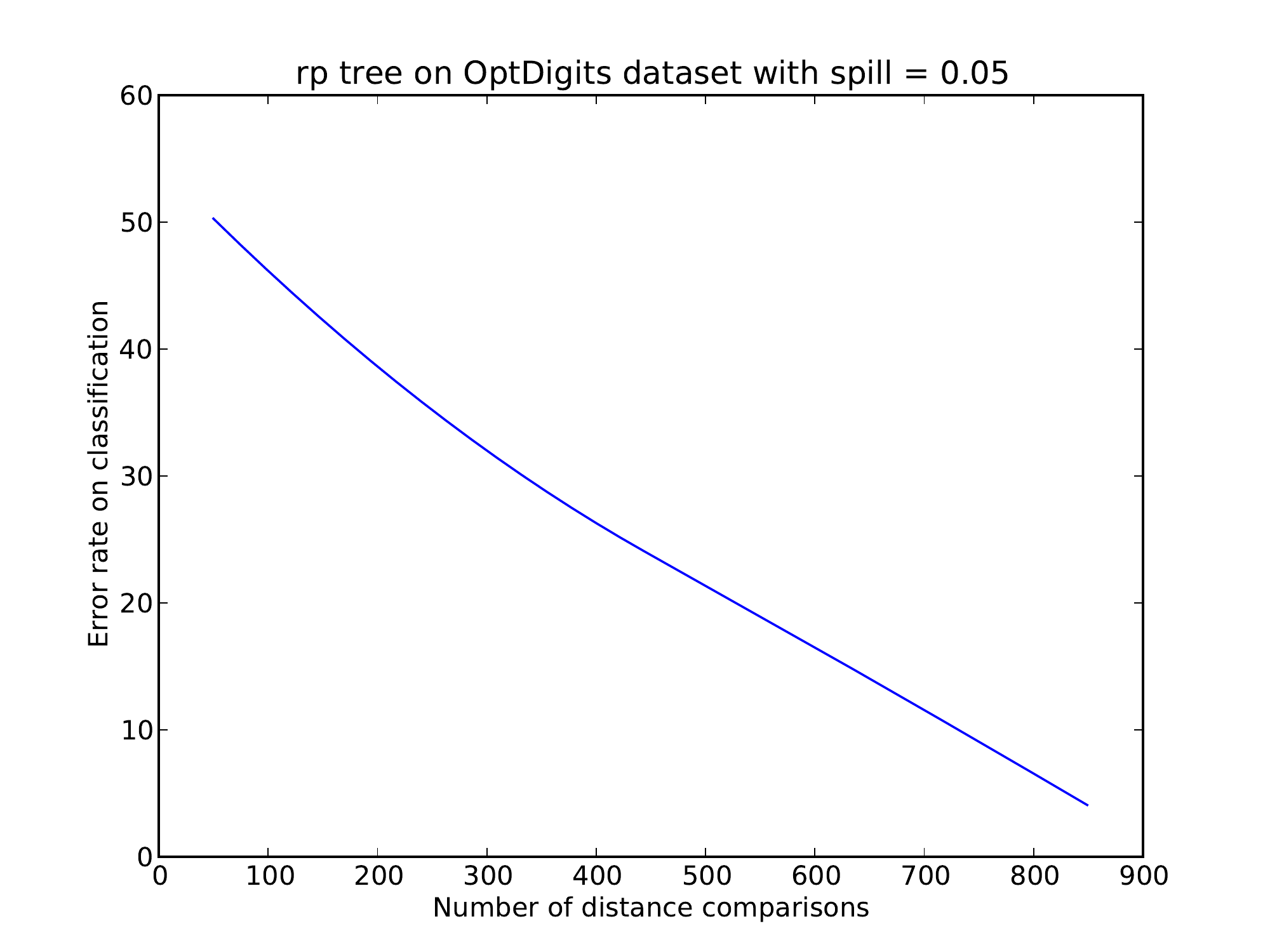}
             
        \end{subfigure}
        ~
        \begin{subfigure}[b]{0.3\textwidth}
                \includegraphics[width=\textwidth]{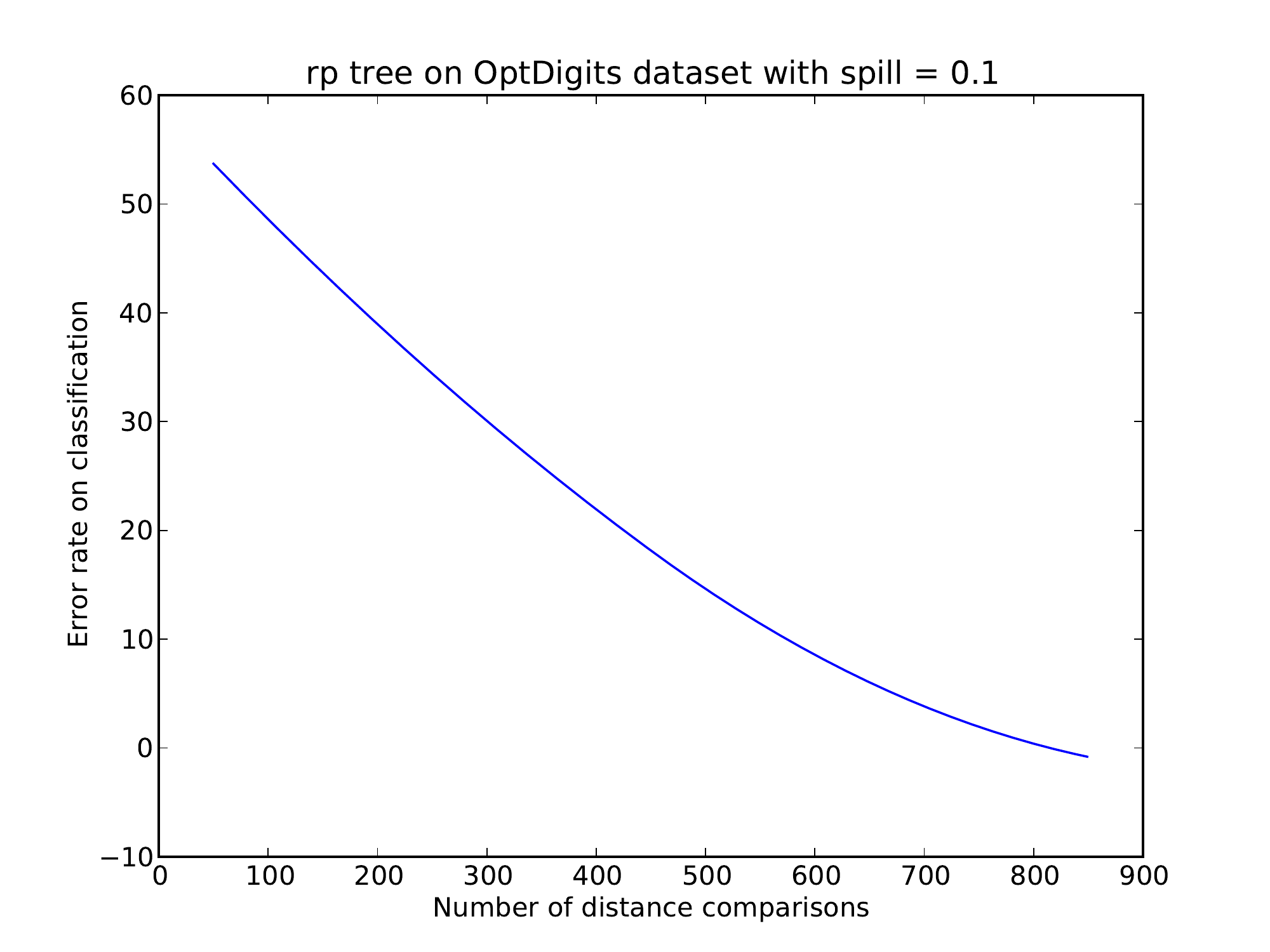}
             
        \end{subfigure}

        \begin{subfigure}[b]{0.3\textwidth}
                \includegraphics[width=\textwidth]{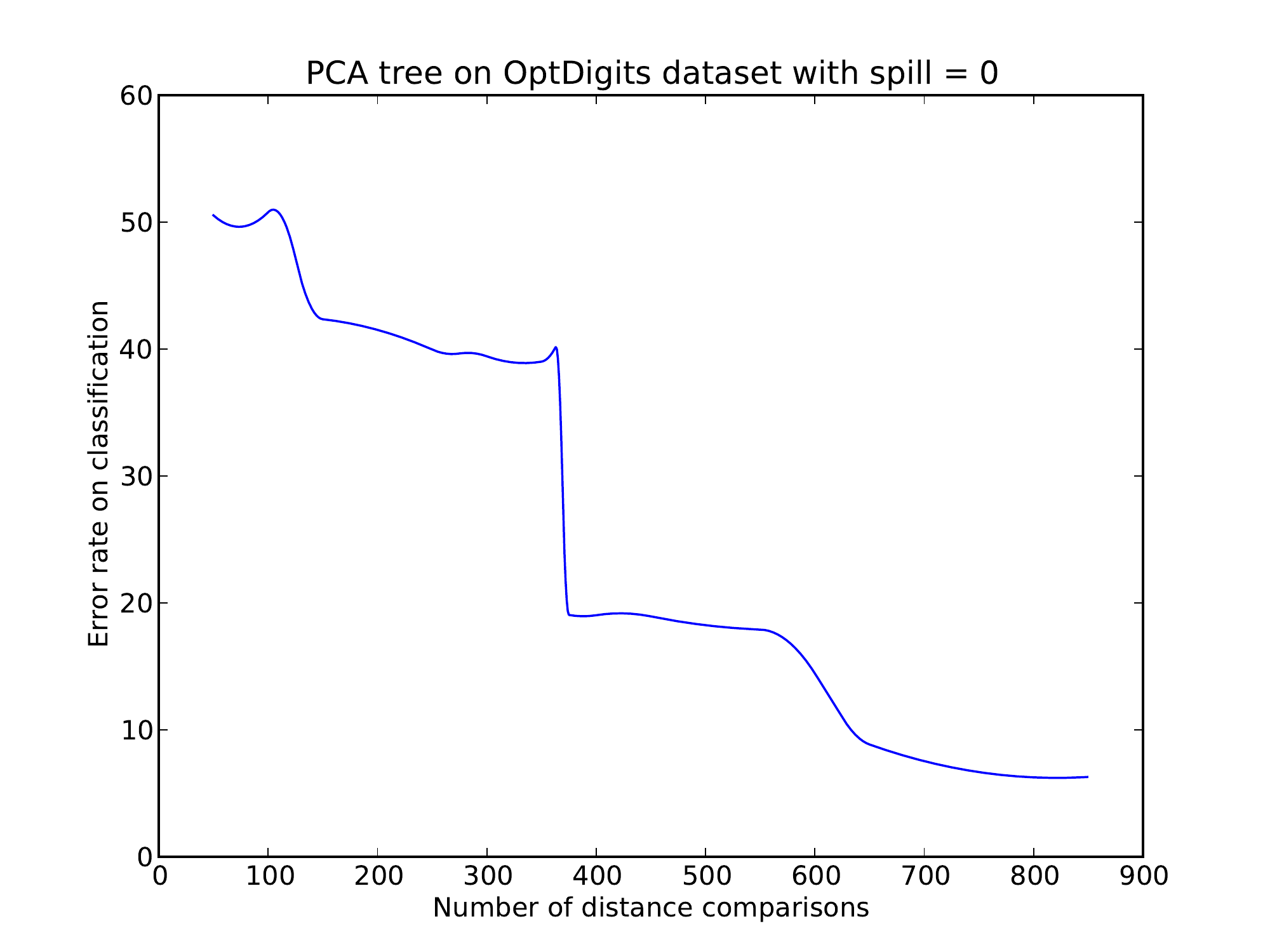}
        \end{subfigure}%
        ~ 
        \begin{subfigure}[b]{0.3\textwidth}
                \includegraphics[width=\textwidth]{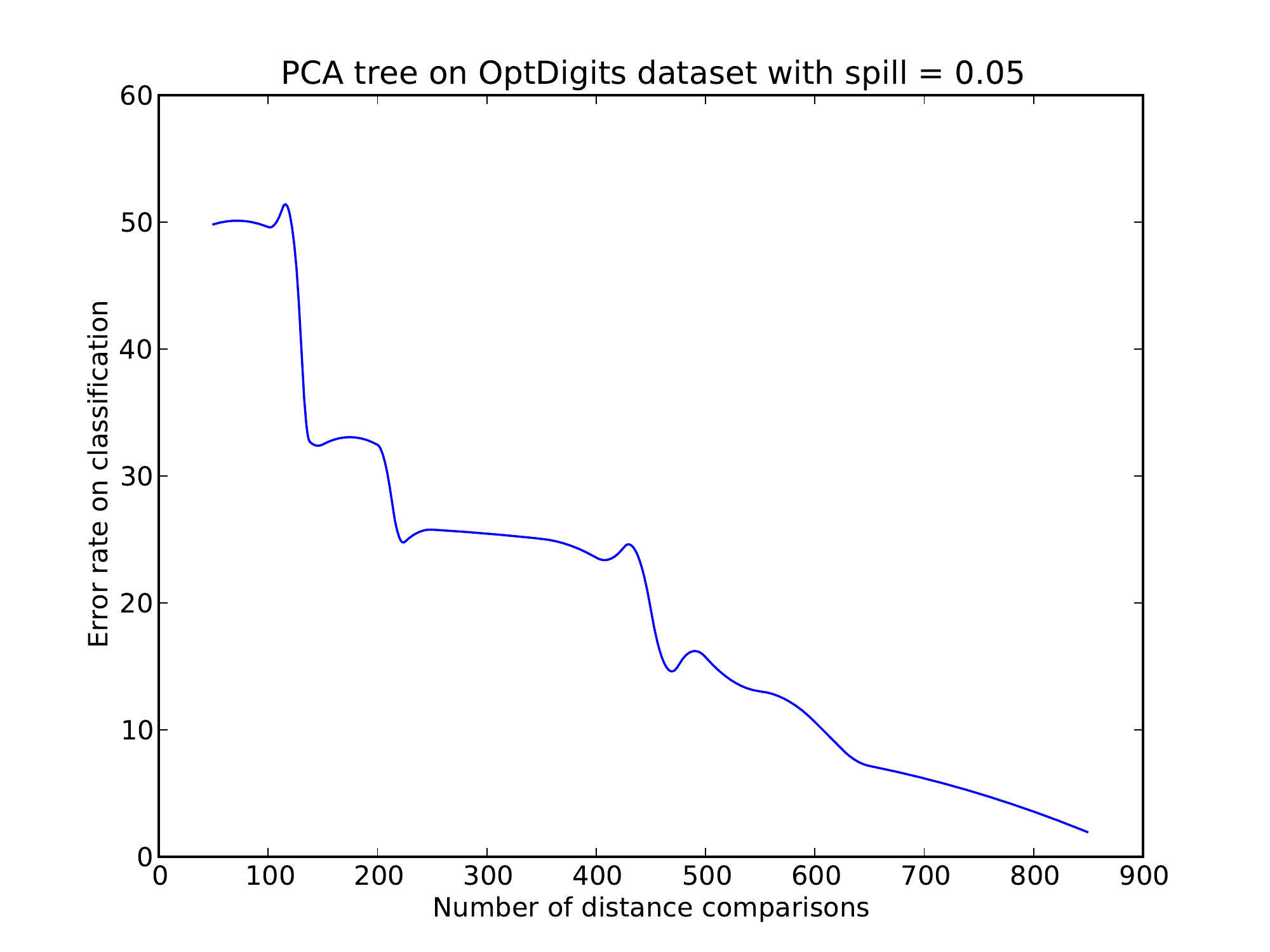}
              
        \end{subfigure}
        ~ 
        \begin{subfigure}[b]{0.3\textwidth}
                \includegraphics[width=\textwidth]{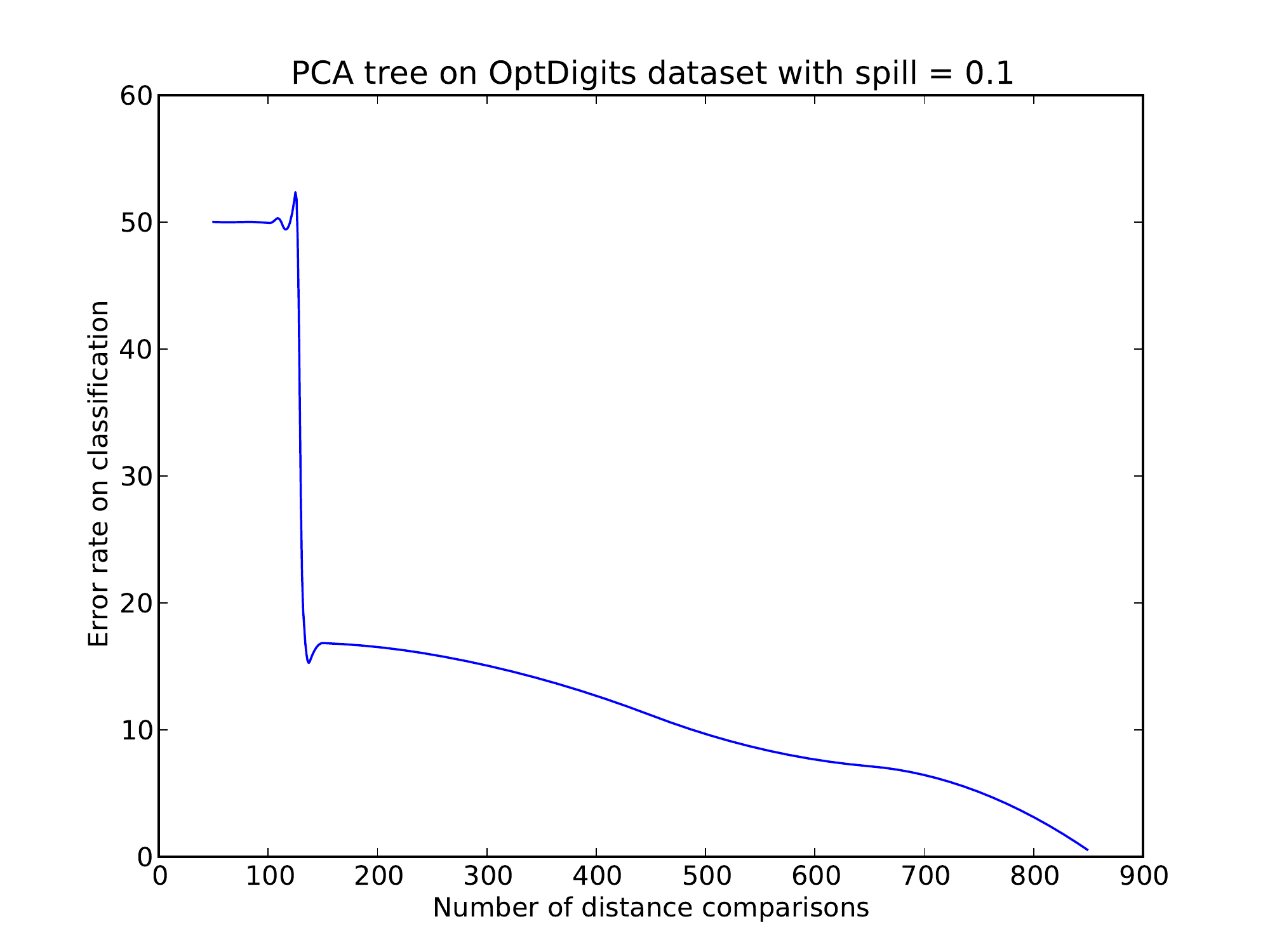}
             
        \end{subfigure}

        \begin{subfigure}[b]{0.3\textwidth}
                \includegraphics[width=\textwidth]{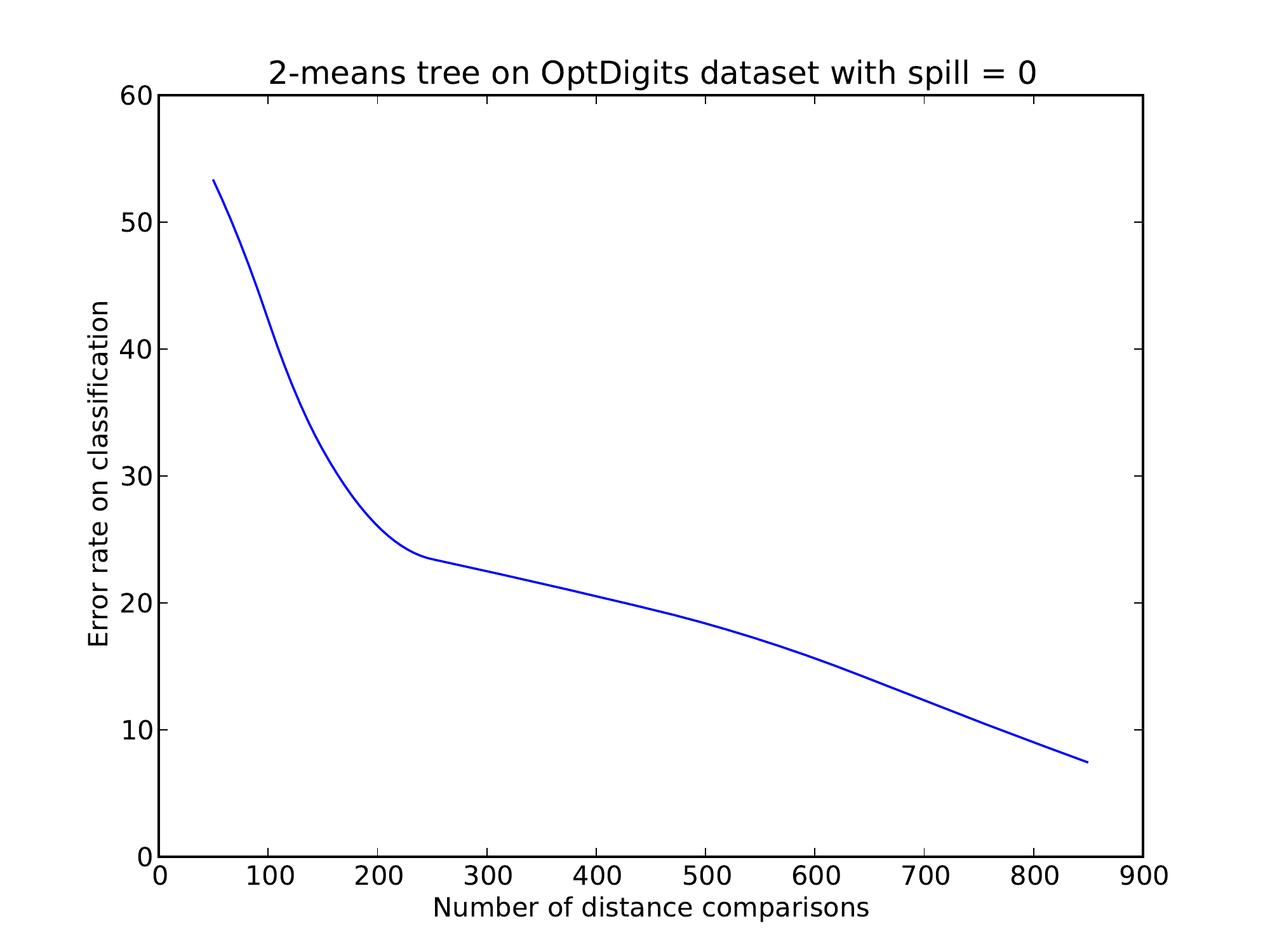}
             
        \end{subfigure}
        ~
        \begin{subfigure}[b]{0.3\textwidth}
                \includegraphics[width=\textwidth]{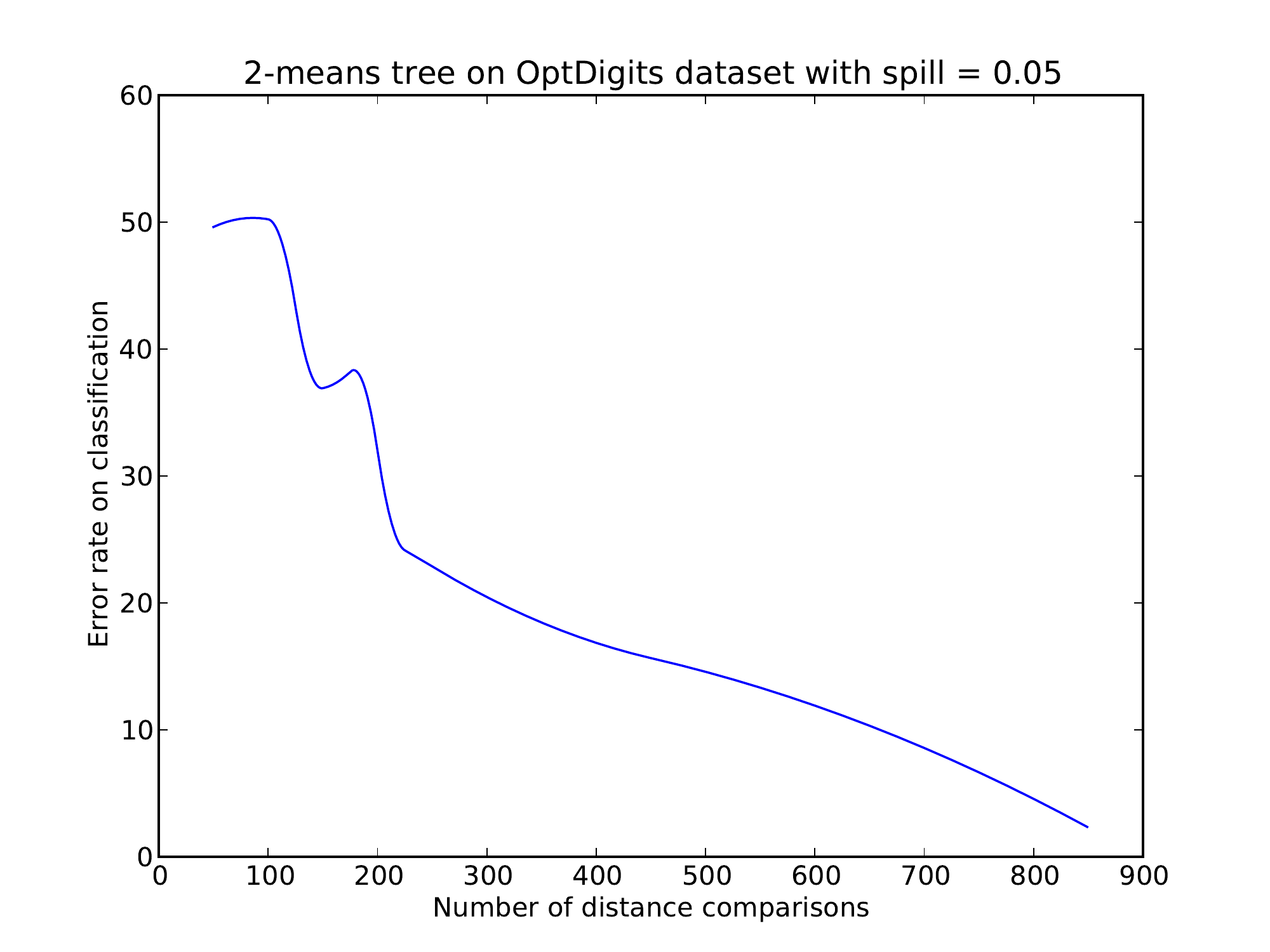}
             
        \end{subfigure}
        ~
        \begin{subfigure}[b]{0.3\textwidth}
                \includegraphics[width=\textwidth]{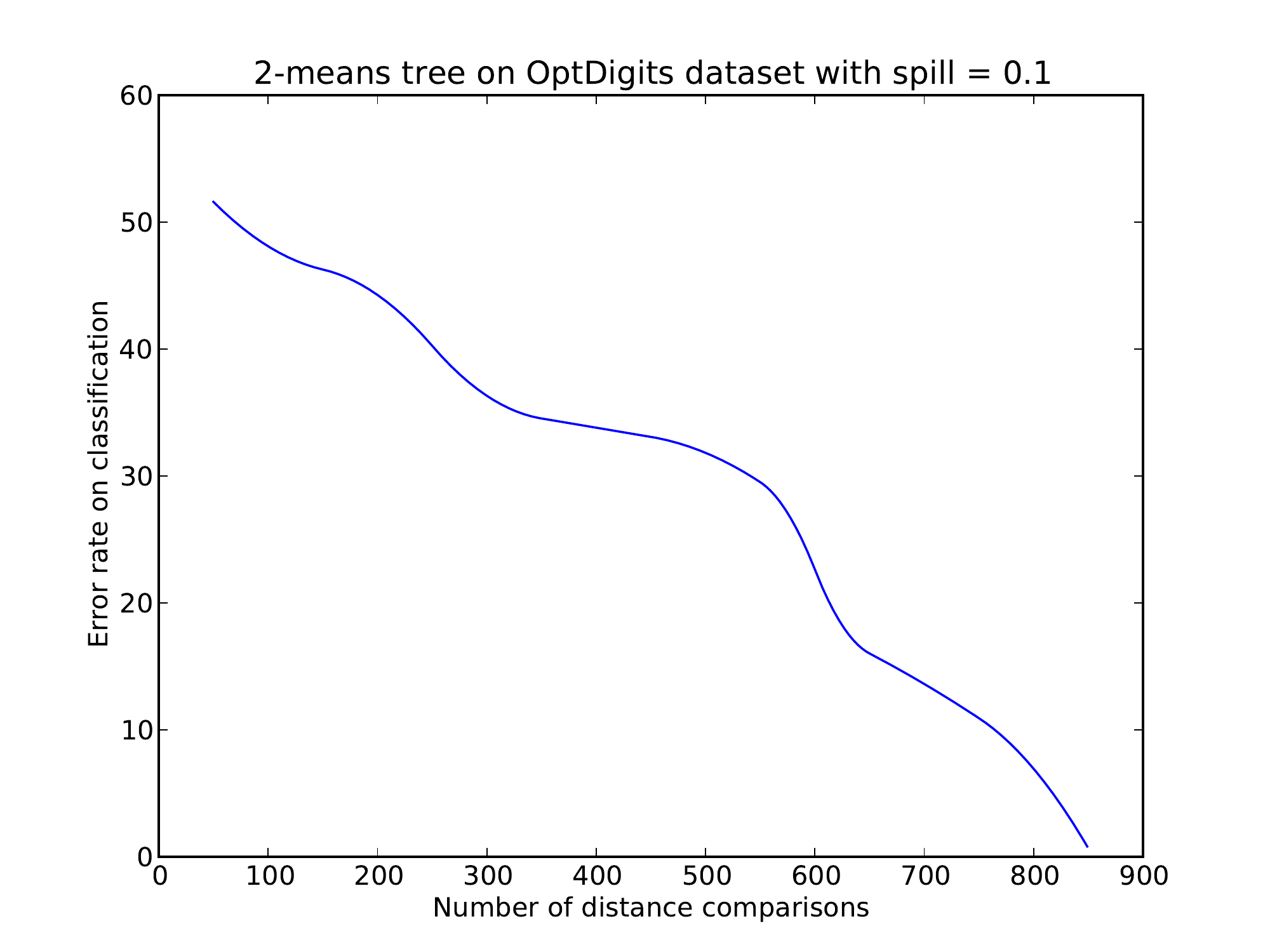}
             
        \end{subfigure}

        \caption{Error Rate on Classification vs number of comparisons for different spatial trees and tree configurations for the OptDigits dataset}
        \label{fig:optc}
\end{figure}

On each dataset we run four different spatial trees $k$-d trees, rp trees, PCA trees and 2-means trees implemented in \cite{ML11}  where we control two parameters the first is spill which we tested for 3 values $0 \%, 0.05 \% , 0.1\% $ and the second is the maximal number of comparisons we will make or the max allowable size of a leaf node. We then plot the number of comparisons made vs the probability of finding the nearest neighbor which we calculate as the ratio of the sum of ranks of the brute force algorithm over the sum of ranks reported by the spatial tree. As an example suppose we're looking for the two nearest neighbors of a query point $q$, brute force search will return the correct ranks 1,2 whose sum is 3. Our datastructure might return the $i$'th and $j$'th nearest neighbor instead. The ratio then becomes $\frac{1 + 2}{i + j}$. We report the results in figure \ref{fig:pima} for the Pima dataset and figure \ref{fig:opt} for the OptDigits dataset. We also judge performance based on classification error using $k$-nn with $k = 10$  vs number of comparisons in figure \ref{fig:pimac} for the Pima dataset and figure \ref{fig:optc} for the OptDigits dataset.

\chapter{Reductions}

This chapter will discuss reductions between a wide class of geometric proximity problems, we will use the notation $P \leq Q$ to say that problem $P$ reduces to $Q$. 

\section{Nearest Neighbor Reductions}
As a general outline, in this section we'll be looking to solve the Euclidean Minimum Spanning Tree problem via a reduction to Bichromatic Nearest Neighbor Search which we prove to be equivalent to Nearest Neighbor Search.
\subsection{Bichromatic Nearest Neighbor}
We introduce the new problem of Bichromatic Bearest Neighbor (BNNS), the setup is very similar to NNS but now every point $x_i$ is also associated with a color $ \chi (x_i) \in \{ 0 , 1 \} $ and we'd like to find the nearest neighbor of a query point $q$ such that the returned point is of a different color from $q$. More formally:

 \[ \argmin_{x \in X \mid \chi(q) \neq \chi(x) } d(p,x) \]

\subsubsection{NNS $ \leq$ BNNS}
The reduction is shown in algorithm \ref{alg:abc}.
\begin{algorithm}

\begin{algorithmic}[1]
       \State X = $\{$ Set $\chi(q) = 0$ and Set $\chi(x) = 1$ for all $x \in X$ s.t $x \neq q$ $\}$
       \State $T_X = BuildTree(X)$
       \State \Return{ $BNNS(q , T_X)$}
\end{algorithmic}
\caption{NNS algorithm via BNNS }
\label{alg:abc}
\end{algorithm}

\subsubsection{BNNS $ \leq$ NNS}
This simple reduction is described in algorithm \ref{alg:abcd}.

\begin{algorithm}

\begin{algorithmic}[1]
  \State $T_X = BuildTree(\{ x \in X : \chi(x) \neq \chi(q) \})$
  \State \Return{ $ NNS(q , T_X ) $}
     
\end{algorithmic}
\caption{BNNS algorithm via NNS}
\label{alg:abcd}
\end{algorithm}

\subsection{Chromatic Nearest Neighbor}
A natural problem that follows from BNNS is Chromatic Nearest Neighbor (CNNS) where we are trying to solve the same problem as BNNS but we could have more than one color or in fact a countably infinite number of colors $\chi(X_i) \rightarrow \mathds{N}$. More practically though the number of colors we can have is bounded by the number of points in our dataset.

\[ \argmin_{x \in X \mid \chi(q) \neq \chi(x) } d(p,x) \]

We will now show an equivalence between the two problems in \ref{alg:cnnsbnns}

\subsubsection{CNNS $\leq$ BNNS}

\begin{algorithm}
\begin{algorithmic}[1]
  \State $T_X = \{ x \in X : \chi(x) \neq \chi(q) \}$
  \State \Return{ $ BNNS(q, T_X) $}
     
\end{algorithmic}
\caption{BNNS algorithm via NNS}
\label{alg:cnnsbnns}
\end{algorithm}

\subsubsection{BNNS $\leq$ CNNS}
This reduction is trivial we just run CNNS, it is shown in \ref{alg:bnnscnns}. CNNS guarantees that $\chi(x) \neq \chi(q)$ no matter how many possible colors we have.

\begin{algorithm}
\begin{algorithmic}[1]
  \State $T_X = BuildTree(X)$
  \State \Return{ $CNNS(q,T_X)$}
     
\end{algorithmic}
\caption{BNNS algorithm via CNNS}
\label{alg:bnnscnns}
\end{algorithm}

\subsection{Euclidean Minimum Spanning Tree}
\begin{figure}[h!]
  \centering
   \includegraphics[scale = 0.5]{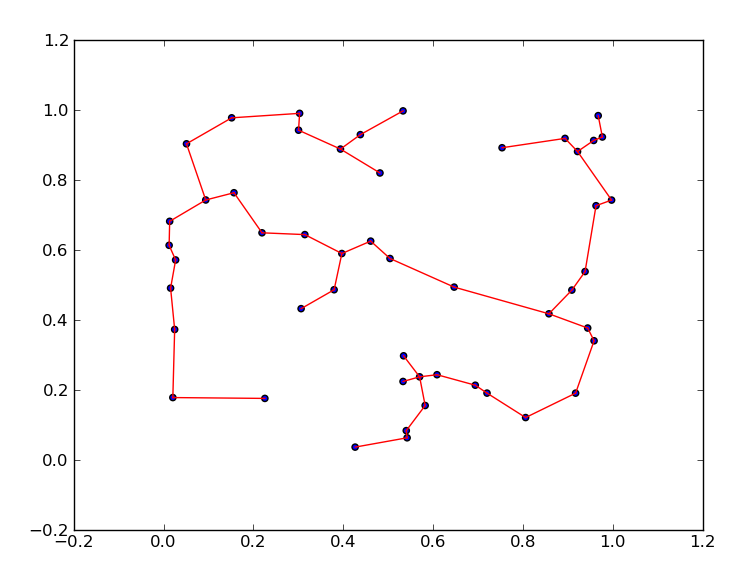}
      \caption{Euclidean MST on random uniform data}
      \label{fig:emst}
\end{figure}

The minimum spanning tree problem is a classic graph problem where given a connected weighted graph $G = (V,E)$ we'd like to find the edges $ST \subset E$ that reach every $v \in V$ such that the weighted sum of the edge $\sum_{e \in ST} w(e)$ is minimized. The Minimum Spanning tree problem is defined over graphs with arbitrary distance functions but we can restrict our attention to the Euclidean $l_2$ norm to recover the Euclidean minimum spanning tree problem. As an example \ref{fig:emst} is a Euclidean MST built on random uniform data.

\subsection{Bor\r{u}vka's algorithm for MST}
In algorithm \ref{alg:boruvka}, we show how to use BNNS to solve the Euclidean Minimum Spanning Tree Problem via Bor\r{u}vka's algorithm. It is also worth noting that we could have adapted the more standard algorithms by Prim or Kruskal like in \cite{Ind00} but the subroutine one would use would be Bichromatic Closest Pair which can be formulated as 
\[
\argmin_{x,x' \in X | \chi(x) \neq \chi(x') }d(x,x') \]
In fact one can trivially solve BCP in $\mathcal{O}(n)$ queries to $BNNS$.

\begin{algorithm}
\caption{Bor\r{u}vka's algorithm for MST}
\label{alg:boruvka}
\begin{algorithmic}[1]
\Procedure{Bor}{$V,E$}
\State $T = (v_1, \dots , v_n)$ for all $v \in V$ \Comment{Initialize $T$ to be the set of one vertex trees}
\While{$|T| > 1$}
  \For{each $C \in$ T} \Comment{$C$ stands for components of $T$}
    \State $S = \phi$ \Comment{$S$ is a set of edges}
    \For{each vertex $v \in C$}
      \State $x' = \argmin_{x \not \in C} d(v,x)$
      \State $S.append(x')$
    \EndFor
    \State $e' = \argmin_{e \in S} w(e)$ \Comment{$w(e)$ is the weight of edge $e$}  
    \State $T.append(e')$

  \EndFor
\EndWhile

\State \Return{$T$} \Comment{$T$ is now the MST of $(V,E)$ }

\EndProcedure

\end{algorithmic}

\end{algorithm}

\begin{theorem}
Euclidean Minimum Spanning Tree can be solved with $\mathcal{O}(n^2)$ queries to Chromatic Nearest Neighbor Search and $\mathcal{O}(n)$ queries to Chromatic Closest Pair with at most $\mathcal{O}(n\log n)$ updates to each data structure.
\end{theorem}

\begin{proof}
To implement the above algorithm we maintain a $CNNS$ structure on our set of points. Whenever we make a call to $x' = \argmin_{x \not \in C} d(v,x)$ we query $CNNS(v,T_X)$ which will guarantee a point from a different cluster and we repeat this $\mathcal{O}|V|$ times. (Of course this last step could also be done with 1 call to a Chromatic Closest Pair Data-Structure). Then whenever we merge two clusters $c_i,c_j \in T$ we have to recolor the points $v \in \argmin_{c_i,c_j} \{|c_i|,|c_j| \}$ i.e we only recolor the points in the smaller of the two clusters. With that trick we will recolor a point at most $\mathcal{O}(\log n)$ times instead of the trivial $\mathcal{O}(n)$. Therefore we will be recoloring all points at most $\mathcal{O}(n \log n)$ times. The correctness of the algorithm follows from the correctness of Bor\r{u}kva's algorithm
\end{proof}

\section{Farthest Neighbor Reductions}

\subsection{Farthest Neighbor Search}

So far we've discussed reductions to nearest neighbor but another natural problem is that of Farthest Neighbor Search. Given $x_1, \dots , x_n = X \subset \mathds{R}^D$ and a query point $q \in \mathds{R}^D$ find its farthest neighbor.

\[ \argmax_{x \in X} d(q,x) \] 

This problem has several curious properties the first is that it is equivalent to finding the minimum enclosing ball centered at $q$, no direct analogue is readily available for the nearest neighbor problem. The minimum enclosing ball problem can be formulated as a convex optimization problem.

\[\min r  \indent \\ s.t \indent d(q,x_i) \leq r \indent  i =  1,\dots, n \]

Farthest Neighbor Search is used as a subroutine for an approximation algorithm for the $k$-centers problem.

\subsection{$k$-centers}

 
We first introduce the $k$-centers problem: Given $n$ i.i.d data points from a set $S \subset X$ where $X$ is a metric space. Find $k$ representative centers of your data set according to the cost function $cost(T) = \max_{x_i} \min_{k \in T}d(x_i,k)$. $k$-center is NP-hard even in $2$-d spaces. The following beautiful  \ref{alg:gonzales} due to Gonzales called farthest  first traversal \cite{Das13} uses furthest neighbor search as a subroutine to approximate the $k$-centers problem.

\begin{algorithm}
\caption{Farthest First Traversal}
\label{alg:gonzales}
\begin{algorithmic}[1]
\State \textbf{Input}: $x_1,\dots,x_n \in R^D$
\State $T = \{ \}$ 
\State pick an arbitrary $x \in X$ 
\Repeat  
\State $z =\arg \max_{x \in S}$  $ d(x,T)$
\State $T = T \cup \{z \}$
\State $x = z$
\Until $|T| = k$
\Return{T} \Comment{List of $k$-centers}
\end{algorithmic}
\end{algorithm}

The algorithm is incredibly simple, pick an arbitrary point and make it a cluster center. Find the farthest point from your original point and make it a cluster center, do this $k$ times and you have a 2-approximation of $k$ center. Finding the farthest neighbor takes $\mathcal{O}(n)$ trivially repeating this $k$ times gives us a running time of $\mathcal{O}(kn)$. Now let's prove why this guarantees a 2-approximation.

\begin{theorem}
Furthest First Traversal is a 2-approximation of the $k$-center problem
\end{theorem}

\begin{proof}
Surprisingly the proof is again very simple. Let's consider what the worst case might look like if we constrain ourselves to only picking cluster centers from our data points. Take two points $x,y$ on a $2D$ plane that are very far away from each other. If we pick a cluster center from our data points then $cost(T) = d(x,y) $ but had we been able to pick a cluster center outside of our data points then the optimal algorithm would have simply set as the cluster center the middle ground between those two points for a $cost(T^*) = d(x,y)/2$. Therefore, $cost(T) \leq 2 cost(T^*)$
\end{proof}

\chapter{Rethinking Nearest Neighbor}

\section{Dual Trees}
The algorithms we discussed above have the same limitation, suppose that we're answering a lot of queries $|Q| = n$ then somehow our pruning rules are redundant (we might re-traverse the tree in a very similar way to find a nearest neighbor). The answer is can we exploit some sort of structure among the points $q \in Q$? The answer is yes we can build another tree for the query points. We call such algorithms dual-tree algorithms and we now show how to use a query tree $T_q$ and a reference tree $T_r$ constructed as is done using cover trees to speedup Nearest Neighbor Search. As an example we mention cover trees \cite{BKL06}, we will not cover the construction of cover trees here but we will mention the interesting invariant that the datastructure guarantees. Given a node at level $j$ we can bound the distance to any of its successors by $\frac{1}{2^j}$. This suggests the following pruning algorithm using cover trees \ref{DualCoverTree}. We denote $N_q$ as nodes in the query tree $T_q$, in the case of cover trees $N_q$ is just a single point $q$. $p \in T_r$ is the best candidate nearest neighbor of $q$ so far and $x$ is the set of all points beneath a given node in $T_r$. 

\begin{algorithm}
\caption{$Prune(N_q,N_r)$}
\label{DualCoverTree}
\begin{algorithmic}[1]

\If{$d(p,q) - \frac{1}{2^j} \leq d(p,x) + \frac{1}{2^j} $ for all $q \in N_q$}
    \State $N_r$ does not need to be explored
\Else 
    \State $Prune(Child(N_q),N_r)$
    \State $Prune(N_q , Child(N_r))$
\EndIf
\end{algorithmic}
\end{algorithm}

This captures the intuition behind if query points are close to one another then they're likely to have similar nearest neighbors, the pruning is conservative so we are sure not to miss any candidates. In what follows we will show that a dual tree can be adapted to any of the tree spatial partitioning algorithms by adding one layer of abstraction in the form of a meta-algorithm.

\section{Meta Algorithms}

In the previous chapter we discussed reductions between geometric proximity problems, the promise is that a good datastructure that performs on any one of these problems could be adapted to work on the others. However, most of the tree datastructures that we've discussed are surprisingly similar and were proposed to be unified under a single meta-tree \cite{CBR13}. They share:

\begin{itemize}
\item A search tree (e.g: $k$-d tree pca tree ...)
\item BaseCase() that determines what is to be done with a combination of points (e.g $\argmin_{x \in X} d(x,q))$
\item Score() that determines whether a certain subtree should be pruned or not (e.g $x_i^d > median^d ... )$
\end{itemize}

Where each node $N$ in the tree contains a convex subset of $S$. As an example, a cover tree falls under this framework because sets consisting of single points are considered to be convex, $k$-d trees also fall under this framework since they partition a space into boxes which are also convex.
\\
The idea here is that if we look at the nearest neighbor literature we notice that there are many different tree based datastructures that were proposed to solve the problem. A typical paper would highlight the shortcomings of some of the past datastructures (more often than not $k$-d trees inadequacy in high dimension) and then propose their own datastructure whose validity they verify with an experimental analysis and a theoretical one. The problem is that there seems to be a lot redundant work and it would be nice to have one meta-algorithm that can reproduce the datastructures we covered in chapter 1 and then analyze this meta-algorithm and have our results automatically carry over to the others. In the two bounds below that draw balls $B$ around a node $N_q$ where $D_q[k]$ is the distance between $q$ and its $k$'th nearest neighbor (so far) and $\mathcal{D}_x^p$ is the set of points that are descendants of the node $N_q$ the $p$ superscript is there to specify that $D_x^p$ is a set of points and not a set of nodes. $\lambda (N_q)$ is the radius of the convex hull of $N_q$ with $2 \lambda (N_q)$ being the diameter i.e the maximum interpoint distance.

\[B_1(N_q) = \max_{x \in \mathcal{D}_x^p} D_q[k] \]

\[B_2(N_q) = \min_{x \in \mathcal{D}_x^p } D_q[k] + 2 \lambda(N_q) \]

It is easy to see why the spatial trees we have so far satisfy these bounds, in particular the cover tree example we covered at the beginning of this chapter uses the second.

\section{Random rp-Tree Forests}
Our experiments show that rp-trees are robust in helping us find the nearest neighbor of a query point $q$ in high dimension. Now instead of just randomly splitting according to a random direction chosen from the unit circle we can build several rp-trees say $k$ of them which will be different with high probability \ref{randomrp}. Then perform $NNS(q,T_i)$ for $i = 1 , \dots , k$ which will each return a leaf node $N_i$ from which we will calculate $\argmin_{x \in \cup_{i = 1}^k N_i} d(q,x)$.

\begin{algorithm}[H]
\caption{Random rp-Tree Forest}
\label{randomrp}
\begin{algorithmic}[1]
\State $i = 0$
\While{$i \leq k$} \Comment{$k$ is the number of trees we will build}
	\State $T_i = BuildRp(X)$
	\State $N_i = NNS(q,T_i)$
	\State $i = i + 1$

\EndWhile
\State $N = \cup_{i = 1}^k N_i$

\State \Return{ $\argmin_{x \in N } d(q,x)$}

\end{algorithmic}
\end{algorithm}

Of course we can also consider cases where every rp-Tree could be a spill tree with a different spill percentage so instead of calling $T_i = BuildRp(X)$ we can call $T_i = BuildRp(X, \alpha_i)$ where $\alpha_i$ is the spill for $T_i$.
The above highlights the power randomness affords us in reasoning about the proximity of points in Euclidean space we summarize these effects below.

\begin{enumerate}
\item Random Projection as a dimensionality reduction preprocessing step via the Johnson-Lindenstrauss Lemma
\item Random Partition of space via Rp Trees
\item Randomness over size of spill
\item Randomness over datastructure via Random Forests
\end{enumerate}
The first point can be understood from \cite{DG02}, the second and third from \cite{DS14}. The fourth introduces some complexity: In fact Rp-Trees only have two hyperparameters to tune, the maximum allowed depth of the tree or the maximum size of a node which is thankfully inversely related to the maximum allowed depth so in practice we only need to tune one of the two parameters. However, with $k$ trees we now have $\mathcal{O}(k)$ more parameters to tune which introduces difficulties both at the theoretical and practical level: its not clear a priori which arrangement of hyperparameters might yield better results. For lack of a clear understanding of the fourth point we leave it here as an open problem.

\section{Bichromatic Closest Pair}
We now introduce the Bichromatic closest pair problem 
\[ \argmin_{x,x' \in X | \chi(x) \neq \chi(x')} d(x,x') \] This problem is central to the EMST reductions in \cite{Ind00} via Kruskal and Prim's. We turn our attention to this problem in this section because it seems like a natural problem for dual trees, we can think of the color of a point $\chi(x) \in \{ 0,1 \}$ as equivalent to setting a point to one of two sets the first a query set $Q$ and the second a reference set $X$.

Below we propose  algorithm \ref{dualBCP} for full BCP , so the algorithm takes as input a list of points $x_1, \dots , x_n \in \mathds{R}^D $ where every point $x_i$ is assigned one of two colors $\chi(x_i) \in \{ 0, 1 \}$ and returns a sorted by value dictionary where the keys are a pair of points and the values are the distances between those two points.

\begin{algorithm}
\caption{Dual Tree BCP}
\label{dualBCP}
\begin{algorithmic}[1]
       \State Set $Q = \{x | \chi(x) = 1 \}$
       \State Set $R =  \{x | \chi(x) = 0 \}$
       \State $L = DualTree(Q,R)$
       \State \Return{ $\argmin_{x,x' \in L} d(x,x')$}
\end{algorithmic}

\end{algorithm}

\section{Batch Nearest Neighbor}
We're again in the setting where we have a set of query points $Q$ and a reference dataset $X$ and for every element $q \in Q$ we'd like to find its nearest neighbor $x \in X$. One of two things could happen, suppose the query points are very well clumped together for any $q_i,q_j \in Q$, $d(q_i,q_j) < \epsilon$ then we obtain an equivalence between batch nearest neighbor and nearest neighbor. The intuition here is that if query points are close to one another then information about the nearest neighbor of one query point implies information about the other points nearest neighbors as well. To see why this is the case, suppose $d(q_i,q_j) < \epsilon$ and $x^* = \argmin_{x \in X}d(q,x) $ this tells us that $d(q_i,q_j) + d(q_i,x^*) \geq d(q_j,x^*)$. If $d(q_i,q_j) < \epsilon \approx 0$ then $d(q_i,x^*) \geq d(q_j,x^*)$.

We can also imagine datasets where the query points might be far apart, in this case query points are unlikely to share nearest neighbors but given that we know the nearest neighbor(s) of a given point $q_i$ and we know that $d(q_i,q_j)$ is large we can prune out the nearest neighbor of $q_i$ when looking for the nearest neighbor of $q_j$. A good approach to pruning here is to draw a ball around every point $q \in Q$ and then given a candidate nearest neighbor that lies outside that ball automatically reject it.

The two points highlighted above will allow us to construct an algorithm for batch nearest neighbor search, the algorithm needs two constants (for the $\beta$ case we can take points that are at least the radius of the convex body away to be very far away so set $\beta = radius(Q)$ the first $\alpha$ sets a threshold where we basically consider two query points $q , q'$ to be nearest neighbor equivalent if $d(q_i,q_j) < \alpha$ and the second $\beta$ that splits one node of the query tree into two subnodes the first comprising of all $q,q'$ s.t $d(q,q') < \beta$ and the second $q,q'$ s.t $d(q,q') \geq \beta$

First we construct the query tree $T_q$ to exploit the two intuitions above in the following way \ref{alg:tqnns}. For example if $\beta = median^i(Q)$ we recover the $k$-d tree.

\begin{algorithm}
\caption{Building a query Tree $BuildT_q(N_q)$}
\label{alg:tqnns}
\begin{algorithmic}[1]

\State Set $root(T_q) = Q$
\While{$| T_q | > constant$}
	\If{ If $d(q,q') < \alpha$}
		\State $Merge(q,q')$ 
	\EndIf
	\State Set $LeftTree = BuildTq(\{q \in Q | d(q,q') < \beta \})$
	\State Set $RightTree = BuildTq(\{q \in Q | d(q,q') \geq \beta \})$ 
\EndWhile
\end{algorithmic}
\end{algorithm}

Now given a query tree $T_q$ we build a reference tree $T_r$ on the dataset $X$ using any reasonable splitting rule and perform nearest neighbor queries on subsets $N_q \subseteq Q$ and $N_r \subseteq X$ \ref{alg:dualnns}. We will have two functions that act on a subset of query points $N_q$

\begin{enumerate}
\item $Merge(N_q)$ which merges all $q \in N_q$ if $Clumped(N_q) = True$
\item $Split(N_q)$ which splits $N_q$ into a right and left subtree if $Spread(N_q) = True$
\end{enumerate}

\begin{algorithm}
\caption{$DualNNS(N_q,N_r)$}
\label{alg:dualnns}
\begin{algorithmic}[1]

\If{$|N_q| = 1$ then return $NNS(N_q,X)$}
\EndIf

\If{$Clumped(N_q) = True$}
	\State  $Merge(N_q)$
\EndIf

\If{ $Spread(N_q) = True$}
	\State
		$Right,Left =  DualNNS((Right,N_r) , DualNNS(Left,N_r))$
\EndIf
\end{algorithmic}
\end{algorithm}

Now let's try to take a closer look at algorithm \ref{alg:dualnns} by introducing a notion of difficulty for batch nearest neighor search.

\section{Measuring the difficulty of Dual Tree NNS}

In a recent paper \cite{DS14} the authors propose a potential function to measure the difficulty of exact NNS. The setting is the usual one we have a query point $q \in \mathds{R}^D$ and a dataset $x_1 , \dots , x_n \in X \subseteq \mathds{R}^D$ and we'd like to find the nearest neighbor to $q$ in $X$
\[ \phi(q, \{ x_1, \dots , x_n \}) = \frac{1}{n} \sum_{i = 2}^n \frac{\| q - x_{(1)} \|}{q - x_{(i)}}  \]
Where $x_{(k)}$ denotes the $k$'th nearest neighbor of $q$. Upon further inspection of this function we can see that when its close to 1 then all the points are more or less the same distance around $q$ and we can expect nearest neighbor queries to be difficult. On the other hand, when $\phi$ is close to 0 then this means that most of the points are far away from the nearest neighbor and intuitevely we'd expect nearest neighbor to become easy. The authors determine in fact that the failure probability of an rp-tree is $\phi \log \frac{1}{\phi}$ and that of a spill tree is $\phi$. Generally though we will be considering nodes $N \subset X$ in our spatial tree so we will add a simple modification:

\[ \phi_m(q, \{ x_1, \dots , x_n \}) = \frac{1}{m} \sum_{i = k + 1}^m \frac{\| q - x_{(1)} \|}{q - x_{(i)}}  \]

 Their results easily extend to the $k$ nearest neigbhor case, a simple modification is made to the potential function $\phi$ which now becomes:

\[ \phi_{k,m}(q, \{ x_1, \dots , x_n \}) = \frac{1}{m} \sum_{i = k + 1}^m \frac{( \| q - x_{(1)} \| + \dots + \| q - x_{(k)})/k \|}{q - x_{(i)}}  \]

The $k$ nearest neighbor expression is unfortunately cluttered so we will drop it and w.l.o.g set $k = 1$ while bearing in mind that the extension does not present any difficulties. For our contribution we first propose an extension to batch nearest neighbor by again simply modifying the potential function $\phi$ where $ x^{q_i} $ denotes the nearest neighbor to $q_i$. We define $x_{(i)}^{q_j}$ as the $i$'th nearest neighbor to query point $q_j$, the potential function for a specific query point is then:

\[  \phi_m(q_i , \{ x_1, \dots , x_n \} ) = \frac{1}{m} \sum_{i = k + 1}^m \frac{\| q_i - x_{(1)}^{q_i} \|}{q_i - x_{(i)}^{q_i}} \]

The potential for $Q$ is just the sum over all the points $q_i \in Q$ (or more generally all query points $q \in N \subset Q$):

\[ \phi_{1}(Q) =   \sum_{q_i \in Q} \phi_{k,m,q_i} \]

The above expression takes into account how easy it is to find the nearest neighbors of a given a set of query points but it does not seem to exploit any structure from the query set $Q$. However, in the dual tree setting we know that the closer query points are the easier batch nearest neighbor queries become and we can represent this intuition as another potential function on the query set $Q$. Here we have two possible candidates either we can look at the average interpoint distance which trivially takes $\mathcal{O}(n^2)$ trivially to compute.

\[ \phi_2(Q) =  \sum_{i = 1}^{n} \sum_{j = i + 1}^{n} \frac{1}{n^2} \| q_{i} - q_{j} \| \]

Alternatively we can use the diameter of $Q$ which again takes $\mathcal{O}(n^2)$ to compute.

\[ \phi_2(Q) = \max_{q_i , q_j} \| q_i - q_j \| \]

We now have all the components we need to rewrite a potential function for exact batch nearest neighbor search.

\[ \phi(Q,X) = \phi_1(Q) \phi_{2}(Q)   \]

The new bound we then propose on batch nearest neighbor search via rp-trees is $\phi_1 \phi_2  \log \frac{1}{\phi_1}  $. $\phi_1$ has a linear dependence with $|Q|$ which is alarming because the previous bounds had no dependence on the number of points in our dataset, but it is intuitive because if $\phi_2$ is large indicates that $Q$ is not well clumped together then we expect the complexity of our problem to increase linearly with the number of query points and should there be a structure to $Q$ our bound will reap its rewards. An algorithm for batch nearest neighbor search would then be one that at every iteration finds
\[\argmin_q \phi_1 \phi_2 \log \frac{1}{\phi_1} \]

and then removes the found query point $q$ from the dataset and performs a linear search instead to find its nearest neighbor(s).

\appendix
\chapter{Mathematical Background}

\begin{definition}
A set $S \subset \mathds{R}^D$ is convex if for any $u_1, \dots , u_k \in S$ and $w_1 , \dots , w_k \geq 0$ s.t $w_1 + \dots + w_k = 1$ 
\[ \sum_{i = 1} ^k u_iw_i \in S \]
\end{definition}

\begin{definition}
The bregman divergence of a function $f$ is
\[d_f(x,y) \equiv f(x) - f(y) - \langle x,y \rangle  \] 
\end{definition}

\begin{lemma}[Johnson-Lindenstrauss]
Given $0< \epsilon < 1$ and a dataset $x_1, \dots , x_n = X \subseteq \mathds{R}^D$, there exists a linear map $f : \mathds{R}^D \rightarrow \mathds{R}^d$ s.t for all $x_i, x_j \in X$

\[(1 - \epsilon) \|x_i - x_j \|^2 < \| f(x_i) - f(x_j)  \|^2 < ( 1 + \epsilon ) \|x_i - x_j \|^2 \]
\end{lemma}

In a nutshell the Lemma says that we can randomly project datapoints on a lower dimensional plane and still not distort interpoint distances too much, a proof of the lemma can be found in \cite{DG02}. The lemma finds its way into many machine learning algorithms as a preprocessing step to limit the curse of dimensionality and spatial trees are no exception.

\bibliographystyle{alpha}

\bibliography{template}


\end{document}